\pdfoutput=1

\documentclass[preprint,3p]{elsarticle}

\usepackage{lineno,hyperref}
\usepackage{bm}
\usepackage{amsmath}
\usepackage{mathtools}
\usepackage{color}
\usepackage{algorithm}
\usepackage{algorithmic}
\usepackage{multirow}
\modulolinenumbers[5]

\usepackage{amssymb}

\usepackage{float}
\usepackage{subfig}
\usepackage{array}
\newcolumntype{C}[1]{>{\centering\arraybackslash}p{#1}}

\usepackage{bigints}

\usepackage{bbding}

\usepackage{units}

\usepackage{xcolor}

\newcommand{\Div}[1]{\nabla \cdot {#1}}
\newcommand{\Grad}[1]{\nabla {#1}}

\usepackage{stmaryrd} 

\newcommand{\avg}[1]{\{\!\{#1\}\!\}}

\newcommand{\jump}[1]{\llbracket {#1} \rrbracket }

\newcommand{\intele}[2]{ \left( {#1},{#2} \right)_{\Omega_{e}} }
\newcommand{\inteleface}[2]{ \left( {#1},{#2} \right)_{\partial\Omega_{e}} }

\usepackage{enumitem}
\setlist[enumerate]{label*=\roman*),ref=\roman*)}

\journal{Journal}

\begin{document}

\begin{frontmatter}

\title{Numerical evidence of anomalous energy dissipation\\ in incompressible Euler flows:\\ Towards grid-converged results for the inviscid Taylor--Green problem}

\author[add1]{Niklas Fehn}
\ead{fehn@lnm.mw.tum.de}
\author[add1]{Martin Kronbichler}
\ead{kronbichler@lnm.mw.tum.de}
\author[add1,add2]{Peter Munch}
\ead{munch@lnm.mw.tum.de, peter.muench@hzg.de}
\author[add1]{Wolfgang A. Wall}
\ead{wall@lnm.mw.tum.de}
\address[add1]{Institute for Computational Mechanics, Technical University of Munich,\\ Boltzmannstr. 15, 85748 Garching, Germany}
\address[add2]{Institute of Materials Research, Materials Mechanics, Helmholtz-Zentrum Geesthacht,\\ Max-Planck-Str.~1, 21502 Geesthacht, Germany}

\begin{abstract}
Providing evidence of finite-time singularities of the incompressible Euler equations in three space dimensions is still an unsolved problem. Likewise, the zeroth law of turbulence has not been proven to date by numerical experiments. We address this issue by high-resolution numerical simulations of the inviscid three-dimensional Taylor--Green vortex problem using a novel high-order discontinuous Galerkin discretization approach. Our main finding is that the kinetic energy evolution does not tend towards exact energy conservation for increasing spatial resolution of the numerical scheme, but instead converges to a solution with nonzero kinetic energy dissipation rate. This implies an energy dissipation anomaly in the absense of viscous dissipation according to Onsager's conjecture, and serves as an indication of finite-time singularities in incompressible inviscid flows. We demonstrate convergence to a dissipative solution for the three-dimensional inviscid Taylor--Green problem with a measured relative~$L_2$-error of~$0.27 \%$ for the temporal evolution of the kinetic energy and~$3.52 \%$ for the kinetic energy dissipation rate.
\end{abstract}


\end{frontmatter}


\section{Motivation}\label{Motivation}
Singularities play a key role in fluid mechanics~\cite{Eggers2018}. While singularities in the form of shocks are well-understood for the compressible Euler equations and the inviscid Burgers equation~\cite{Burgers1948} as a simplified model, the occurrence of singularities for the incompressible Euler equations in three space dimensions that develop in finite time is discussed controversially. The occurrence of finite-time singularities is strongly related to anomalous dissipation of kinetic energy in three-dimensional incompressible Euler flows according to the pioneering work by Onsager~\cite{Onsager1949}, which is well-documented in the review articles~\cite{Eyink2006,Eyink2008} and in the recent essay~\cite{Dubrulle2019}. Due to the relation between singularities and dissipation, we distinguish between (i) a \textit{direct} approach to identify finite-time singularities for incompressible Euler flows, e.g., by showing that the vorticity blows up in finite time through different methods (e.g., an analysis of the kind~$\Vert \boldsymbol{\omega} \Vert_{\infty} \sim (t_* - t)^{-\gamma}$ according to the Beale--Kato--Majda theorem~\cite{Beale1984} trying to identify~$t_*$ and~$\gamma$ from numerical results), and (ii) an \textit{indirect} approach proposed in the present work that provides indications of finite-time singularities by observing an ``anomalous'' dissipative behavior in the kinetic energy evolution. As detailed below, most approaches in the literature can be identified as belonging to the first category, while we here focus on a new technique related to the second category. To complement these results, we additionally show numerical results related to the direct identification approach such as the temporal evolution of the maximum vorticity~$\Vert \boldsymbol{\omega} \Vert_{\infty}$ and the enstrophy~$\mathcal{E}$. A strategy to identify potential singularities directly are time series expansions shown in~\cite{Brachet1983, Pelz1997, Taylor1937, Morf1980}, but it was found that numerical inaccuracies prevent a definite answer when using this technique. Numerical investigations by means of PDE solvers have played the most dominant role in the exploration of finite-time singularities as detailed below.

\subsection{State-of-the-art and limitations in tracing finite-time singularities}
A difficulty in identifying singularities with the direct approach is the inherent conflict that arbitrarily small structures can not be resolved with a numerical simulation of finite resolution, and therefore renders this problem one of the most challenging topics in computational fluid dynamics. Much work has been done in this field. In the 1980s and 1990s, several early works on direct numerical simulation of both inviscid and high-Reynolds-number viscous incompressible flows reported indications of finite-time singularities for the incompressible Euler equations~\cite{Brachet1983, Kerr1989, Kerr1993, Brachet1992, Boratav1994}. Symmetry in the initial conditions plays an important role as specifically mentioned and addressed by some works~\cite{Boratav1994, Pelz1997, Pelz2001}, raising the question whether singularities are possible for problems that are not perfectly symmetric. However, these works do not allow a definite answer to the question of finite-time singularities, see also the review articles on this topic~\cite{Gibbon2008, Hou2008}. One of the main reasons why the results of these studies have been inconclusive is that the spatial resolution has been limited due to the computational power and computational approaches available at the time. Numerical results shown in~\cite{Hou2008} suggest that dynamic depletion of vortex stretching could be a mechanism that prevents a finite-time blow-up, but the same authors report evidence for a finite-time singularity for a different flow configuration with solid boundaries in a later work~\cite{Luo2014}.  In terms of the flow configuration being studied, numerical investigations on finite-time singularities can be categorized as follows: The Taylor--Green vortex has been analyzed in~\cite{Brachet1983, Brachet1992, Brachet1991, Shu2005, Cichowlas2005, Bustamente2012} and for a regularized problem considering the Euler--Voigt equations in~\cite{Larios2018}, the high-symmetry Kida--Pelz initial condition in~\cite{Hou2008, Cichowlas2005, Grafke2008}, colliding Lamb dipoles in~\cite{Orlandi2012}, and other perturbed cylindrical vortex tubes in~\cite{Kerr1989, Kerr1993, Hou2008, Grauer1998, Kerr2013}. Most studies used spectral methods as discretization schemes. For the direct numerical simulations, common approaches to trace singularities are monitoring the maximum vorticity~$\Vert \boldsymbol{\omega} \Vert_{\infty}$ over time, see the Beale--Kato--Majda theorem~\cite{Beale1984}, and the ``analyticity strip'' method, see~\cite{Sulem1983}, which aims at capturing the smallest scales of the flow. The width of the analyticity strip~$\delta (t)$, obtained from fitting the energy spectrum to~$E(k,t) = C(t) k^{-n(t)}\exp(-2 k \delta (t))$, is monitored over time for successively finer spatial resolutions up to a resolution for which extrapolations of~$\delta (t)$ allow conclusions whether~$\delta (t)$ reaches~$0$ in a finite-time (finite-time singularity) or whether~$\delta (t)$ decreases only exponentially in time (regularity at all times). Numerical results for the three-dimensional inviscid Taylor--Green vortex shown in~\cite{Brachet1983, Cichowlas2005} indicate only an exponential decay, but this might be due to the limited spatial resolution and also due to the fact that only small times of the TGV flow have been considered, so that a finite-time singularity can not be excluded from these results. In a later work~\cite{Bustamente2012}, a change in regime indicating potentially faster than exponential decay is reported and the results are ``not inconsistent with the occurrence of a singularity'', but again resolutions higher than the maximum one of~$4096^3$ would be required for definite answers. The development of pancake-like structures with exponentially growing vorticity during the early development of turbulence from smooth initial data is studied in~\cite{Agafontsev2015}. In~\cite{Kerr2013}, a new kind of analysis based on rescaled vorticity moments is proposed studying anti-parallel vortex tubes, and only double-exponential growth in vorticity is observed as opposed to the singular behavior suspected in a previous work~\cite{Kerr1993}. In~\cite{Cichowlas2005} it is estimated that conclusions regarding finite-time singularities using the analyticity strip method would require spatial resolutions of~$(16\mathrm{k})^3$ to~$(32\mathrm{k})^3$ for the Kida--Pelz initial data. A recent study~\cite{Campolina2018} suggests that the resolution available via classical DNS is not sufficient to investigate blow-up. A model describing a cascade of transformations between vortex filaments and sheets potentially explaining the mechanism of singularity formation in the Euler equations is proposed in~\cite{Brenner2016}.

\subsection{Energy dissipation anomaly or the zeroth law of turbulence}
We now focus onto the evolution of kinetic energy in incompressible Euler flows. Of particular interest is the question whether inviscid flows are able to dissipate energy, and if so, by which mechanism such a behavior can be explained, given that no viscous effects are present. The kinetic energy dissipation equation valid for incompressible viscous~($\nu > 0$) flows with continuously differentiable solution on a domain with periodic boundaries reads~\cite{Onsager1949,Eyink2006}
\begin{align*}
\frac{\partial E(t,\nu)}{\partial t} = - \int_{\Omega} \nu \nabla \bm{u} : \nabla \bm{u} \;\mathrm{d}\Omega \ .
\end{align*}
From phenomenological descriptions of turbulence, there is empirical evidence that the dissipation rate does not tend to zero in the limit~$\mathrm{Re}\rightarrow \infty$ or~$\nu \rightarrow 0$ but takes a value independent of~$\nu$, which is known as dissipation anomaly or the zeroth law of turbulence~\cite{Eyink2008,Dubrulle2019}. As noted in~\cite{Eyink2008}, this has first been observed by Taylor~\cite{Taylor1935}, and also Kolmogorov's similarity theory of turbulence~\cite{Kolmogorov1941} is based on the assumption of a non-vanishing energy dissipation rate in the inviscid limit. Numerical evidence that the dissipation rate is independent of~$\nu$ for large~$\mathrm{Re}$ is for example given in~\cite{Sreenivasan1998,Kaneda2003,Orlandi2012}, and experimental evidence for example in~\cite{Dubrulle2019}. It can then be conjectured that the dissipation rate in the inviscid limit equals the viscous dissipation rate in the limit~$\nu \rightarrow 0$
\begin{align*}
\frac{\partial E_{\nu=0}(t)}{\partial t} \overset{?}{=} \lim_{\nu \rightarrow 0} \frac{\partial E(t,\nu)}{\partial t} = \lim_{\nu \rightarrow 0} -  \nu \int_{\Omega} \nabla \bm{u}^{\nu} : \nabla \bm{u}^{\nu} \;\mathrm{d}\Omega = - D(t) \overset{?}{\leq} 0 \ ,
\end{align*}
with anomalous energy dissipation for~$D(t) > 0$. The question marks indicate that opinions in the literature are ambivalent regarding these questions. As a consequence of the above equation when taken as valid, the enstrophy would be inversely proportional to the viscosity for large Reynolds numbers in the case of anomalous dissipation. The question that then comes up is whether there is a theory available that can explain energy dissipation in the absence of viscosity. According to Onsager~\cite{Onsager1949}, energy dissipation in three-dimensional incompressible flows can take place in the absence of viscosity by the formation of singularities, with the cascade from large to (arbitrarily) small scales taking place in finite time, see also~\cite{Eyink2006}. According to Onsager's conjecture, energy is conserved if the velocity is H\"older continuous with exponent~$> 1/3$. Hence, the effect explaining the occurrence of kinetic energy dissipation in the limit~$\nu \rightarrow 0$ is that the velocity gradient might tend to infinity in this limit.~\footnote{Literally, Onsager wrote~\cite{Onsager1949}: ``In the absence of viscosity, the standard proof of the conservation of energy does not apply, because the velocity field does not remain differentiable!''. Interestingly, Onsager did not consider the energy-dissipating behavior of inviscid flows (``ideal turbulence'') an anomalous behavior but rather a matter of fact.} As noted in~\cite{Dubrulle2019}, the original Kolmogorov cascade picture implies irregularities of the velocity field (at least locally) with H\"older exponent~$\leq 1/3$, but it was Onsager who established the link between energy dissipation and irrelgularities of the velocity field for the Euler equations. For mathematical literature dealing with proofs of Onsager's conjecture we refer to~\cite{Buckmaster2016,Buckmaster2017} and references therein. The above considerations might explain why this phenomenon is denoted as ``kinetic energy dissipation anomaly'', an alternative term used e.g.~in~\cite{Dubrulle2019} is ``inertial dissipation'' (as opposed to viscous dissipation). The one-dimensional inviscid Burgers equation~\cite{Burgers1948} with formation of a shock and the associated dissipation of energy serves as a prominent and well-understood example, and is for example discussed in~\cite{Sulem1983} in the context of finite-time singularities and in~\cite{Dubrulle2019} in the context of inertial energy dissipation. For the two-dimensional incompressible Euler equations, it is known that these equations can not develop singularities in finite time from smooth initial data~\cite{Eyink2006}. Regarding three-dimensional turbulent flows, Onsager's conjecture appears to be widely accepted by now with the occurrence of singularities representing a building block of modern understandings of turbulence~\cite{Dubrulle2019}.

\subsection{Interplay between physics and numerics}\label{sec:IntroInterplayPhysicsNumerics}

We turn back to the numerical solution of turbulent flows and particularly the inviscid limit. Having a closer look at numerical simulations of the inviscid Taylor--Green vortex flow which we focus on in the present work, it can be observed that these simulations have been performed mainly for small times up to~$t\approx 5$ (up to~$t=4$ in~\cite{Brachet1983,Brachet1992,Cichowlas2005, Bustamente2012} and up to~$t\leq 6$ in~\cite{Shu2005,Chapelier2012}), but not beyond the time at which finite-time singularities have been suspected, especially not up to the time at which the transition to a fully turbulent state takes place, with maximum kinetic energy dissipation rate at time~$t \approx 9$ (expected from high-Reynolds number viscous simulations, see~\cite{Brachet1983}), and subsequent decaying turbulence. As mentioned by some works, one reason for this is that the results for a specific resolution are no longer reliable at later times once the flow becomes under-resolved. Related to this aspect are computational resources which limit the spatial resolution no matter how powerful a computer is. A plausible explanation might also be a lack of robustness of the numerical discretization scheme for this test case, see for example~\cite{Chapelier2012,Winters2018},  and the incompressible Euler equations in general, which are very challenging in terms of energy stability and numerical blow-up of the discretization scheme. Tightly coupled to this aspect is a lack of understanding of what is to be expected in terms of the temporal evolution of the kinetic energy from a physical perspective (energy dissipation anomaly discussed above), and from a numerical perspective. In~\cite{Moura2017, Moura2017setting, Piatkowski2019}, it is argued that a fundamentally different behavior in terms of energy dissipation and time reversibility is expected between viscous flows in the limit~$\mathrm{Re}\rightarrow \infty$ and inviscid flows at~$\mathrm{Re} = \infty$. Especially in numerical studies, it is often expected that energy conservation holds for an exact solution of the Euler equations~(since $\nu = 0$), see for example~\cite{Shu2005, Bustamente2012, Grauer1998, Chapelier2012, Winters2018, Schroeder2019, Krais2020} and the recent review article~\cite{Coppola2019} to mention just a few. It is therefore often considered a desirable quality criterion if a numerical method preserves the kinetic energy exactly in the inviscid limit~$\nu = 0$. Numerical schemes that are exactly energy conserving can indeed be constructed and are widely used in this context. Inviscid TGV simulations performed in~\cite{Schroeder2019} using exactly divergence-free, energy-conserving discretization methods result in an exact conservation of energy, and the results are considered superior as compared to simulations with upwind fluxes that show a dissipative behavior. The results shown in the present study point in a fundamentally different direction as discussed in more detail in Sections~\ref{sec:IntroEnergyConsiderations} and~\ref{Results} below. If Onsager's hypothesis is true, it is clear that energy-conserving numerical methods would result in an~$\mathcal{O}(1)$ error in the case of inviscid flows with anomalous/inertial energy-dissipation (due to non-smooth problems with singularities in the velocity gradient or vorticity according to Onsager's conjecture). In terms of the kinetic energy spectrum, an energy conserving numerical scheme can be expected to lead to an unphysical equipartitioning of energy when simulating beyond the time of the finite-time singularity~\cite{Orlandi2009, Orlandi2012}. Due to this hypothetical inconsistency between the physical behavior and what such a numerical method can provide under Onsager's conjecture, one can conclude that the application of energy-conserving numerical methods is only reasonable for times~$t < t_*$ before the singularity forms and where no dissipation is expected physically. The situation is less complicated for two-dimensional Euler flows that are non dissipative. In that case, it can be expected that the kinetic energy dissipation rate converges to zero under mesh refinement for a consistent and energy-stable discretization scheme, and that there is no conflict with physics if an energy conserving scheme is applied. For the numerical investigation of three-dimensional problems with the open problem of finite-time singularities and the mechanism of energy dissipation, the dissipation behavior of a discretization scheme is therefore of outmost importance as a means to numerically represent the anomalous dissipation and to obtain results that are physically meaningful.

\subsection{Identifying finite-time singularities by energy considerations}\label{sec:IntroEnergyConsiderations}

\begin{figure}[t]
 \centering
	\includegraphics[width=0.49\textwidth]{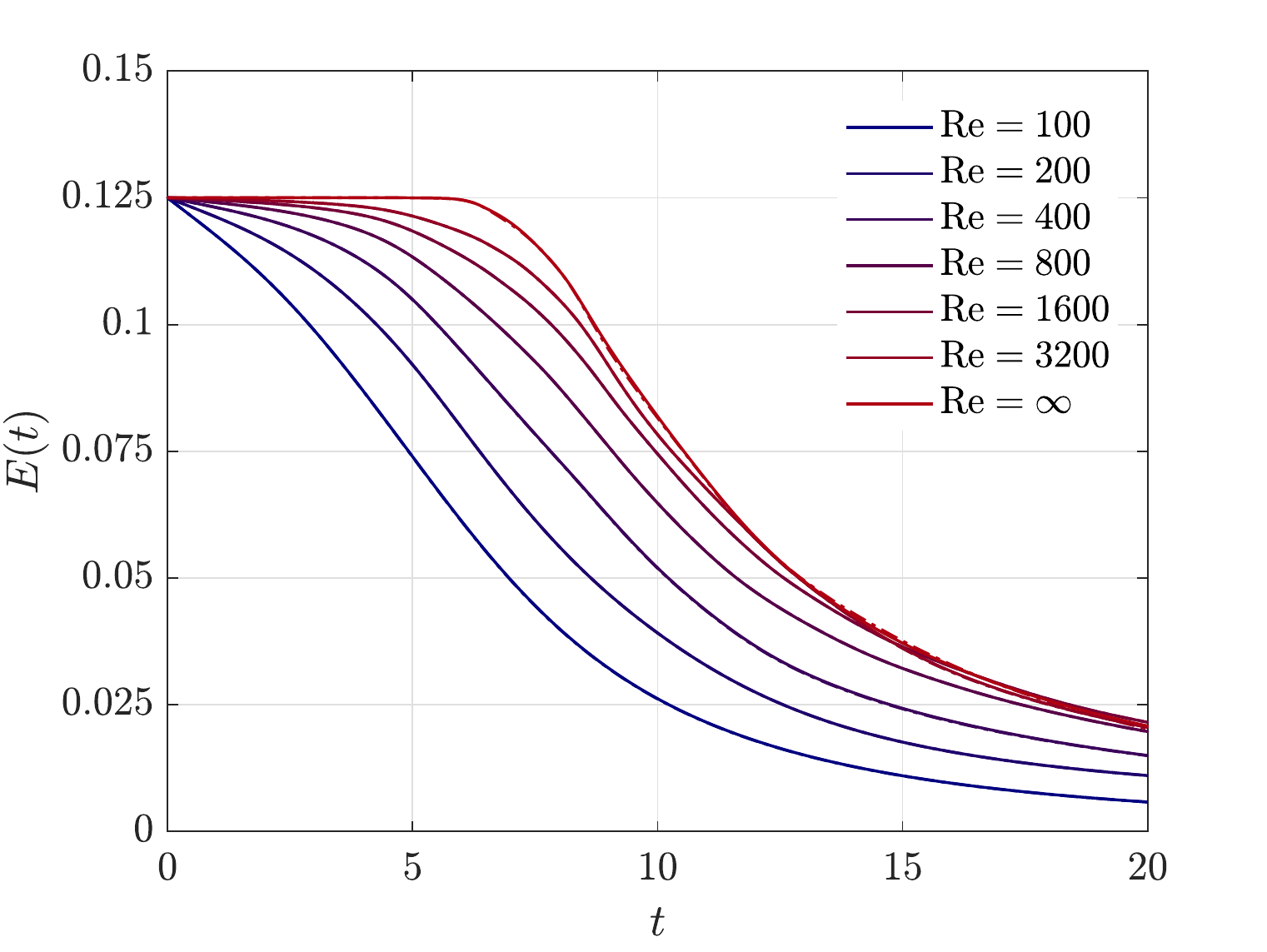}
	\includegraphics[width=0.49\textwidth]{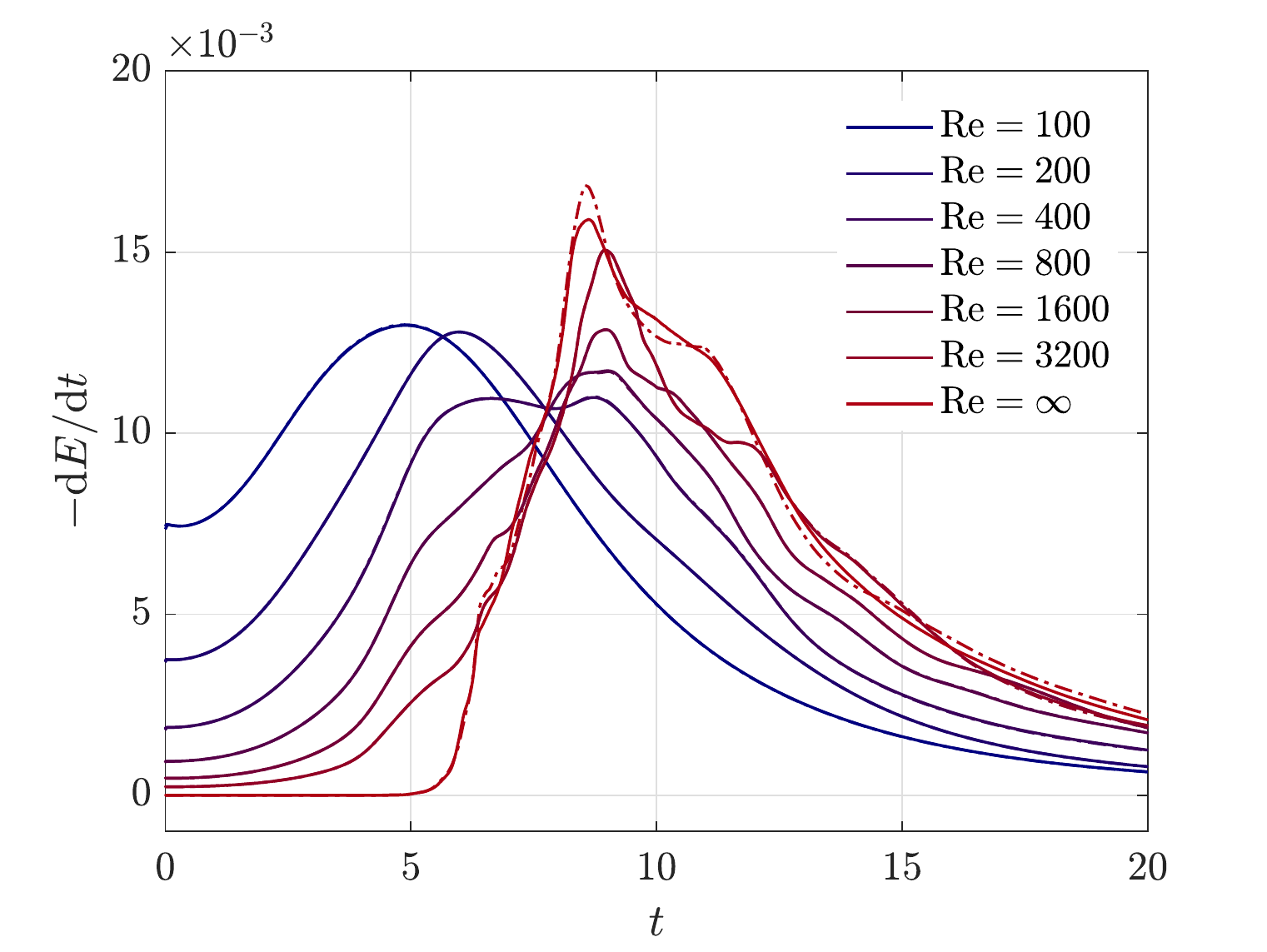}
\caption{Temporal evolution of kinetic energy and kinetic energy dissipation rate for the three-dimensional Taylor--Green vortex problem for increasing Reynolds number of~$\mathrm{Re}=100,200,400,800,1600,3200,\infty$. For each~$\mathrm{Re}$, results are shown for two mesh resolutions (fine mesh as solid line, coarse mesh as dashed-dotted line). The effective resolutions (see Section~\ref{Results} for a definition) are~$64^3, 128^3$ for~$\mathrm{Re}=100$,~$128^3, 256^3$ for~$\mathrm{Re}=200, 400$,~$256^3, 512^3$ for~$\mathrm{Re}=800$,~$1024^3, 2048^3$ for~$\mathrm{Re}=1600$,~$2048^3, 4096^3$ for~$\mathrm{Re}=3200$, and~$4096^3, 8192^3$ for~$\mathrm{Re}=\infty$. Note that while for small and moderate Reynolds numbers it appears to hold that~$E_{\mathrm{Re}_1}(t) \leq E_{\mathrm{Re}_2}(t) \, \forall \, 0 \leq t \leq 20$ if~$\mathrm{Re}_1 < \mathrm{Re}_2$, we obtain converged results demonstrating that this does no longer hold for the pair~$\mathrm{Re}_1=1600,\mathrm{Re}_2=3200$. More importantly, the results suggest that the kinetic energy reduces to a value as low as approximately~$0.02$ at time~$t=20$ for large Reynolds numbers, and that a similar amount of energy dissipation takes place even in the inviscid limit.}
\label{fig:3d_tgv_Re_study}
\end{figure}

With the energy dissipation anomaly in mind and its connection to finite-time singularities, the \textit{indirect} approach to identify finite-time singularities proposed here is to capture the temporal evolution of the kinetic energy by a numerical method with appropriate inbuilt dissipation mechanisms. The idea behind this approach is that if mesh-convergence in the kinetic energy evolution can be demonstrated numerically for a dissipative solution, this is a clear indication of finite-time singularities. For a solution that is non-dissipative from a theoretical perspective, dissipation in the numerical simulation can only originate from under-resolution and will tend to zero when increasing the resolution. From numerical simulations of viscous problems at finite Reynolds number, there are indications that the zeroth law of turbulence holds for the Taylor--Green vortex problem. Numerical results for the kinetic energy dissipation rate for increasing Reynolds numbers up to~$\mathrm{Re} = 3000$  shown in~\cite{Brachet1983}, additional results for a higher Reynolds number of~$\mathrm{Re} = 5000$ shown in~\cite{Brachet1991},~$\mathrm{Re}=10000$ in~\cite{Arndt2020}, and~$\mathrm{Re}=20000$ in~\cite{Lamballais2019}  strongly suggest that the function~$D(t)$ does not tend to zero in the limit~$\nu \rightarrow 0$. This argument is summarized in Figure~\ref{fig:3d_tgv_Re_study} showing results for viscous and inviscid simulations of the TGV obtained with the present discretization approach. For each Reynolds number, results are shown for two resolutions of the numerical discretization approach to judge whether the results are mesh-independent. We achieve converged results for all finite Reynolds numbers shown in Figure~\ref{fig:3d_tgv_Re_study}. In the inviscid limit, the temporal evolution of the kinetic energy is almost indistinguishable for the two finest resolutions, while small differences are still visible in the energy dissipation rate that slightly differs between the two resolutions~$4096^3$ and~$8192^3$ at later times around the dissipation maximum and beyond. However, the onset of dissipation around~$t\approx 6$ appears to be converged also for this challenging inviscid simulation, a result that has not been obtained to date, see the discussion in Section~\ref{sec:IntroInterplayPhysicsNumerics}. A key motivation to address the question regarding the occurrence of finite-time singularities by the indirect approach is therefore that it might not be necessary to resolve the smallest scales of the flow to demonstrate grid-convergence in the temporal evolution of the kinetic energy. This is based on the observation that resolving the vorticity field (direct approach) typically requires significantly finer resolutions than resolving the kinetic energy (indirect approach). Hence, this work proposes an alternative with the goal to obtain a clearer indication of finite-time singularities than what is currently possible with the direct approach. Obviously, resolving the smallest scales is not possible in the inviscid limit at infinite Reynolds number for finite spatial resolutions. Consequently, one might argue that a single simulation will not serve as numerical evidence of a finite-time singularity, since the vorticity or enstrophy will take finite values at all times. We instead believe that a numerical analysis in terms of rigorous mesh refinement studies can provide a trend underpinning such investigations. Even if the flow is under-resolved at later times of the simulation and even if none of the simulations provides the ultimate resolution required to resolve the smallest (local) flow features, conclusions can be drawn from a series of simulations by studying the convergence behavior. To capture all interesting flow regimes, the inviscid Taylor--Green vortex is therefore simulated up to~$t=20$ in the present work as shown in Figure~\ref{fig:3d_tgv_Re_study}. Let us note that a similar idea to identify singularities experimentally based on energy arguments has been used in~\cite{Saw2016,Kuzzay2017} by calculating the inertial dissipation at scale~$l$ from PIV measurements. To the best of the authors' knowledge, the present study is the first to propose the use of energy arguments for singularity detection in numerical simulations.

\subsection{Outline}

The rest of this article is organized as follows. We describe the mathematical model of the incompressible Navier--Stokes or Euler equations and its numerical discretization in space and time in Section~\ref{Numerics}. Section~\ref{Results} shows results for the one-dimensional Burgers equation with formation of a shock, a two-dimensional shear layer problem with a numerical investigation of the kinetic energy dissipation in the limit~$\nu \rightarrow 0$, and finally the three-dimensional Taylor--Green vortex problem that has been suspected to exhibit finite-time singularities in the inviscid limit. In Section~\ref{Discussion}, we summarize our results, draw conclusions, and raise  questions based on the present observations.

\section{Numerical methods}\label{Numerics}
We seek (weak) numerical solutions to the incompressible Euler equations solved on a domain~$\Omega \subset \mathbb{R}^d$ in~$d=2,3$ space dimensions. These have their origin in the equations for viscous fluids with kinematic viscosity~$\nu$ described by the incompressible Navier--Stokes equations
\begin{align}
\frac{\partial \bm{u}}{\partial t} + \nabla \cdot (\bm{u} \otimes \bm{u}) - \nu \nabla^2 \bm{u} + \nabla p &= \bm{0} \ , \label{eq:MomentumEquation}\\
\nabla \cdot \bm{u} &= 0  \label{eq:ContinuityEquation} \ ,
\end{align}
where~$\bm{u}$ denotes the~$d$-dimensional velocity vector and~$p$ the kinematic pressure. The Euler equations are recovered by setting~$\nu=0$. This system of partial differential equations does not extend to~$d=1$ in a meaningful way, since the incompressibility constraint~$\partial u /\partial x = 0$ would imply~$u = \text{const}$.~\footnote{However, the one-dimensional Burgers equations serves as a simplified mathematical model for more complex higher-dimensional problems. While results for the inviscid Burgers equation are presented in Section~\ref{Results}, this section deals with discretizations of the incompressible Navier--Stokes equations for~$d=2,3$.} Let us note that the temporal and spatial discretization schemes discussed below are generic and hold for both two- and three-dimensional problems, but that there are major differences in terms of the flow physics and the mechanisms that make up the nature of turbulence such as the energy transfer to small scales according to a turbulence cascade in three dimensions~\cite{Onsager1949}. This fundamentally different behavior is attributed to the vortex stretching term in the vorticity form of the Euler equations~\cite{Gibbon2008}
\begin{align*}
\frac{\mathrm{D}\boldsymbol{\omega}}{\mathrm{D}t} = \left(\boldsymbol{\omega}\cdot \nabla\right) \bm{u} \ ,
\end{align*}
where the vortex stretching term on the right-hand side vanishes in two dimensions since the vorticity~$\boldsymbol{\omega}$ is perpendicular to the velocity~$\bm{u}$ in that case.

The following two subsections detail the temporal and spatial discretization of the incompressible Navier--Stokes equations~\eqref{eq:MomentumEquation} and~\eqref{eq:ContinuityEquation}. Discretization in time is based on projection methods that solve for velocity and pressure unknowns in different sub-steps of a time step. Discretization in space is based on high-order discontinuous Galerkin methods with suitable stabilization techniques that render the method robust for under-resolved, high-Reynolds number flows. The present discretization approach has been developed recently and is documented in a series of publications~\cite{Krank2017,Fehn2017,Fehn2018a,Fehn2019Hdiv}, with a focus on the stability of projection methods in~\cite{Fehn2017}, and a focus on the stability and dissipation characteristics of discontinuous Galerkin discretizations for under-resolved turbulence in~\cite{Fehn2018a,Fehn2019Hdiv}. For reasons of brevity and to focus on the main aspects, we avoid technical aspects related to the imposition of boundary conditions in the present work, but refer to the original publications. We also do not mention aspects related to the implementation of the method and its computational costs. The computational efficiency of the present discretization methods in terms of fast implementations and fast iterative solvers is discussed in~\cite{Fehn2018b,Fehn2020hybrid, Kronbichler2019fast,Arndt2020}.

\subsection{Temporal discretization -- high-order projection method}
Discretization in time is based on projection methods which aim at obtaining computationally efficient incompressible flow solvers by decoupling the velocity and pressure unknowns~\cite{KarniadakisSherwin2005}. A convection--diffusion type problem is solved for the velocity and a Poisson equation for the pressure with a subsequent projection of the velocity onto the space of solenoidal vector fields according to the Helmholtz decomposition. We use the high-order dual splitting scheme proposed in~\cite{Karniadakis1991} which consists of the following four sub-steps
\begin{align}
\frac{\gamma_0^n\hat{\bm{u}}-\sum_{i=0}^{J-1}\alpha_i^n\bm{u}^{n-i}}{\Delta t_n} &=
- \sum_{i=0}^{J-1}\beta_i^n \Div{\left(\bm{u}^{n-i} \otimes \bm{u}^{n-i}\right)}\; ,\label{eq:DualSplitting_ConvectiveStep}\\
-\nabla^2 p^{n+1} &= -\frac{\gamma_0^n }{\Delta t_n}\Div{\hat{\bm{u}}} \; ,\label{eq:DualSplitting_PressureStep}\\
\hat{\hat{\bm{u}}} &= \hat{\bm{u}} - \frac{\Delta t_n}{\gamma_0^n} \Grad{p^{n+1}}\; ,\label{eq:DualSplitting_ProjectionStep}\\
\frac{\gamma_0^n}{\Delta t_n} \bm{u}^{n+1}  -   \nu \nabla^2 \bm{u}^{n+1} &=
\frac{\gamma_0^n}{\Delta t_n}\hat{\hat{\bm{u}}} \; ,\label{eq:DualSplitting_ViscousStep}
\end{align}
where~$n$ denotes the current time step in which the equations are integrated from time~$t_n$ to time~$t_{n+1} = t_n + \Delta t_n$. The temporal discretization is based on a BDF scheme of order~$J$ with coefficients~$\gamma_0^n$ and~$\alpha_i^n, i=0,\hdots,J-1$. In the first sub-step, equation~\eqref{eq:DualSplitting_ConvectiveStep}, the convective term is treated explicitly in time by using a high-order extrapolation scheme of order~$J$ with coefficients~$\beta_i^n, i=0,\hdots,J-1$. In the next two sub-steps, a pressure Poisson equation, equation~\eqref{eq:DualSplitting_PressureStep}, is solved and a divergence-free velocity is obtained in the projection step, equation~\eqref{eq:DualSplitting_ProjectionStep}. Finally, the viscous term is taken into account in the last sub-step, equation~\eqref{eq:DualSplitting_ViscousStep}, which can be omitted in case of the Euler equations,~$\bm{u}^{n+1} = \hat{\hat{\bm{u}}}$. The explicit treatment of the convective term implies a restriction of the time step sizes according to the Courant--Friedrichs--Lewy (CFL) condition. For reasons of computational efficiency, we use adaptive time stepping with variable time step sizes~$\Delta t_n$ by readjusting the time step size after each time step in a way to maximize the time step size and operate close to the CFL stability limit. We introduce the CFL condition below since it also depends on the spatial discretization scheme. A second-order accurate time integration scheme with~$J=2$ is used in the present work.

\subsection{Spatial discretization -- high-order discontinuous Galerkin method}
We assume a computational domain~$\Omega_h= \bigcup_{e=1}^{N_{\text{el}}} \Omega_{e} \in \mathbb{R}^d$ consisting of conforming quadrilateral or hexahedral elements~$\Omega_e$,~$e=1,\hdots, N_{\mathrm{el}}$. A common abstraction of finite element methods is to define a mapping~$\bm{x}^e(\boldsymbol{\xi})$ from a reference element~$\tilde{\Omega}_e = \left[0,1\right]^d$ with Cartesian coordinates~$\boldsymbol{\xi}$ to element~$\Omega_e$ in physical space with coordinates~$\bm{x}$, and approximate the solution within each element by polynomials defined in reference coordinates. Here, the numerical solution in~$d=2,3$ is represented by a tensor product of one-dimensional Lagrange polynomials with a Legendre--Gauss--Lobatto point distribution, and is allowed to exhibit discontinuities between elements in an~$L^2$-conforming sense. The spaces of shape functions are then given as
\begin{align*}
\mathcal{V}^{u}_{h} &= \left\lbrace\bm{u}_h\in \left[L^2(\Omega_h)\right]^d\; : \; \bm{u}_h\left(\bm{x}^e(\boldsymbol{\xi})\right)\vert_{\Omega_{e}}= \tilde{\bm{u}}_h^e(\boldsymbol{\xi})\vert_{\tilde{\Omega}_{e}}\in \mathcal{V}^{u}_{h,e}=[\mathcal{Q}_{k}(\tilde{\Omega}_{e})]^d\; ,\;\; \forall e=1,\ldots,N_{\text{el}} \right\rbrace\;\; ,\\
\mathcal{V}^{p}_{h} &= \left\lbrace p_h\in L^2(\Omega_h)\; : \; p_h\left(\bm{x}^e(\boldsymbol{\xi})\right)\vert_{\Omega_{e}} = \tilde{p}_h^e(\boldsymbol{\xi})\vert_{\tilde{\Omega}_{e}}\in \mathcal{V}^{p}_{h,e}=\mathcal{Q}_{k-1}(\tilde{\Omega}_{e})\; ,\;\; \forall e=1,\ldots,N_{\text{el}} \right\rbrace\; .
\end{align*}
Here,~$\mathcal{Q}_{k}$ denotes the space of tensor-product polynomials of degree~$k$, and we highlight the ambiguity in notation related to the spatial wavenumber~$k$ in the context of energy spectra. The polynomial degree of the velocity shape functions is~$k$, while it is~$k-1$ for the pressure for reasons of inf--sup stability. In this work, we only consider problems with Cartesian meshes so that the mapping~$\bm{x}^e(\boldsymbol{\xi})$ is an affine transformation. At the interface between two elements~$e^-$ and~$e^+$, the numerical solution is not unique and we use the superscripts~$-$ and~$+$ to denote the solution from the two sides of an interface. When referring to element~$e$, we denote interior information by~$-$, and the outward pointing unit normal vector by~$\bm{n}=\bm{n}^-$. To define numerical fluxes, we introduce the average operator~$\avg{u}=(u^- + u^+)/2$ and the jump operator~$\jump{u}=u^- \otimes \bm{n}^- + u^+ \otimes \bm{n}^+$, where~$\bm{n}^+ = - \bm{n}^-$. We apply the usual abbreviations of integrals,~$\intele{v}{u} = \int_{\Omega_e} v \odot u \; \mathrm{d}\Omega$ for volume integrals and~$\inteleface{v}{u} = \int_{\partial \Omega_e} v \odot u \; \mathrm{d} \Gamma$ for surface integrals with~$\odot$ denoting an inner product. We now state the weak formulation for all sub-steps of the projection scheme. Obtain the numerical solutions~$\hat{\bm{u}}_h,\hat{\hat{\bm{u}}}_h, \hat{\hat{\hat{\bm{u}}}}_h ,\bm{u}_h^{n+1}\in\mathcal{V}^u_h$ and~$p_h^{n+1}\in\mathcal{V}^p_h$ by testing with all test functions~$\bm{v}_h \in \mathcal{V}^{u}_{h,e}$,~$q_h \in \mathcal{V}^{p}_{h,e}$ for all elements~$e=1,...,N_{\text{el}}$:

In the first step, the convective term is discretized by the local Lax--Friedrichs flux
\begin{align}
\begin{aligned}
\intele{\bm{v}_h}{\frac{\gamma_0^n \hat{\bm{u}}_h-\sum_{i=0}^{J-1}\alpha_i^n\bm{u}^{n-i}_h}{\Delta t_n}}
=\\
- \sum_{i=0}^{J-1} \beta_i^n \left(-\intele{\Grad{\bm{v}_h}}{\left(\bm{u}_h \otimes \bm{u}_h\right)^{n-i}} +
\inteleface{\bm{v}_h}{\left(\avg{\bm{u}_h \otimes \bm{u}_h} + \frac{\Lambda}{2}\jump{\bm{u}_h} \right)^{n-i}  \cdot \bm{n}} \right) \; ,
\end{aligned}\label{eq:DualSplitting_ConvectiveStep_WeakForm}
\end{align}
where the stabilization parameter is chosen as~$\Lambda = \max \left( \vert \bm{u}_h^{-} \cdot \bm{n}\vert , \vert \bm{u}_h^{+} \cdot \bm{n}\vert\right)$.
The pressure Poisson operator is discretized by the symmetric interior penalty Galerkin (SIPG) method
\begin{align}
\begin{aligned}
\intele{\Grad{q_h}}{\Grad{p^{n+1}_h}}
-\inteleface{\Grad{q_h}}{\frac{1}{2}\jump{p^{n+1}_h}}
- \inteleface{q_h}{\avg{\Grad{p^{n+1}_h}}\cdot\bm{n}}
+ \inteleface{q_h}{\tau\jump{p^{n+1}_h}\cdot\bm{n}}\\
= - \frac{\gamma_0^n}{\Delta t_n} \left( -\intele{\Grad{q_h}}{\hat{\bm{u}}_h}+\inteleface{q_h}{\avg{\hat{\bm{u}}_h}\cdot\bm{n}} \right)\; ,
\end{aligned}\label{eq:DualSplitting_Pressure_WeakForm}
\end{align}
where a central flux is used for the velocity divergence operator on the right-hand side of the pressure Poisson equation. A central flux is also used for the pressure gradient term in the projection step
\begin{align}
\begin{aligned}
\intele{\bm{v}_h}{\hat{\hat{\bm{u}}}_h} = \intele{\bm{v}_h}{\hat{\bm{u}}_h}-\frac{\Delta t_n}{\gamma_0^n} \left( -\intele{\Div{\bm{v}_h}}{p^{n+1}_h}+\inteleface{\bm{v}_h}{\avg{p^{n+1}_h}\bm{n}} \right)\; .
\end{aligned}\label{eq:DualSplitting_Projection_WeakForm}
\end{align}
The viscous term is also discretized by the SIPG method
\begin{align}
\begin{aligned}
\intele{\bm{v}_h}{\frac{\gamma_0^n}{\Delta t_n} \hat{\hat{\hat{\bm{u}}}}_h}
+ \intele{\Grad{\bm{v}_h}}{\nu \Grad{\hat{\hat{\hat{\bm{u}}}}_h}}
- \inteleface{\Grad{\bm{v}_h}}{\frac{\nu}{2} \jump{\hat{\hat{\hat{\bm{u}}}}_h}}\\
- \inteleface{\bm{v}_h}{\nu \avg{\Grad{\hat{\hat{\hat{\bm{u}}}}_h}}\cdot\bm{n}}
+ \inteleface{\bm{v}_h}{\nu\tau \jump{\hat{\hat{\hat{\bm{u}}}}_h}\cdot\bm{n}}
= \intele{\bm{v}_h}{\frac{\gamma_0^n}{\Delta t_n}\hat{\hat{\bm{u}}}_h} \; ,
\end{aligned}\label{eq:DualSplitting_ViscousStep_WeakForm}
\end{align}
while this step is skipped,~$\hat{\hat{\hat{\bm{u}}}}_h = \hat{\hat{\bm{u}}}_h$, in the inviscid limit when solving the incompressible Euler equations. In a final postprocessing step, consistent divergence and continuity penalty terms are applied to weakly enforce the incompressibility constraint and normal continuity of the velocity field~\cite{Fehn2018a, Fehn2020ALE}
\begin{align}
\begin{aligned}
\intele{\bm{v}_h}{\bm{u}_h^{n+1}} + \intele{\Div{\bm{v}_h}}{\tau_{\mathrm{D}}\Div{\bm{u}_h^{n+1}}}\Delta t_n
+ \inteleface{\bm{v}_h\cdot \bm{n}}{\tau_{\mathrm{C}}\left(\bm{u}_h^{-} - \bm{u}_h^{+}\right)^{n+1}\cdot \bm{n}}\Delta t_n  
= \intele{\bm{v}_h}{\hat{\hat{\hat{\bm{u}}}}_h} \, . 
\end{aligned}\label{eq:DualSplitting_Penalty_WeakForm}
\end{align}
The above postprocessing step is specifically related to the type of spatial discretization used in this work, i.e., the~$L^2$-conforming discontinuous Galerkin method, and disappears in the continuous case. It was found that the divergence and normal-continuity penalty terms are crucial in terms of mass conservation and energy stability in order to obtain a method that is robust for large Reynolds numbers and coarse spatial resolutions, and we refer to~\cite{Fehn2018a, Fehn2019Hdiv} for detailed numerical justification and validation of this approach. In this context, we note that alternatives to this stabilized approach exist, e.g., by using tailored finite element function spaces. For example, function spaces can be used that result in an~$H(\text{div})$-conforming (normal-continuous) velocity field and that is divergence-free in every point of the computational domain. The present~$L^2$-conforming approach does not fulfill these two properties exactly, but in a weak finite element sense. Definitions of the penalty parameters~$\tau, \tau_{\mathrm{D}}, \tau_{\mathrm{C}}$ are given in~\cite{Fehn2018a}, where the default penalty factor~$\zeta=1$ is used in the present work unless specified otherwise. In terms of the occurrence of finite-time singularities and anomalous energy dissipation, the present work makes use of the argument that -- due to the weighted residual formulation -- the present discretization can be applied to problems which lack regularity and for which the differential form of the equations is no longer an appropriate description. With respect to the implementation of the method, integrals in the weak form are evaluated by means of Gaussian quadrature, with consistent integration according to the~$3/2$-rule for the non-linear convective term (also denoted as polynomial dealiasing). Since we consider uniform Cartesian meshes~(with elements of size~$h$ in all coordinate directions) in the present work, integrals are calculated exactly. To obtain the size of the next time step within the adaptive time stepping scheme, we use the following local CFL condition evaluated at all quadrature points~$q$ of element~$e$
\begin{align}
\Delta t = \min_{e=1,...,N_{\mathrm{el}}} \left( \min_{q} \frac{\mathrm{Cr}}{k^{1.5}}\left.\frac{h}{\Vert \bm{u}_h \Vert }\right\vert_{q,e} \right)\; ,\label{eq:local_CFL_Condition}
\end{align}
ensuring stability of the time integration scheme by selecting a Courant number~$\mathrm{Cr} < \mathrm{Cr}_{\mathrm{crit}}$. We point out that -- due to the CFL condition -- an increase in spatial resolution by reducing~$h$ or increasing~$k$ automatically implies an increase in temporal resolution through a reduced time step size. The present solver is implemented in~\texttt{C++} and makes use of the finite element library~\texttt{deal.II}~\cite{dealII, dealII90}.

\section{Results}\label{Results}

This section presents numerical results for three test cases in~$d=1,2$, and~$3$ space dimensions. The one-dimensional problem is the well-known inviscid Burgers equation developing a singularity in finite-time for appropriate initial conditions. The two-dimensional example is a shear layer roll-up problem. The particular example used for the~$d=2$ investigations is not of primary importance as it is known theoretically that regularity is expected for two-dimensional Euler flows when starting from regular initial data. Instead, the aim of these one- and two-dimensional examples is an investigation to which extent the numerical discretization scheme is able to mimic physical behavior with potentially singular solutions. Having validated the numerical method for these well-understood problems, it is applied to the three-dimensional inviscid Taylor--Green problem, for which the physical understanding in terms of the occurrence of finite-time singularities and the related aspect of anomalous energy dissipation is speculative at present. As a further preparation, we summarize the quantities of interest in the following subsection.

\subsection{Quantities of interest}
Since turbulent flows in three space dimensions are our primary interest, we restrict the discussion in this subsection to the three-dimensional case implying extensions of certain relations to one- and two-dimensional problems only where possible. Of primary importance for the present study is the temporal evolution of the kinetic energy
\begin{align*}
E(t) = \frac{1}{V_{\Omega}} \int_{\Omega} \frac{1}{2} \bm{u}(\bm{x},t) \cdot \bm{u}(\bm{x},t) \ \mathrm{d}\Omega \ ,
\end{align*}
and its dissipation rate~$\mathrm{d} E / \mathrm{d} t$. The kinetic energy is normalized by the volume~$V_{\Omega} = \int_{\Omega} 1 \mathrm{d}\Omega$ of the computational domain. The integrals are evaluated numerically by means of Gaussian quadrature with~$k+1$ quadrature points in each coordinate direction. The time derivative used to obtain the dissipation rate is computed numerically from the kinetic energy at discrete instances of time via a second order finite difference formula for variable time step sizes with first order approximations at the end points. If anomalous dissipation~($\mathrm{d} E / \mathrm{d} t < 0$) occurs, the temporal evolution of the enstrophy~$\mathcal{E}$,
\begin{align*}
\mathcal{E}(t) = \frac{1}{V_{\Omega}} \int_{\Omega} \frac{1}{2} \boldsymbol{\omega}(\bm{x},t) \cdot \boldsymbol{\omega}(\bm{x},t)\ \mathrm{d}\Omega \ ,
\end{align*}
is expected to exhibit a singularity~$\mathcal{E} \rightarrow \infty$ in finite time. Integrals are computed by Gaussian quadrature, which is exact down to round-off errors due to polynomial integrands and Cartesian meshes. A related local quantity is the maximum vorticity
\begin{align*}
\Vert \boldsymbol{\omega} \Vert_{\infty} (t) \ ,
\end{align*}
where we take the maximum over all quadrature points over all elements in the discrete case and monitor its evolution over time with the interest in a detection of potentially singular behavior in finite time. Although the maximum vorticity will remain finite for every numerical simulation of finite resolution, a mesh refinement study may give hints on the expected behavior if the resolution was further increased. Finally, we consider kinetic energy spectra by transformation into wavenumber space~$\bm{k}$~\cite{Cichowlas2005,Bustamente2012}
\begin{align*}
E(k, t) = \frac{1}{2} \lim_{\Delta k \rightarrow 0} \frac{\int_{\Vert \bm{k} \Vert=k}^{\Vert \bm{k} \Vert=k+\Delta k}\Vert \hat{\bm{u}}(\bm{k},t)\Vert^2 \mathrm{d} \bm{k}}{\Delta k} \hspace{0,5cm}\overset{\text{DFT}}{\approx} \sum_{\substack{\bm{k} \in \mathbb{Z}^3 \\ k-\frac{1}{2} \leq \Vert \bm{k} \Vert < k+\frac{1}{2}}}  \frac{1}{2} \Vert \hat{\bm{u}}_{\mathrm{DFT}}(\bm{k},t)\Vert^2 \ ,
\end{align*}
where~$\hat{\bm{u}}(\bm{k},t)$ denotes the Fourier transform of the velocity, which only exists at discrete wavenumber vectors~$\bm{k}$ in case of a discrete Fourier transformation (DFT) obtained from sampled data of a discrete velocity field. The solution is first interpolated onto~$N$ equidistant points per element and per coordinate direction~\footnote{The number of sampling points is chosen as~$N=k+1$ in the present work, i.e., equal to the number of nodal points of the discontinuous Galerkin discretization, where~$k$ is the polynomial degree of the shape functions and should not be mixed up with the wavenumber~$k$ typically used in the context of energy spectra. The equidistant interpolation points all lie within the element away from the boundaries where the solution is discontinuous. If interpolation points on the boundary are used, one typically takes the average of the solution from neighboring elements.} to which the discrete Fourier transformation is applied, using the library~\texttt{FFTW} in the present work~\cite{FFTW}. Note that considering~$E(k,t)$ as a function of a scalar wavenumber~$k$ as well as the summation over spheres of radius~$k$ introduces the assumptions of homogeneity and isotropy. In case of anomalous energy dissipation, the enstrophy is expected to become infinite. Hence, exploiting the relation ~$\mathcal{E}(k,t) = k^2 E(k,t)$ between the enstrophy and energy spectra and further assuming a power law behavior for the kinetic energy spectrum of the form~$E(k,t) = C(t) k^{-n (t)}$, a singularity at time~$t=t_*$ with~$\mathcal{E}(t_*) = \int_0^{\infty} \mathcal{E}(k,t_*) \mathrm{d}k = \infty$ would correspond to a decay with slope~$n(t_*)=3$ in the energy spectrum, see also~\cite{Orlandi2012,Orlandi2009}. We use this criterion as a further validation of the results in case other quantities give hints of a potentially singular behavior. Apart from that, energy spectra are typically investigated to assess the well-known~$k^{-5/3}$ Kolmogorov spectrum for fully-developed, homogeneous isotropic turbulence. We apply this property to investigate whether the numerical results match expected physical behavior obtained from classical cascade pictures in case of inviscid flows and beyond the time of potential singularities,~$t > t_*$.

\subsection{One-dimensional inviscid Burgers equations}
We begin with studying the one-dimensional inviscid Burgers equation~\cite{Burgers1948}
\begin{align*}
\frac{\partial u}{\partial t} + \frac{\partial }{\partial x} \frac{u^2}{2} = 0 \ ,
\end{align*}
as a simplified model for the incompressible Euler equations. It is well known that this equation develops singularities in finite time, see for example~\cite{Sulem1983,Dubrulle2019}. It is therefore particularly interesting to study the behavior of a discretization scheme for this problem first. The spatial discretization is based on a discontinuous Galerkin scheme very similar to the one described in Section~\ref{Numerics} for the two- and three-dimensional case, i.e., the convective term is discretized with a local Lax--Friedrichs flux. Gaussian quadrature with a~$3/2$-overintegration rule is used as in the higher dimensional case due to the quadratic nonlinearity of the convective term, but intentionally no additional measure such as limiting, filtering, or other Riemann fluxes are taken to specifically address the jump that forms in the solution. We also emphasize that no artificial viscosity approach is used to deal with the singularity. For time integration, the classical explicit fourth-order Runge--Kutta method is used with a Courant number of~$\mathrm{Cr}=0.4$.

\begin{figure}[t]
 \centering
	\includegraphics[width=0.49\textwidth]{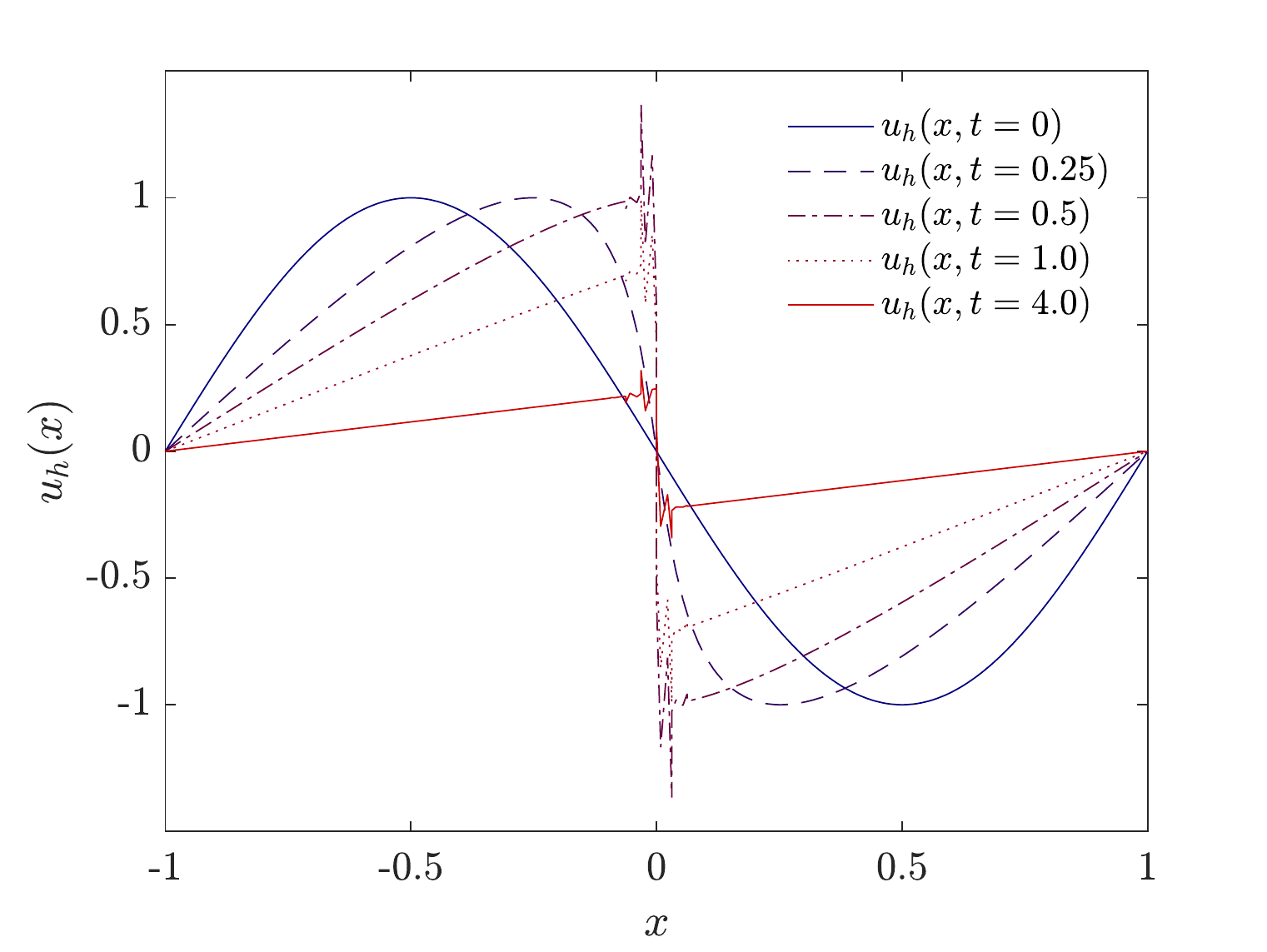}
	\includegraphics[width=0.49\textwidth]{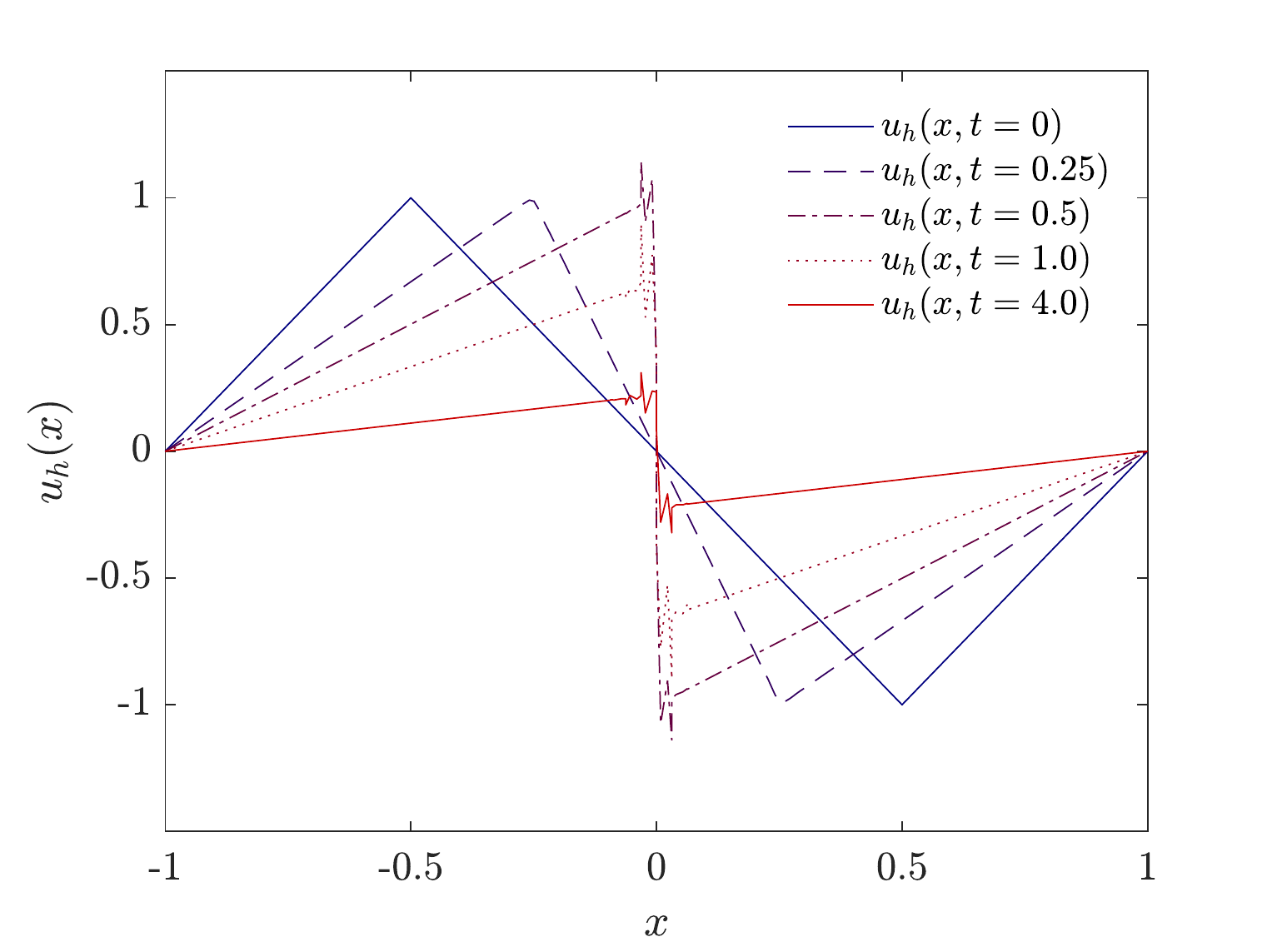}
\caption{One-dimensional inviscid Burgers equations for two different initial solutions that form a singularity: On the left, the initial condition is a sine function, while it is a simple hat function that is piecewise linear on the right. The spatial resolution used for the computations corresponds to refinement level~$l=6$ and polynomial degree~$k=3$, resulting in an effective resolution of~$256^1$.}
\label{fig:1d_burgers_sine_versus_hat}
\end{figure}

\begin{figure}[t]
 \centering
	\includegraphics[width=0.49\textwidth]{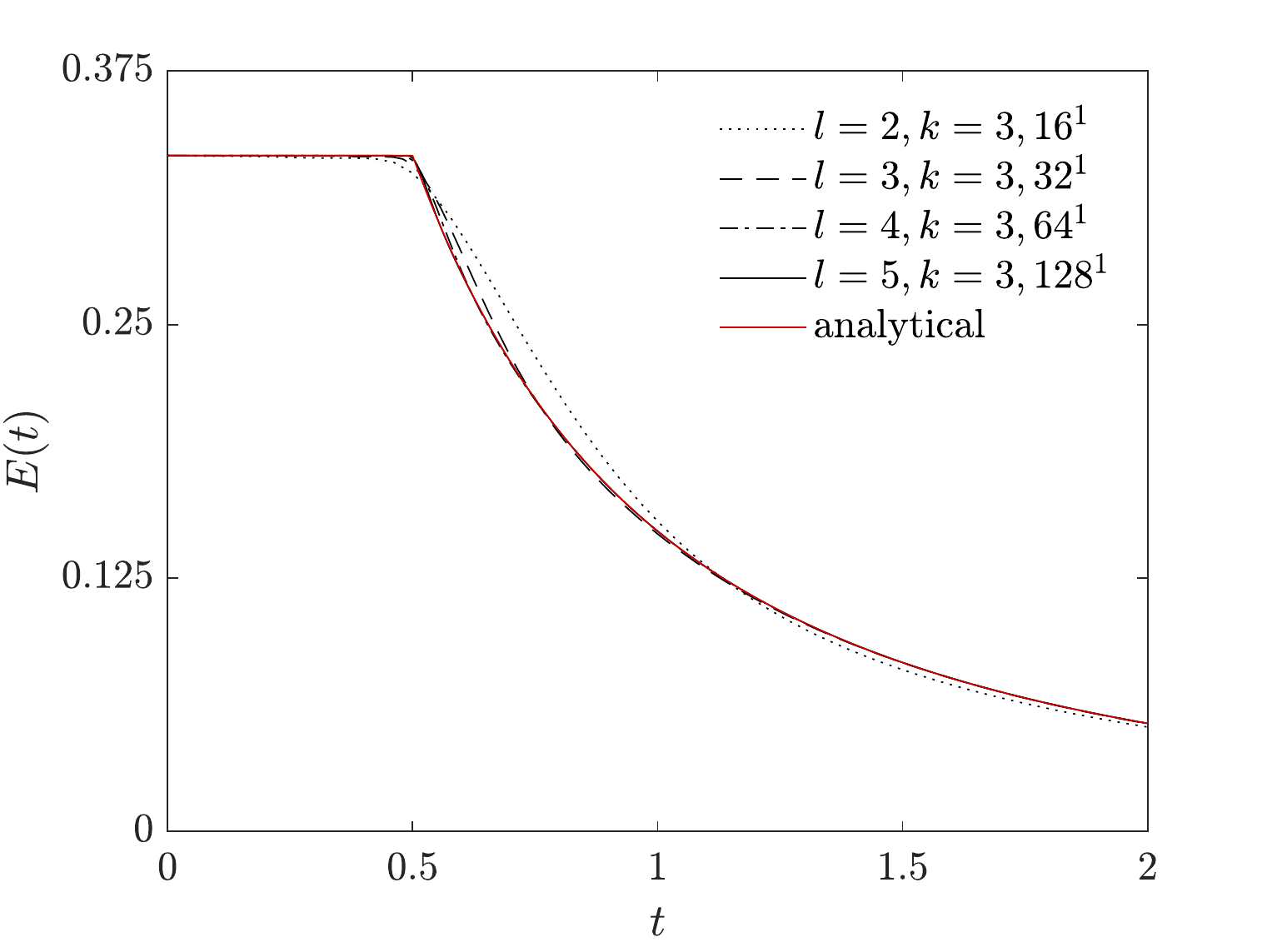}
	\includegraphics[width=0.49\textwidth]{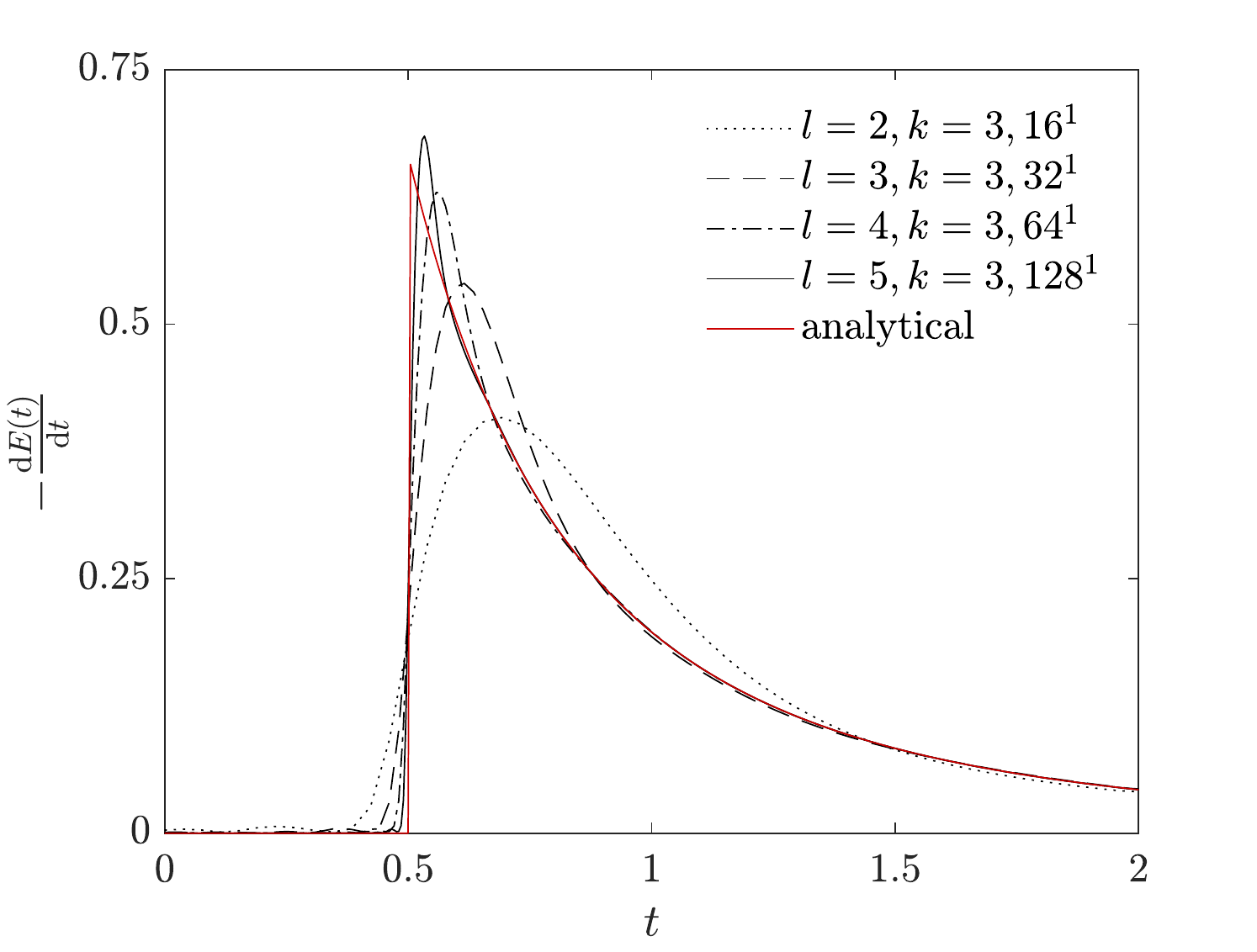}
\caption{One-dimensional inviscid Burgers equations with hat function as initial condition: Temporal evolution of kinetic energy as well as dissipation rate and convergence towards analytical profile for a mesh refinement study with refine levels~$l=2,\hdots,5$ and polynomial degree~$k=3$, resulting in effective resolutions of~$16^1,\hdots,128^1$.}
\label{fig:1d_burgers_convergence_energy_and_dissipation_rate}
\end{figure}

\begin{figure}[t]
 \centering
	\includegraphics[width=0.49\textwidth]{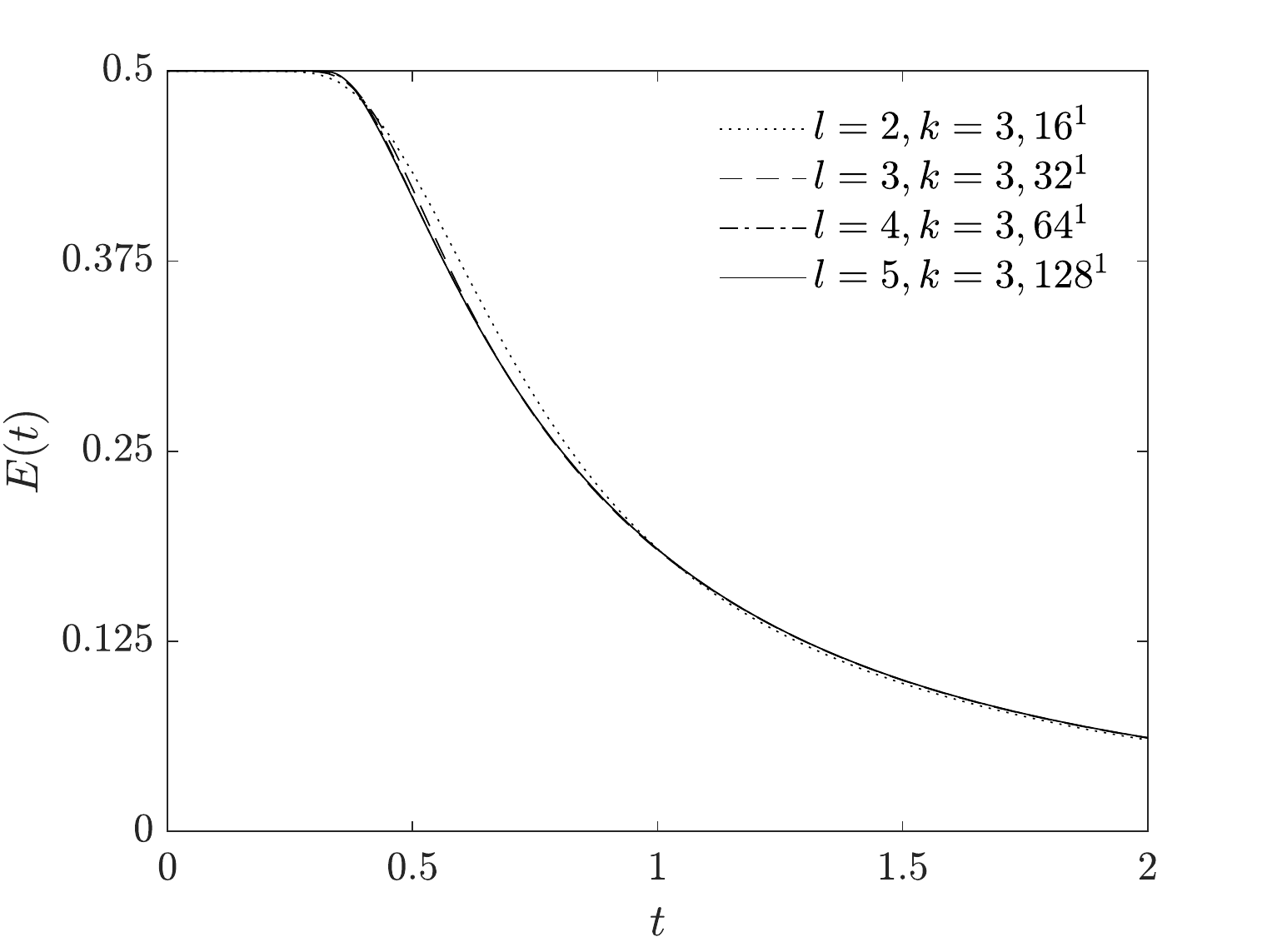}
	\includegraphics[width=0.49\textwidth]{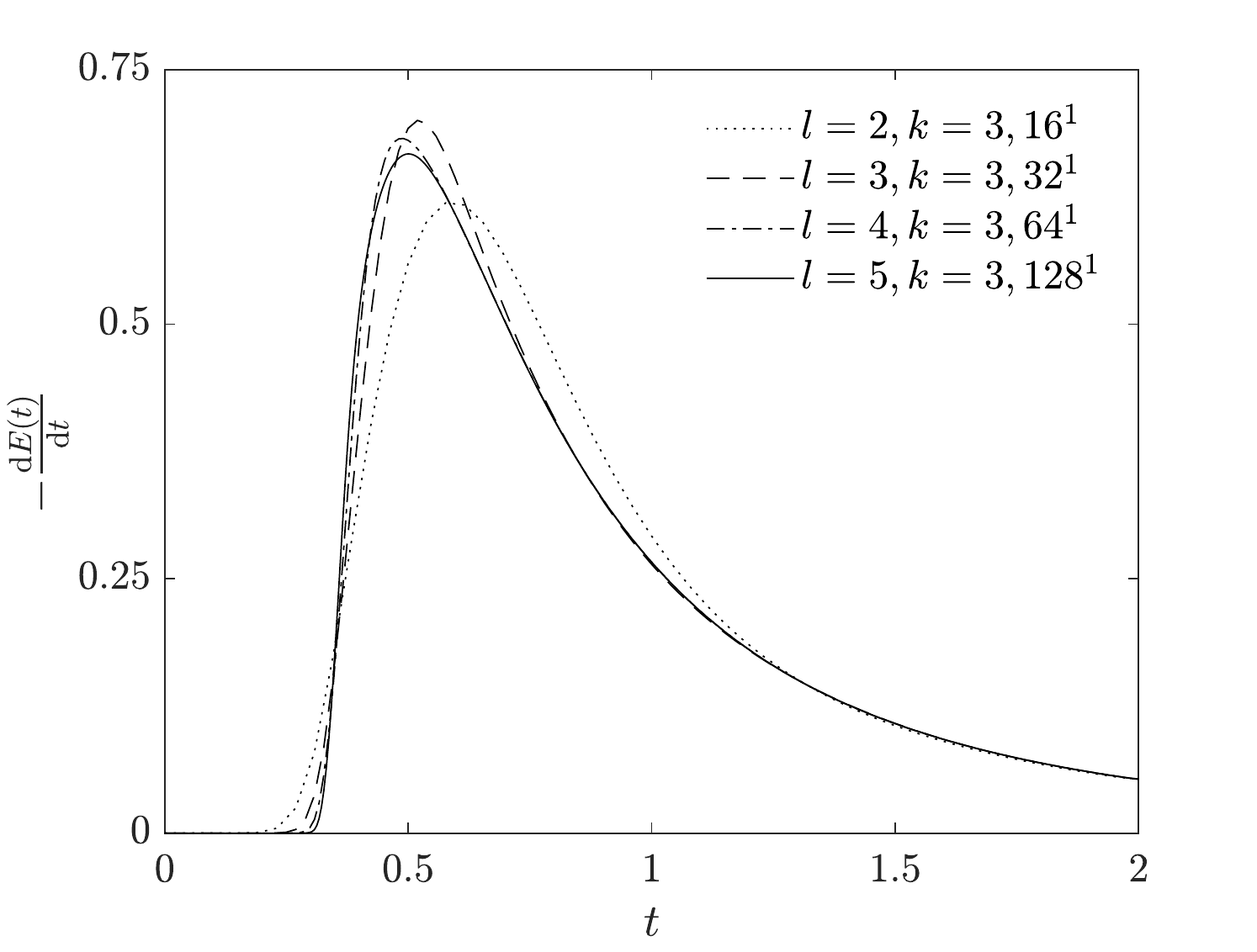}
\caption{One-dimensional inviscid Burgers equations with sine function as initial condition: Temporal evolution of kinetic energy as well as dissipation rate and mesh refinement study with refine levels~$l=2,\hdots,5$ and polynomial degree~$k=3$, resulting in effective resolutions of~$16^1,\hdots,128^1$.}
\label{fig:1d_burgers_convergence_energy_and_dissipation_rate_sine}
\end{figure}

Figure~\ref{fig:1d_burgers_sine_versus_hat} shows the numerical solution~$u_h(x)$ at various instances of time and the formation of a shock. The problem is solved on the domain~$\Omega = \left[-1, 1\right]$ with Dirichlet boundary conditions prescribed at both boundary points of the one-dimensional domain. An equidistant grid with~$2^l$ elements is used where~$l$ denotes the level of refinement. For polynomial approximations of degree~$k$, the effective resolution becomes~$(k+1)2^l$. We exemplarily select two different initial solutions, a sine function,~$u_h(x, t=0) = - \sin (\pi x)$, and a hat function,~$u_h(x, t=0) = - 2 \vert x + 0.5 \vert + 1$ for~$x < 0$ and~$u_h(x, t=0) = 2 \vert x - 0.5 \vert -1$ for~$x \geq 0$. Due to the chosen initial conditions with~$u>0$ for~$x<0$ and vice versa, the solution piles up in the middle of the domain and a singularity ($\partial u /\partial x \rightarrow \infty$) forms at~$x=0$ in both cases. The oscillating behavior of the numerical solution around the singularity could be improved by the advanced discretization techniques mentioned above. From the results shown in Figure~\ref{fig:1d_burgers_sine_versus_hat} it is plausible that the kinetic energy is conserved until the formation of the shock and that energy will be dissipated at later times. For the hat function chosen as initial condition, it is straight-forward to derive an analytical expression for the temporal evolution of the kinetic energy as well as its dissipation rate, which is why we consider this setup in more detail in the following. According to the method of characteristics it follows that the shock forms at time~$t_*=0.5$. From that time on, the solution can be written as
\begin{align*}
u(x,t) = f(t) \left(x - \text{sign} (x) \right) \ ,
\end{align*}
where~$\text{sign} (x)$ takes values of~$\pm 1$ depending on the sign of the argument. The temporal evolution part~$f(t)$ which describes the absolute value of~$u$ taken to the left and right of the origin of the coordinate system at~$x=0$ can be obtained from the following consideration
\begin{align*}
f(t + \mathrm{d}t) = f(t) - \underbrace{\left.\frac{\partial u (x,t)}{\partial x}\right\vert_{x=0^-}}_{=f(t)} \underbrace{\mathrm{d}x}_{ \substack{= u(x=0^-,t) \mathrm{d}t \\ = f(t)\mathrm{d}t} } = f(t) \left(1 - f(t) \mathrm{d}t\right) \ ,
\end{align*}
i.e., the solution at~$x=0^-$ at time~$t+\mathrm{d}t$ equals the solution at position~$-\mathrm{d}x = -u(x=0^-,t) \mathrm{d}t$ at time~$t$.
Separation of variables and integration yields the result~$f(t) = 1 /(t+t_*)$. The kinetic energy~$E(t) = \int_{\Omega} \frac{1}{2} u^2(x,t)\mathrm{d}x$ is therefore given as
\begin{align*}
E(t) = \int_{-1}^{1} \frac{1}{2} f^2(t) \left(x - \text{sign} (x) \right)^2 \mathrm{d}x = \frac{f^2(t)}{3} \; .
\end{align*}
The kinetic energy dissipation rate is obtained by differentation which yields a~$(t+t_*)^{-3}$ decay for times~$t\geq t_*$. The dissipation rate is~$\Delta u^3/12$ when expressed in terms of the jump~$\Delta u$ of the solution, in agreement with the result in~\cite{Dubrulle2019} where it is noted that this inviscid dissipation is identical to the dissipation of the viscosity solution in the limit~$\nu \rightarrow 0$. In Figure~\ref{fig:1d_burgers_convergence_energy_and_dissipation_rate}, we show results for both the kinetic energy and the dissipation rate for a sequence of mesh refinement levels of~$l=2,\hdots, 5$ with degree~$k=3$, resulting in effective resolutions of~$16^1,\hdots,128^1$. For increasing spatial resolution, the numerical results converge to the analytical profiles. It can be seen that achieving grid-convergence for the dissipation rate requires higher spatial resolutions as compared to the temporal evolution of the kinetic energy itself. This is expected since the dissipation rate contains a temporal derivative that results in a higher sensitivity with respect to deviations (here numerical discretization error) from the exact solution. Figure~\ref{fig:1d_burgers_convergence_energy_and_dissipation_rate_sine} shows the same results for the problem with sine function as initial condition. We observe that the onset of energy dissipation is smooth as opposed to the hat function where the kinetic energy exhibits a kink and the dissipation rate a jump at the time of the singularity. In other words, the occurrence of a finite-time singularity does not imply an instantaneous onset of dissipation. We keep this in mind when considering the three-dimensional inviscid Taylor--Green problem which is a problem that also starts from sine-like initial data.

The point that we want to make with this example is that a discretization scheme that involves purely numerical mechanisms of dissipation can provide the physically correct amount of dissipation for a sufficiently fine spatial resolution, see also the discussion in the introduction. Note that this is fundamentally different from viscosity solutions~$u^{\nu}$ for small~$\nu > 0$, for which the required dissipation is realized by the additional viscous term in the equations and for which the dissipation stemming from the numerical discretization scheme tends to zero if the mesh resolves the viscosity solution~$u^{\nu}$ exhibiting steep but finite gradients. Although the one-dimensional Burgers equation can not reflect the complexity of three-dimensional turbulent flows, these results put confidence in numerical discretization schemes to also predict the solution in a physically correct way for the higher-dimensional problems studied below.

\subsection{Two-dimensional shear layer problem}

We consider the two-dimensional shear layer roll-up problem~\cite{Brown1995} where the initial velocity is given as
\begin{align*}
\bm{u}(\bm{x},t=0) = \left( \tanh \left(\rho (0.25 - \vert x_2-0.5 \vert) \right), \delta \sin \left( 2\pi x_1\right) \right)^{\mathsf{T}} .
\end{align*}
Following~\cite{Brown1995}, we set the two parameters~$\rho, \delta$ to~$\rho = 30$ and~$\delta = 0.05$. The problem is solved on the domain~$\Omega = \left[0,1\right]^2$ with periodic boundaries in both directions. In the following, different viscous simulations with viscosities~$\nu = 2.5 \cdot 10^{-3}, 10^{-3}, 10^{-4}$ are considered, as well as the inviscid limit with~$\nu=0$. The mesh is uniform Cartesian with~$(2^l)^2$ elements for refinement level~$l$, the polynomial degree of the shape functions is~$k=7$, resulting in the effective resolution of~$\left( (k+1) 2^l \right)^2$. The simulations are run for the time interval~$0 \leq t \leq 4$. The time step size is adapted dynamically with a Courant number of~$\mathrm{Cr}=0.25$.

\begin{figure}[t]
 \centering 
 \subfloat[Velocity magnitude for~$\nu=2.5 \cdot 10^{-3}, 10^{-3}, 10^{-4}, 0$ (from left to right).]{
    \includegraphics[width=0.24\textwidth]{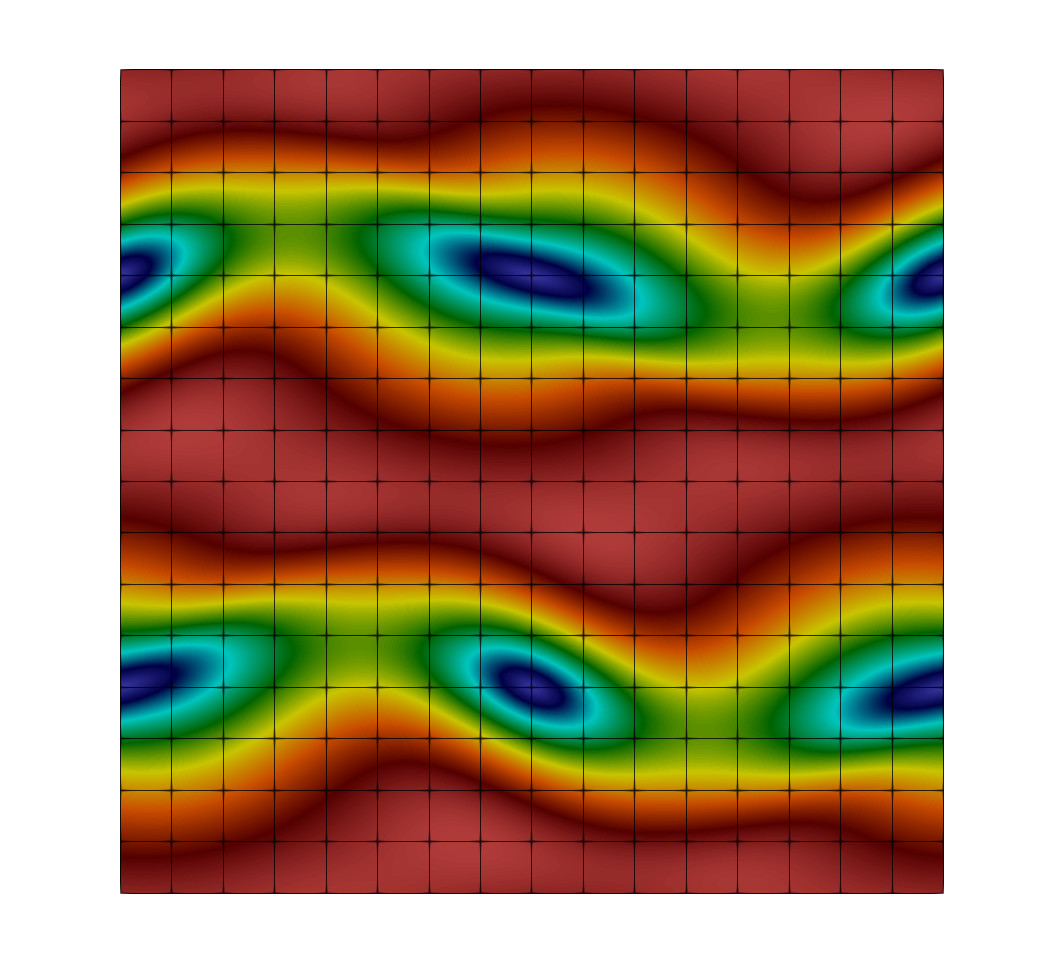}
    \includegraphics[width=0.24\textwidth]{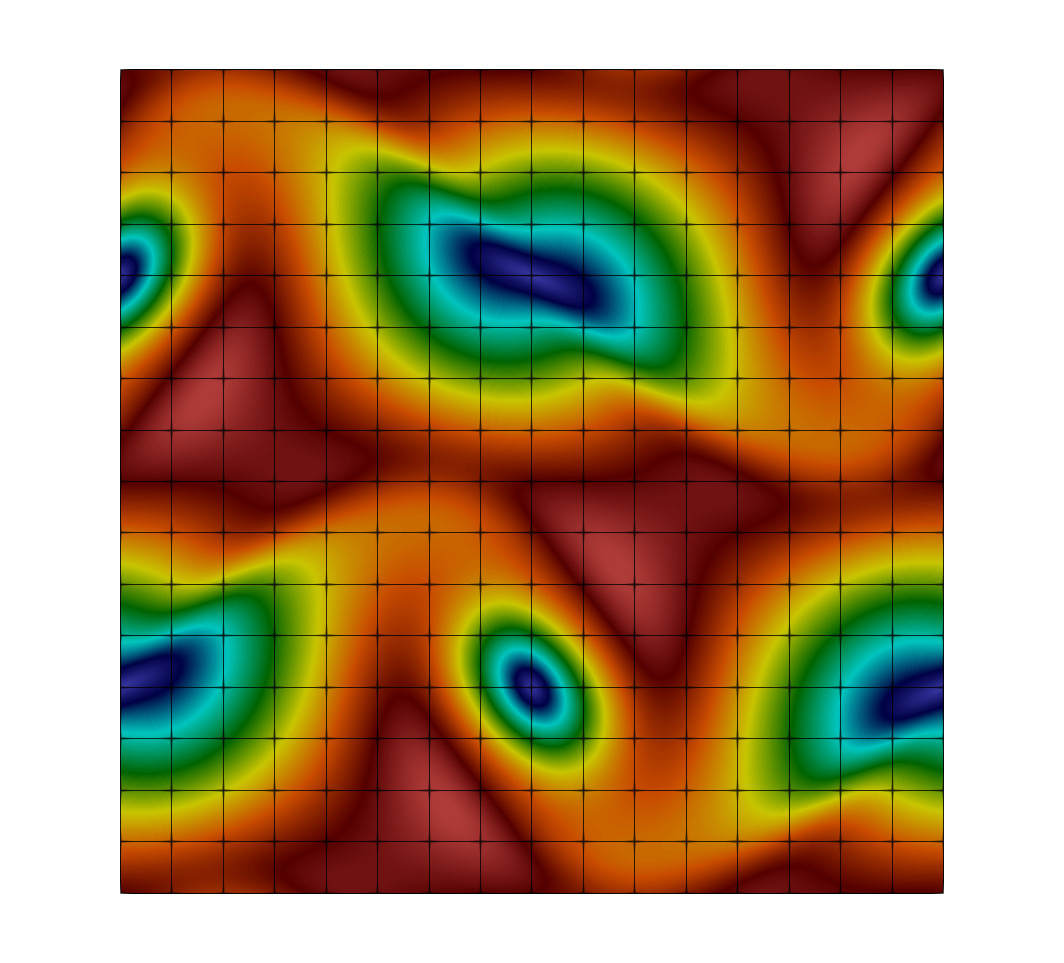}
    \includegraphics[width=0.24\textwidth]{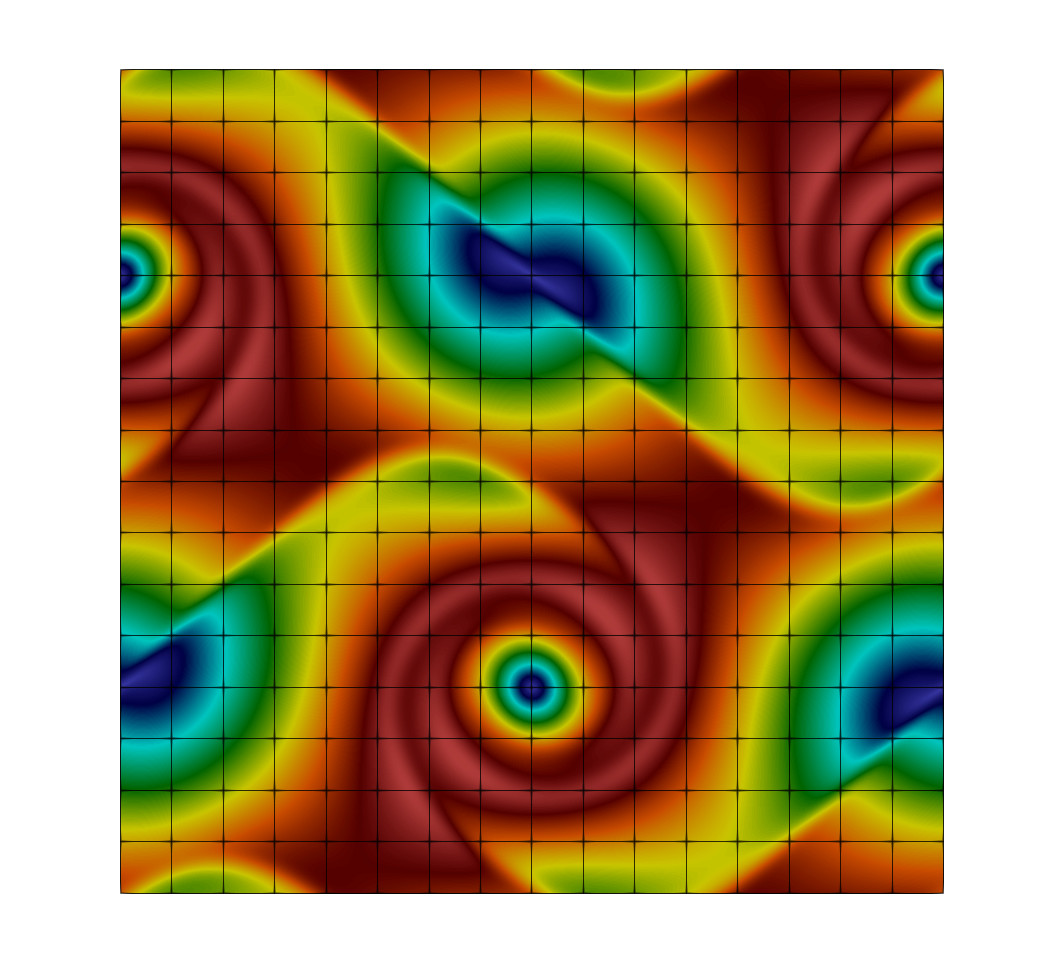}
    \includegraphics[width=0.24\textwidth]{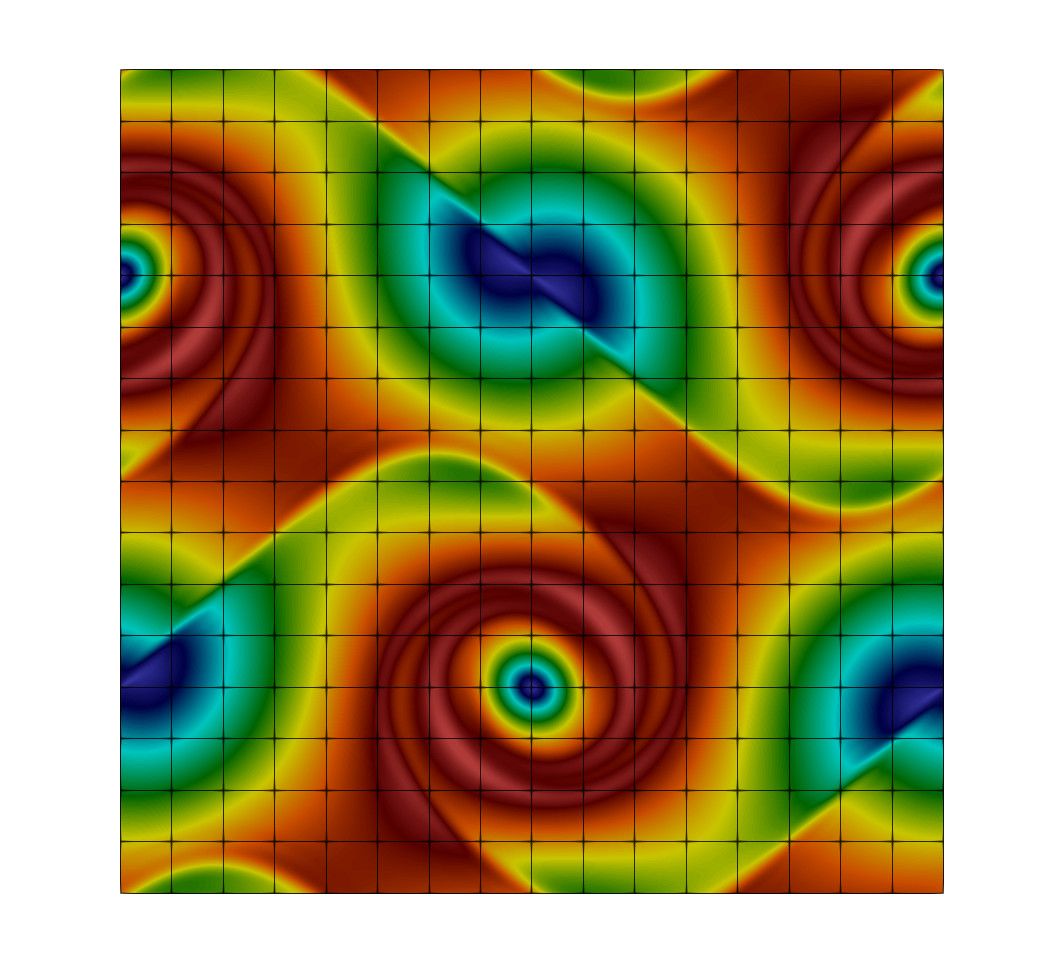}} 
    
\subfloat[Vorticity magnitude for~$\nu=2.5 \cdot 10^{-3}, 10^{-3}, 10^{-4}, 0$ (from left to right).]{
    \includegraphics[width=0.24\textwidth]{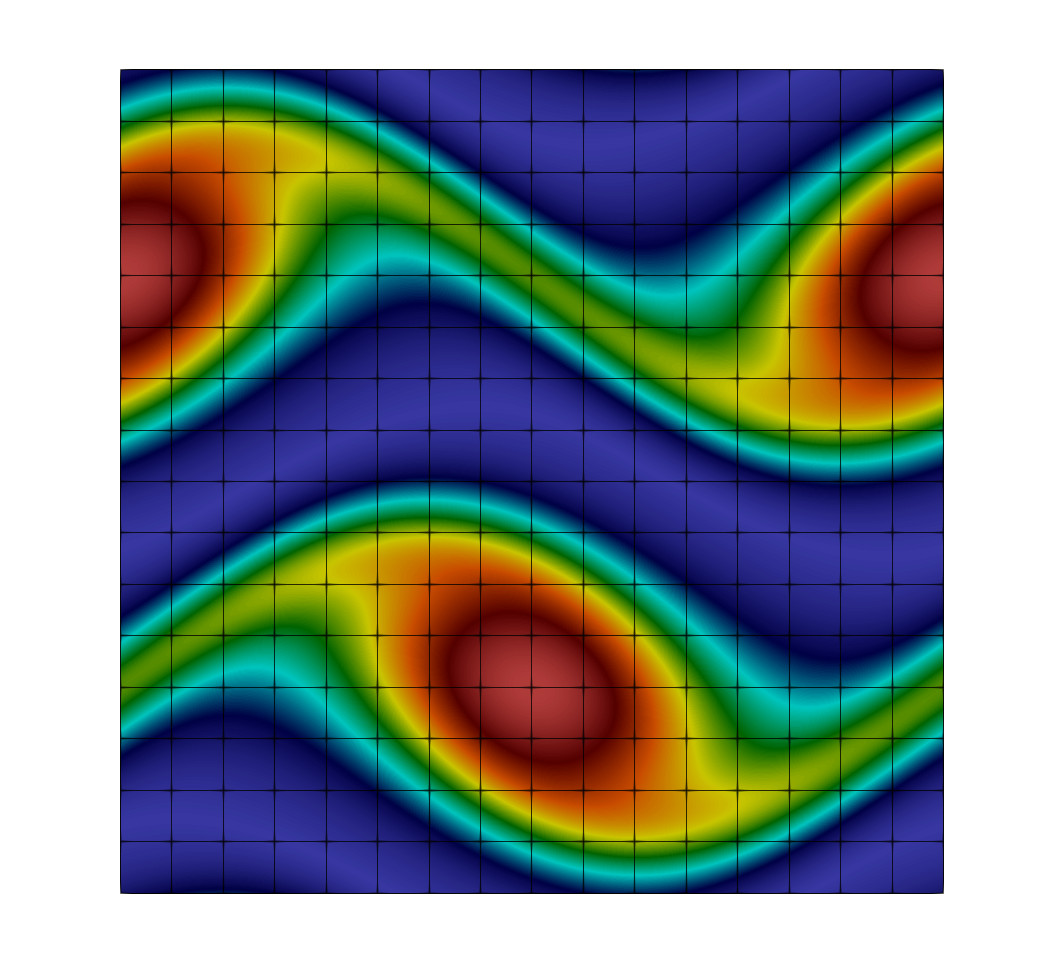}
    \includegraphics[width=0.24\textwidth]{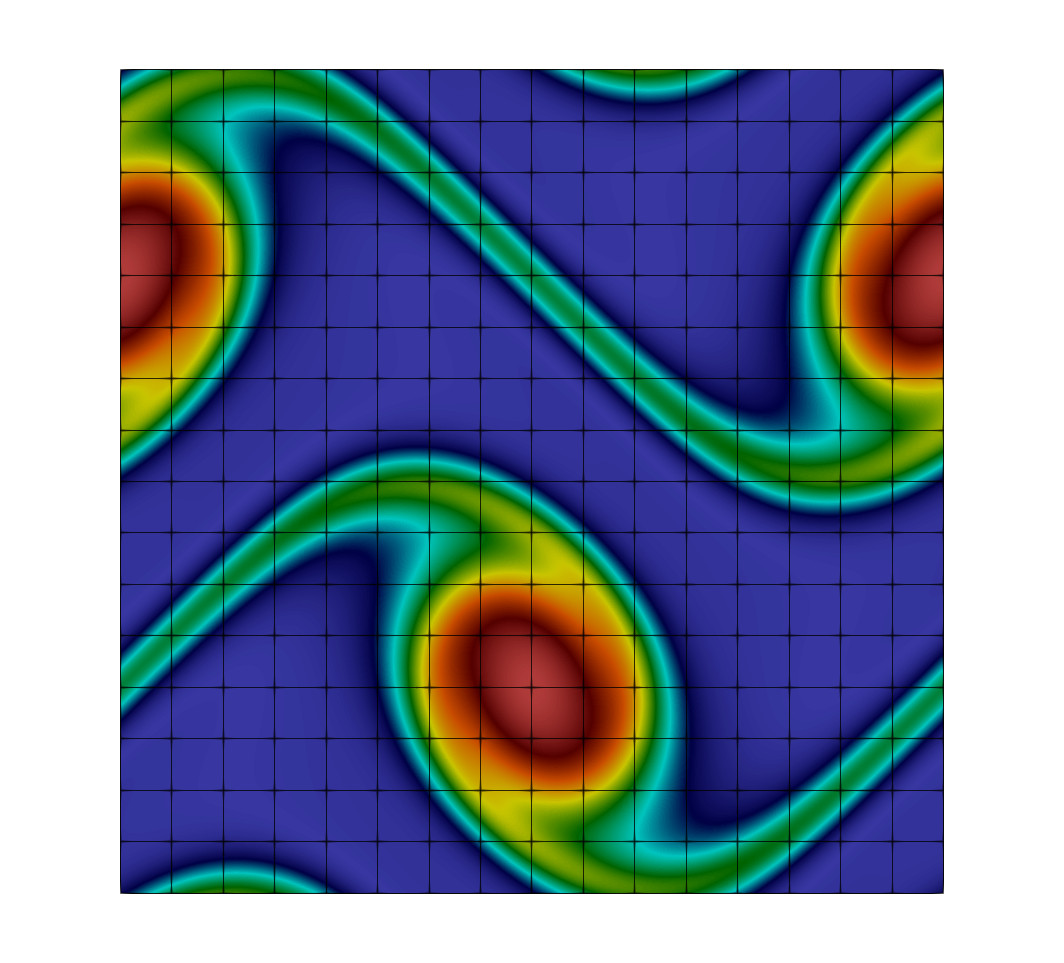}
    \includegraphics[width=0.24\textwidth]{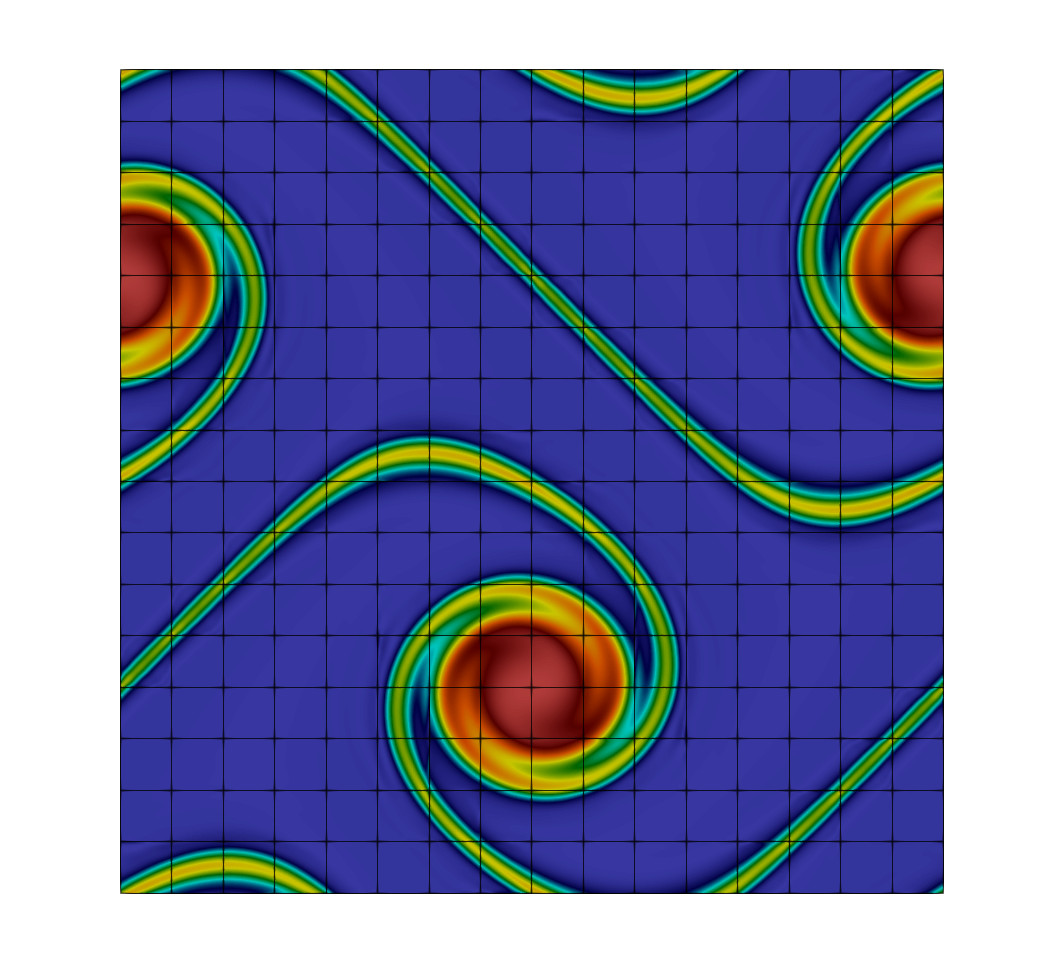}
    \includegraphics[width=0.24\textwidth]{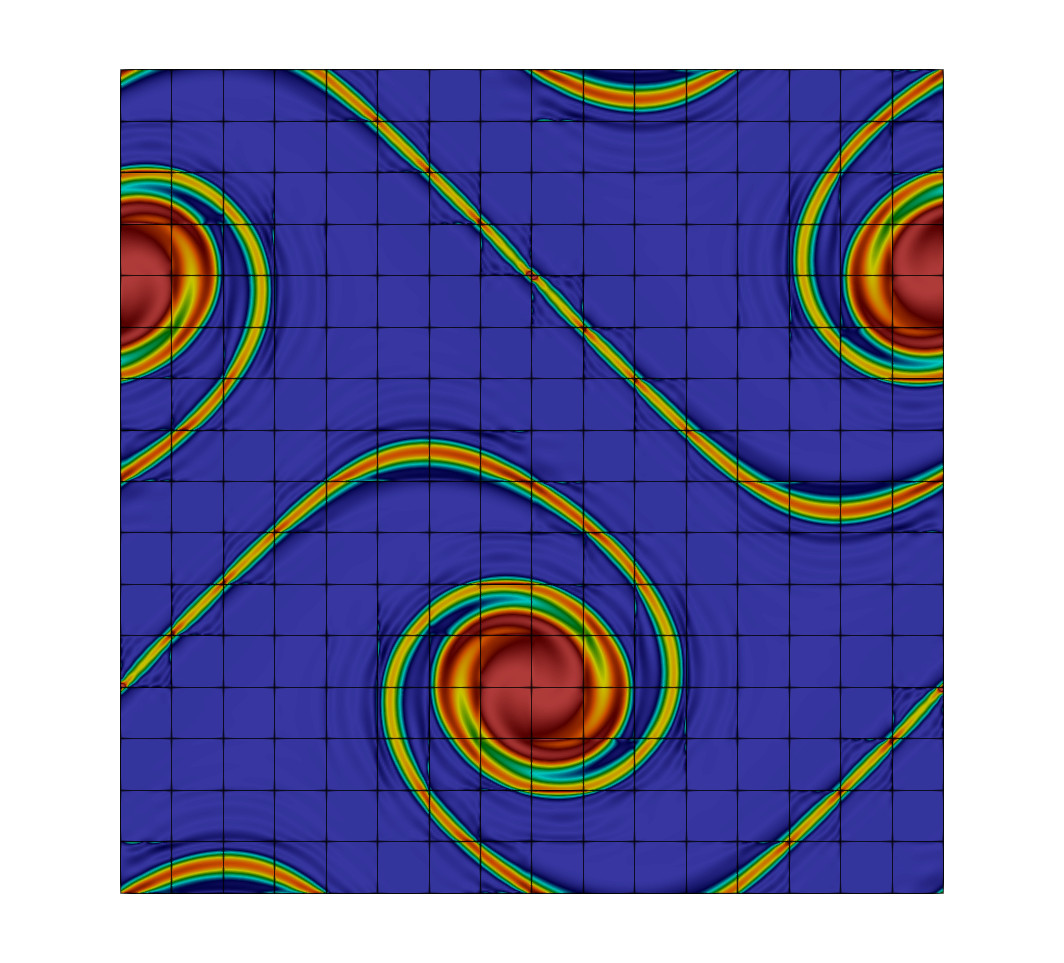}} 

\caption{Two-dimensional shear layer roll-up problem: Contour plots of velocity magnitude and vorticity magnitude at time~$t=1.2$ for four different values of the viscosity (blue indicates low value and red high value). The results shown correspond to a mesh with~$16^2$ elements with a polynomial degree of the shape functions of~$k=7$ (effective resolution~$128^2$).}
\label{fig:2d_shear_layer}
\end{figure}

\begin{figure}[t]
 \centering
	\includegraphics[width=0.49\textwidth]{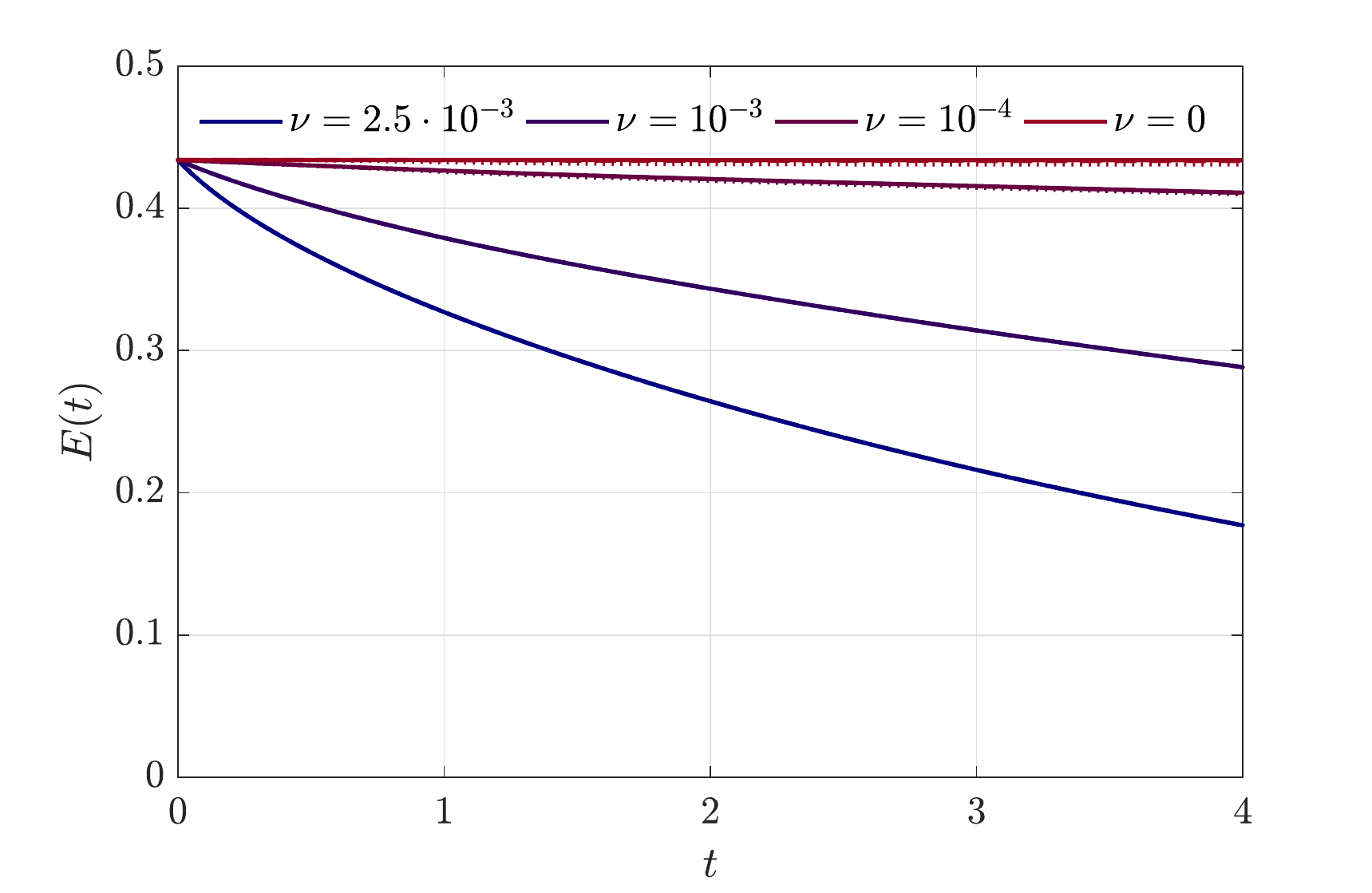}
	\includegraphics[width=0.49\textwidth]{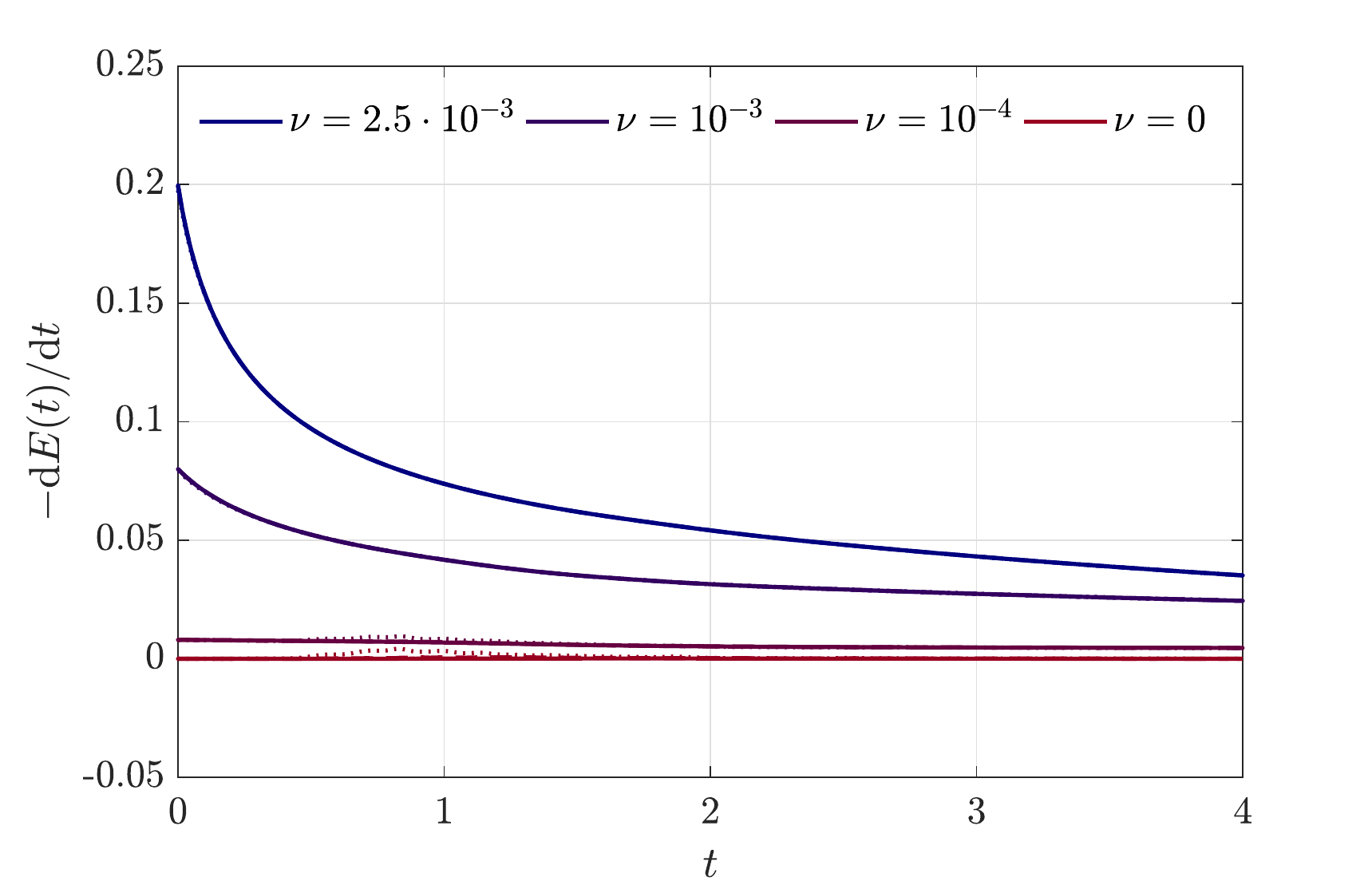}
\caption{Two-dimensional shear layer roll-up problem: Temporal evolution of kinetic energy and kinetic energy dissipation rate for decreasing viscosity values of~$\nu = 2.5 \cdot 10^{-3}, 10^{-3}, 10^{-4}$, and~$0$. For each visosity, results are shown for three different effective resolutions of~$32^2$ (dotted lines),~$64^2$ (dashed lines), and~$128^2$ (solid lines) corresponding to meshes with~$4^2, 8^2$, and~$16^2$ elements with polynomial degree~$k=7$.}
\label{fig:2d_shear_layer_energy}
\end{figure}

The aim of this example is to verify the robustness and accuracy of the present high-order discontinuous Galerkin discretization for a simple two-dimensional example. As mentioned in the introduction, the energy is conserved for the two-dimensional incompressible Euler equations, and this property should be preserved by a consistent discretization scheme for sufficiently fine spatial resolutions. Figure~\ref{fig:2d_shear_layer} shows contour plots of velocity magnitude and vorticity magnitude at time~$t=1.2$ for a mesh with~$16^2$ elements (refinement level~$l=4$) for different values of the viscosity. 
In Figure~\ref{fig:2d_shear_layer_energy}, we show the temporal evolution of the kinetic energy and the kinetic energy dissipation rate for the different viscosity values. For each viscosity, results obtained on three meshes of increasing resolution with~$4^2, 8^2$, and~$16^2$ elements are shown. For large viscosities,~$\nu = 2.5 \cdot 10^{-3}$ and~$10^{-3}$, the results for the temporal evolution of the kinetic energy and dissipation rate coincide for all meshes. Also for the smallest viscosity of~$\nu = 10^{-4}$ and the inviscid limit~$\nu=0$ the results obtained on the two finest meshes coincide and only minor deviations can be observed for the coarsest mesh. This is in qualitative agreement with the contour plots for the velocity magnitude in Figure~\ref{fig:2d_shear_layer} which demonstrate that the velocity field is smooth and well-resolved on the finest mesh for all viscosities. The resolution requirements are higher for the vorticity containing spatial derivatives of the velocity field. As already noted in~\cite{Brown1995}, the vorticity field is still not well-resolved even if convergence has already been achieved for the velocity or kinetic energy. It can be seen from Figure~\ref{fig:2d_shear_layer} that the vorticity field is well-resolved for the viscous cases~$\nu = 2.5 \cdot 10^{-3}, 10^{-3}, 10^{-4}$, but shows grid-dependence with unphysical elevations of the vorticity at the element corners especially in the thin shear layer that is most difficult to resolve. In agreement with what is expected physically, the kinetic energy dissipation rate tends to zero for~$\nu \rightarrow 0$ and the kinetic energy is conserved in the inviscid limit~$\nu=0$.

An important aspect concerns the numerical robustness of the discretization scheme. In~\cite{Chalmers2019}, instabilities are reported for the same shear layer problem with viscosity~$\nu=0$ for a discontinuous Galerkin discretization with polynomial degree~$k=7$ and refinement level~$l=4$. This originates from the fact that the advanced discretization techniques developed in~\cite{Krank2017, Fehn2018a} that render the discretization robust in under-resolved scenarios and that are used in the present work have not been applied in that study. No robustness problems have been observed for the present discretization scheme even for the coarsest resolutions, e.g.~refinement levels of~$l=0,1$ not shown explicitly here, which is a key requirement to obtain a feasible incompressible flow solver for three-dimensional turbulent problems, which are significantly more challenging in terms of the stability of a numerical discretization scheme compared to this two-dimensional example. Instabilities have also been reported for continuous spectral element discretizations for this two-dimensional shear layer problem, where filtering techniques can be used to recover stability~\cite{Fischer2001}, at the cost of introducing new parameters into the discretization scheme. A discretization technique with properties similar to the present stabilized DG discretization in terms of robustness and accuracy are exactly divergence-free $H$(div)-conforming discretizations, see for example~\cite{Schroeder2018} where two-dimensional examples such as the Kelvin--Helmholtz instability problem are discussed which are comparable in nature to the shear layer problem considered here, and the recent comparative study~\cite{Fehn2019Hdiv} analyzing three-dimensional turbulent flow problems for under-resolved scenarios.

\subsection{Three-dimensional Taylor--Green vortex problem}
We consider the 3D Taylor--Green vortex problem~\cite{Taylor1937} defined by the following initial velocity field
\begin{align*}
\bm{u}(\bm{x},t=0) = \left( \sin x_1 \cos x_2 \cos x_3, - \cos x_1 \sin x_2 \cos x_3, 0 \right)^{\mathsf{T}}  .
\end{align*}
Results of viscous simulations for increasing Reynolds number~$\mathrm{Re}=\frac{1}{\nu}$ are given in Figure~\ref{fig:3d_tgv_Re_study}. In the following, the focus in entirely on the inviscid limit~$\nu =0$. The simulations are run over the time interval~$0\leq t \leq T = 20$ to cover the different flow regimes of laminar flow, transition to turbulence, and decaying turbulence. To reduce computational costs for fixed resolution of the flow (or to increase the effective resolution for a given amount of computational costs) it is common practice to exploit the symmetry of the Taylor--Green problem and simulate the flow on the impermeable box~$\Omega = \left[0 , \pi \right]^3$ with symmetry boundary conditions on all boundaries~\cite{Brachet1983}, i.e.,
\begin{align*}
\bm{v}\cdot \bm{n} = 0 \;  ,
\frac{\partial \bm{v}}{\partial n} = 0 \; ,
\end{align*} 
as opposed to the periodic box~$\Omega = \left[-\pi , \pi \right]^3$ that is also used in computational studies. This optimization allows to reduce computational costs by a factor of~$8$ for the present DG discretization. Further symmetries can be exploited by spectral methods leading to the so-called fundamental box~\cite{Brachet1983,Kida1985}.

The computational domain~$\Omega = \left[0 , \pi \right]^3$ is discretized using a uniform Cartesian grid consisting of~$(2^l)^d$ elements, where~$l$ denotes the level of refinement. The number of unknowns is given as~$N_{\mathrm{DoFs}}=(2^l)^d(d(k_u+1)^d+(k_p+1)^d)=(2^l)^d(d(k+1)^d+k^d)$. It is common practice in the literature to express the effective mesh resolution in terms of the periodic box to obtain comparability between different discretization techniques that exploit different levels of symmetry. Hence, we define the effective spatial resolution as~$ (2^{l+1}(k+1))^d$, e.g., the effective resolution is~$64^3$ for refine level~$l=3$ and polynomial degree~$k=3$. Absolute tolerances of~$10^{-12}$ and relative tolerances of~$10^{-6}$ are used for the iterative linear solvers, where relative tolerance means that the residual is reduced by a factor of~$10^{-6}$ compared to the initial residual that uses as initial guess a high-order extrapolation of the solution from previous time steps. The polynomial degree used for the TGV simulations is~$k=3$ and adaptive time stepping with~$\mathrm{Cr}=0.25$ is used for all simulations. The penalty factors of the divergence and continuity penalty terms are chosen as defined in~\cite{Fehn2018a}, except for the finest resolution of~$8192^3$ where the penalty factors have been increased by a factor~$\zeta=2$ compared to the standard definition. For this fine resolution, the simulation remained stable also for the default value of~$\zeta = 1$, but we observed oscillations in the maximum vorticity at early times, given an indication of the need for a slightly larger penalization of the divergence-free constraint and normal continuity of the velocity field. Since these oscillations disappeared when increasing the penalty factors by a factor of~$2$, this value has finally been used for this highest resolution simulation. The highest resolution of~$8192^3$ has~$N_{\mathrm{DoFs}}=2.35 \cdot 10^{11}$ unknown degrees of freedom, and $2.27 \cdot 10^{5}$ time steps have been solved during this simulation. The computations have been performed on a large supercomputer using almost~$100\mathrm{k}$ cores for the highest resolution, requiring a run time of approximately~$8.4$ days.

\begin{figure}[t]
 \centering 
 \subfloat[Velocity magnitude on plane~$x=\pi$ at times~$t=0,1,2,3,4$ (from left to right).]{
    \includegraphics[width=0.19\textwidth]{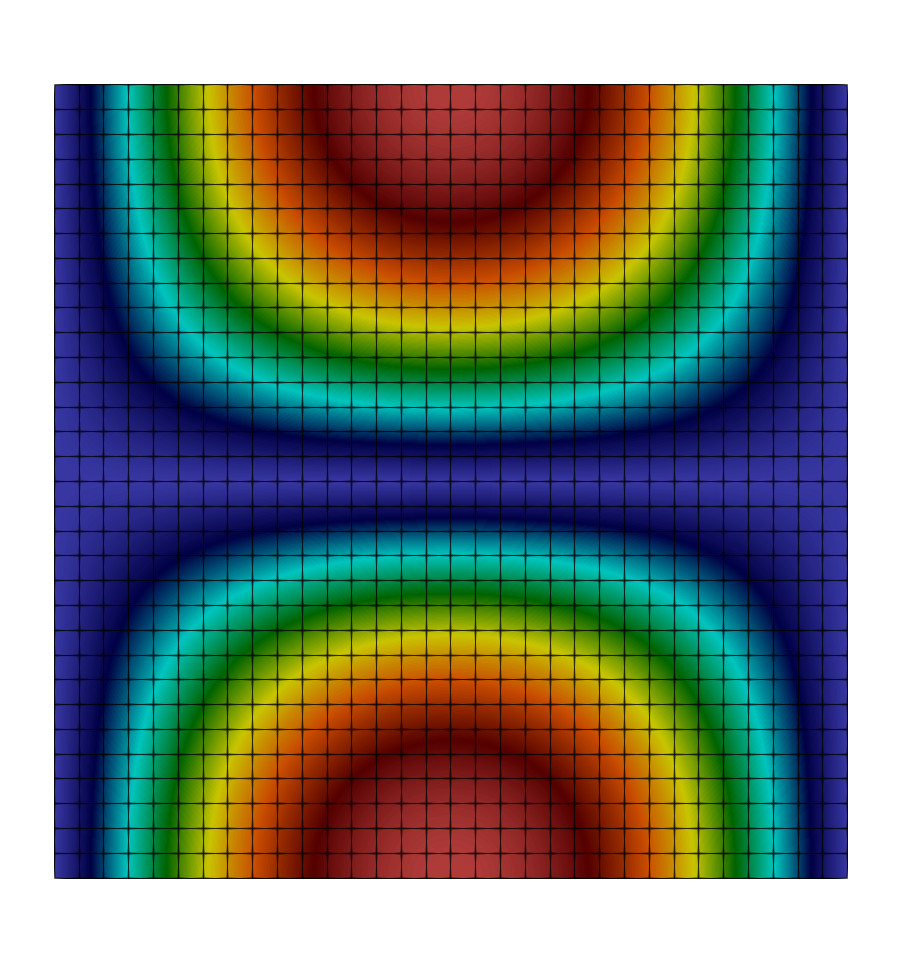}
    \includegraphics[width=0.19\textwidth]{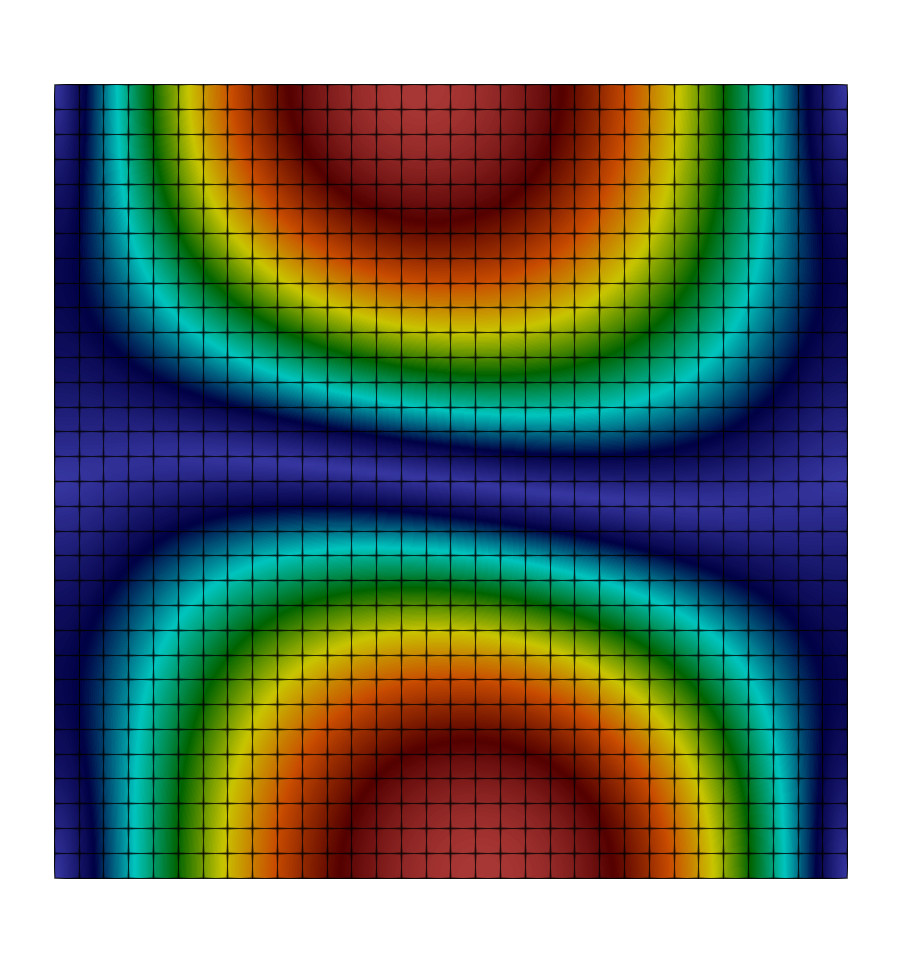}
    \includegraphics[width=0.19\textwidth]{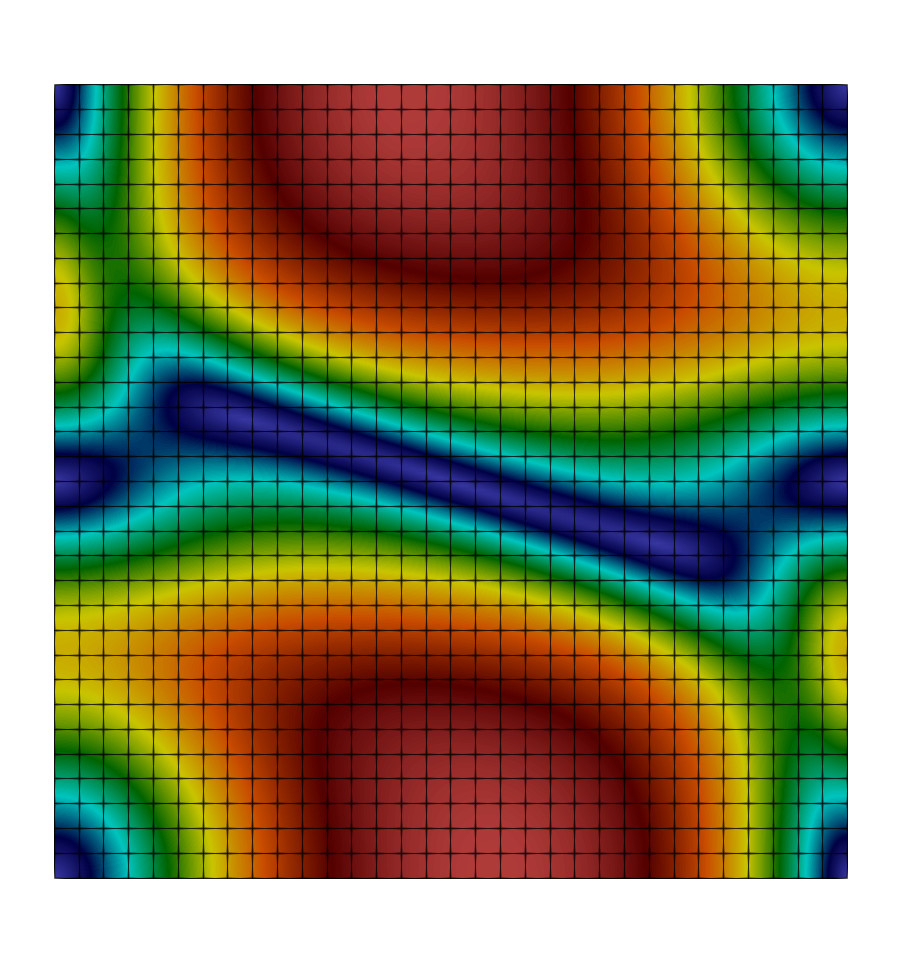}
    \includegraphics[width=0.19\textwidth]{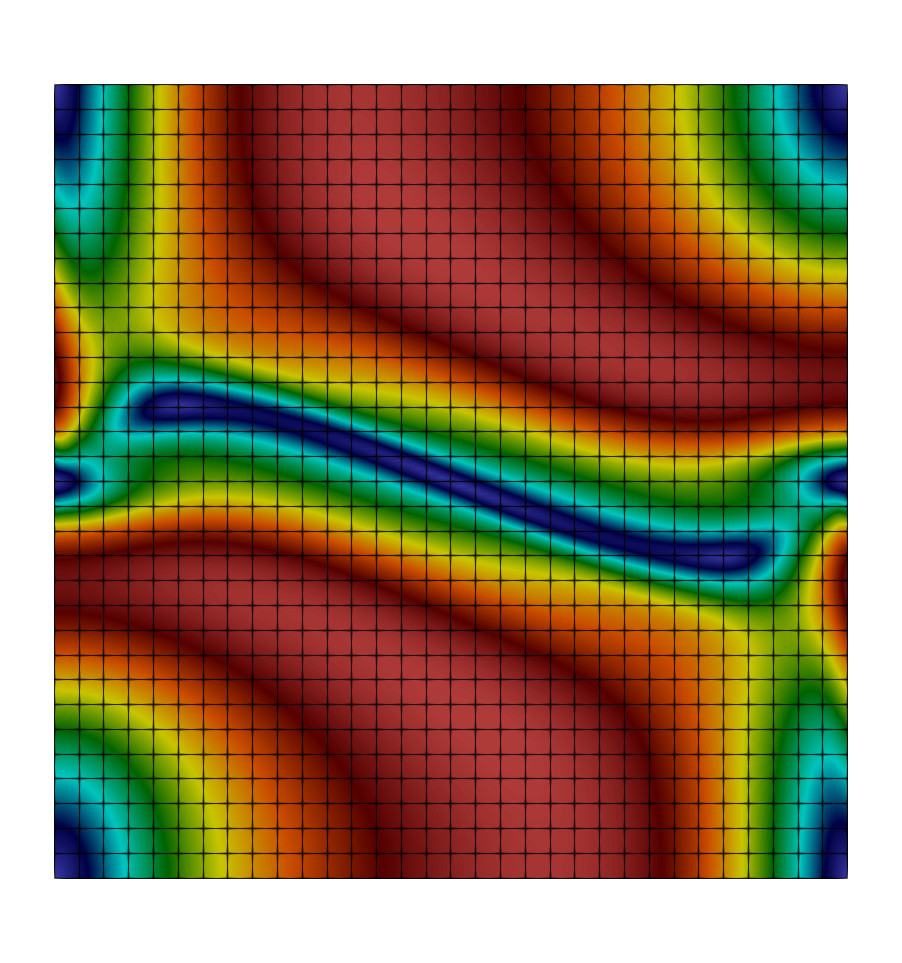}
    \includegraphics[width=0.19\textwidth]{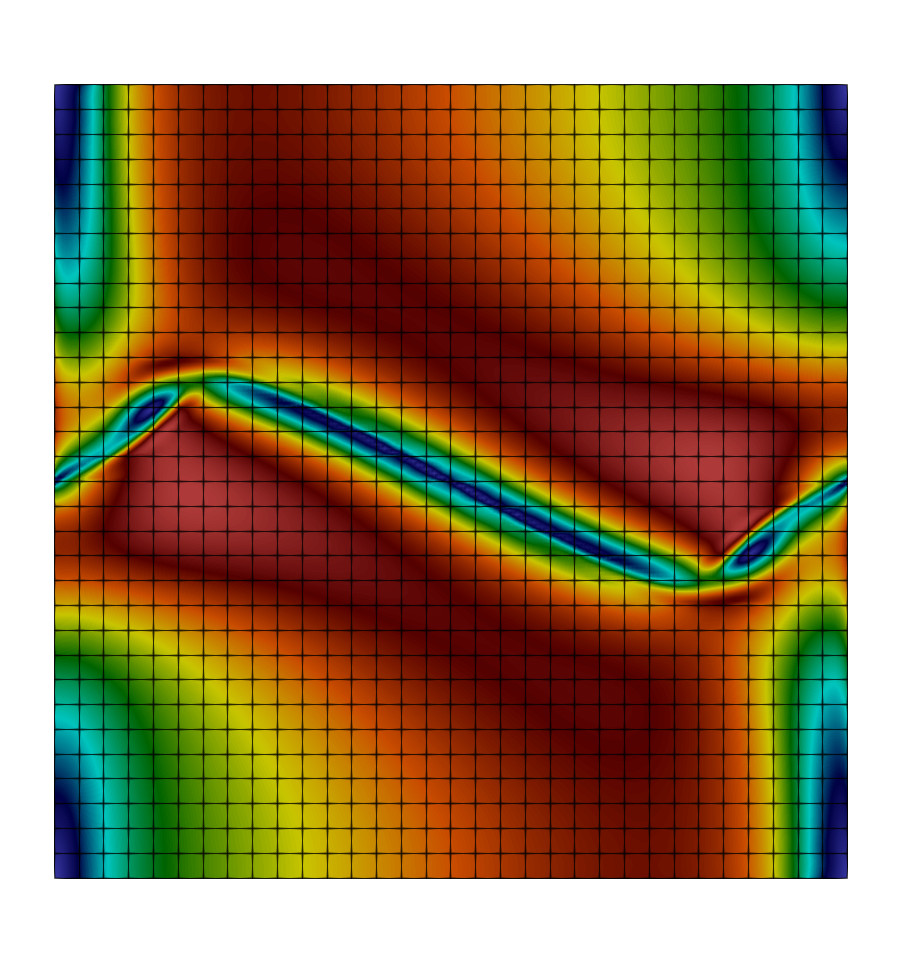}}
    
\subfloat[Vorticity magnitude on plane~$x=\pi$ at times~$t=0,1,2,3,4$ (from left to right).]{
    \includegraphics[width=0.19\textwidth]{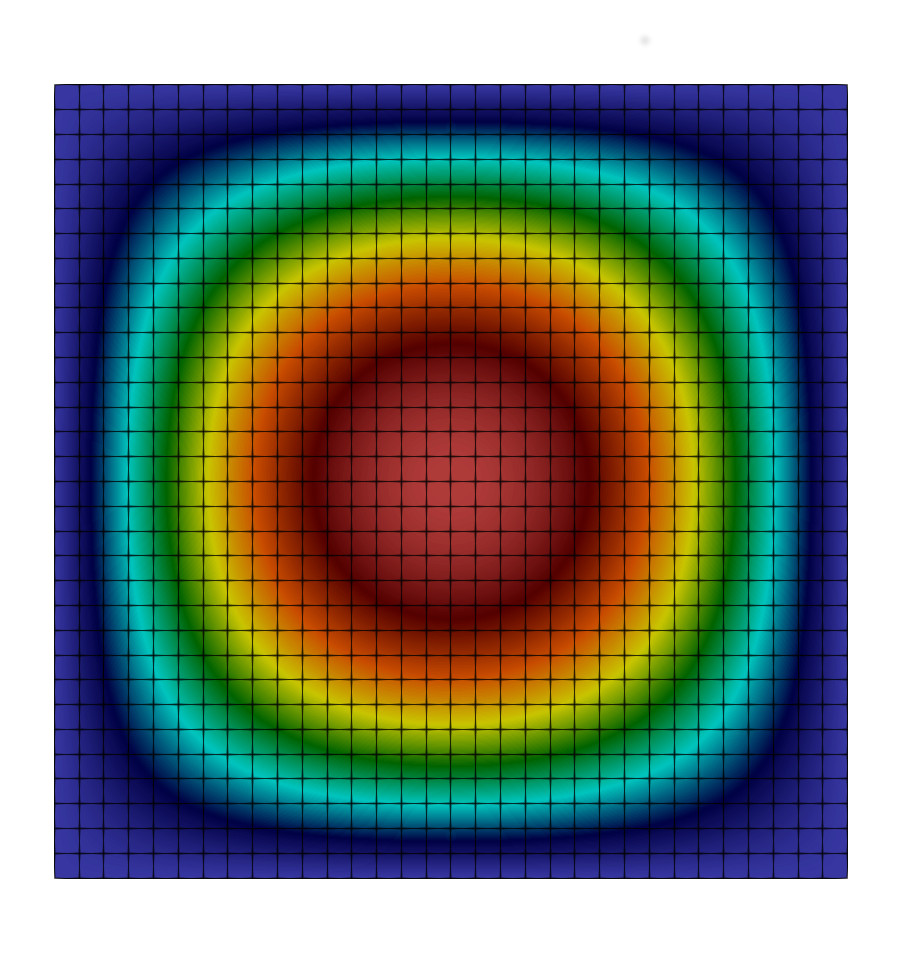}
    \includegraphics[width=0.19\textwidth]{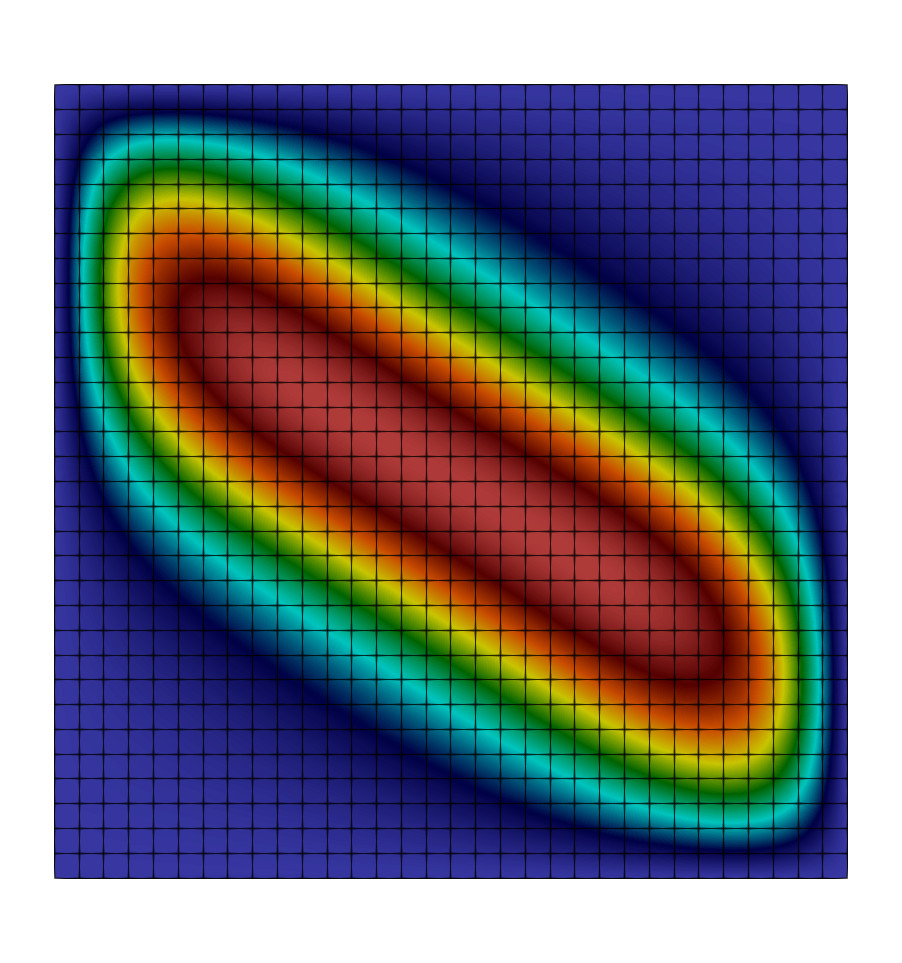}
    \includegraphics[width=0.19\textwidth]{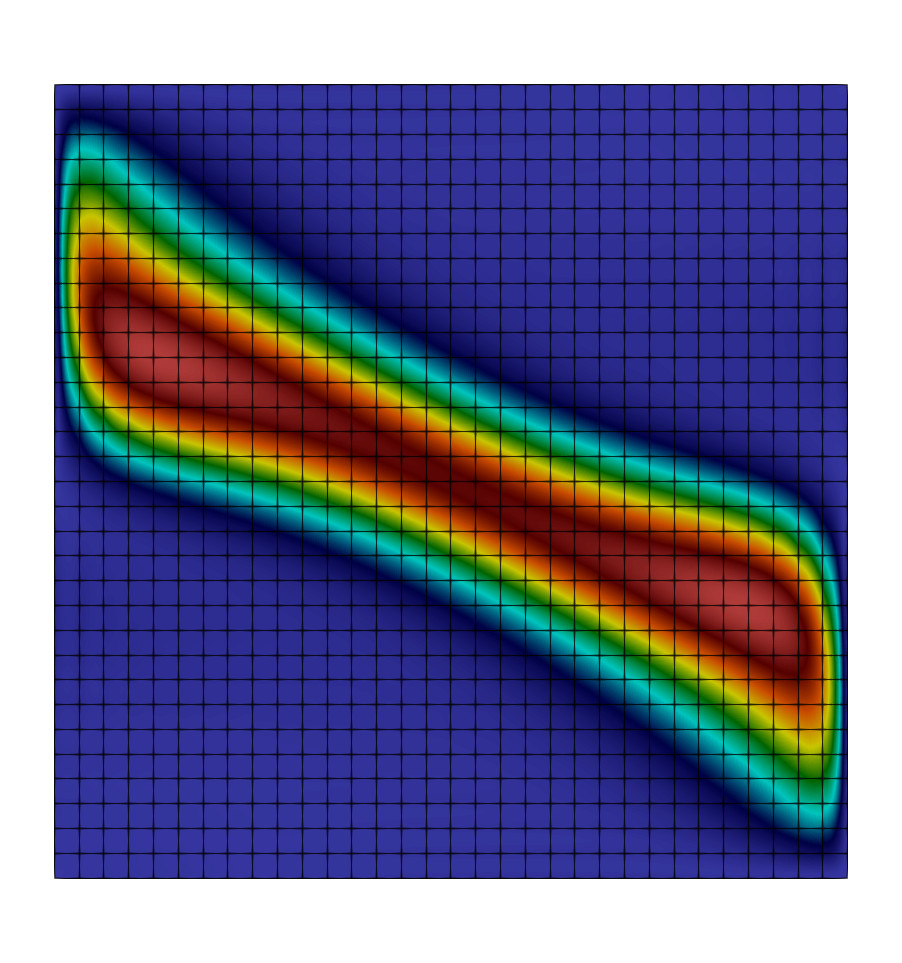}
    \includegraphics[width=0.19\textwidth]{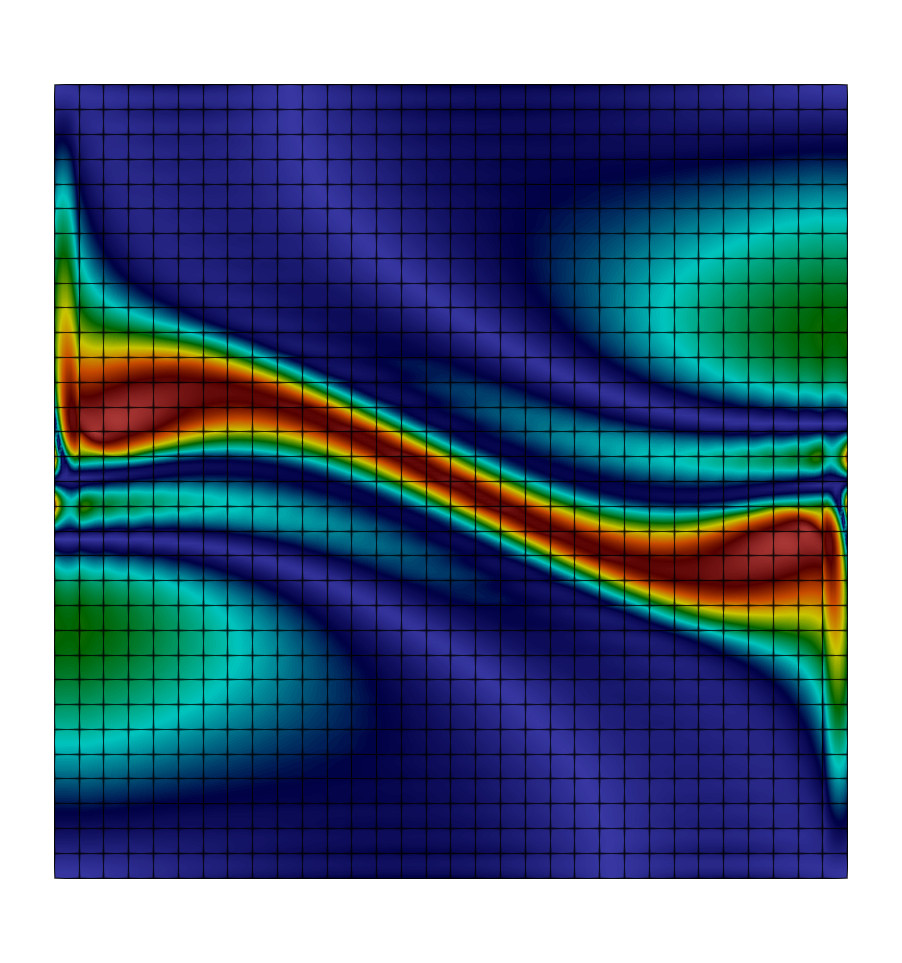}
    \includegraphics[width=0.19\textwidth]{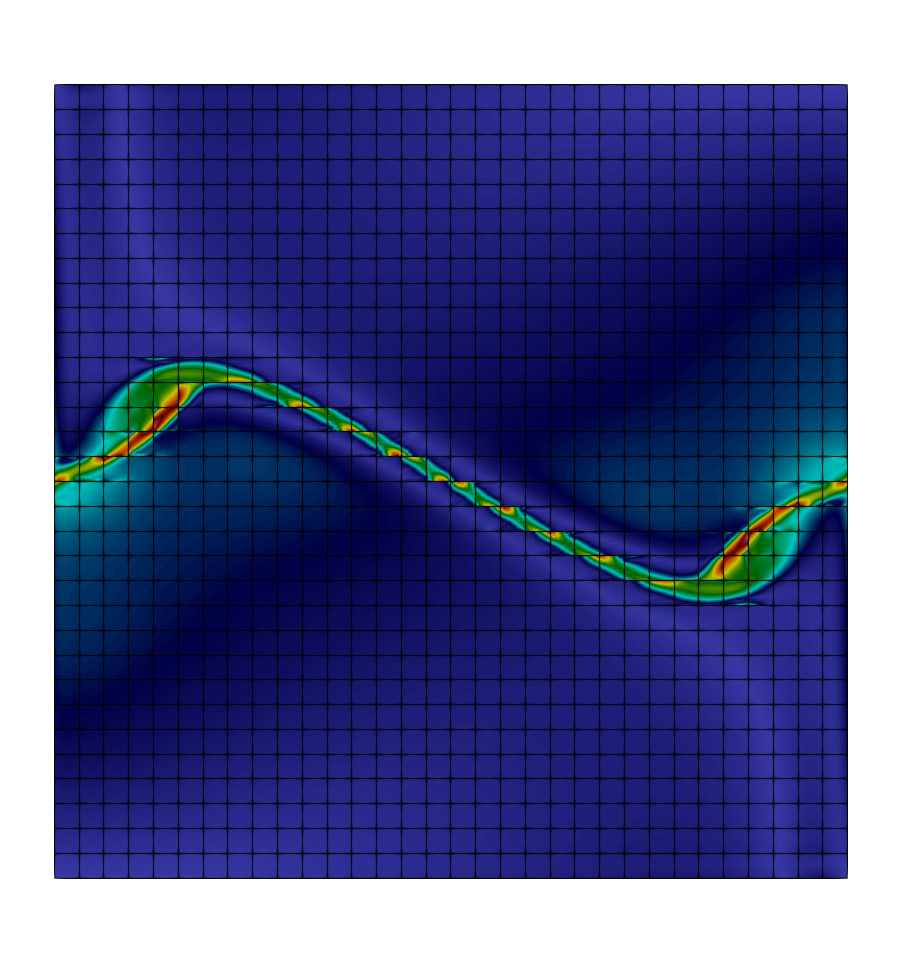}}

\caption{Three-dimensional inviscid Taylor--Green problem: Contour plots of velocity magnitude and vorticity magnitude (blue indicates low value and red high value) on plane~$x=\pi$ of impermeable box. The results shown correspond to a mesh with~$32^3$ elements with a polynomial degree of the shape functions of~$k=3$ (effective resolution~$256^3$).}
\label{fig:3d_tgv_visualization}
\end{figure}

We briefly summarize the type of discretization, the maximal effective resolutions, and the final time~$T$ of the simulations considered in previous numerical studies for the three-dimensional inviscid Taylor--Green vortex problem. In~\cite{Brachet1983}, a spectral method with maximum resolution of~$256^3$ (exploiting symmetry) has been used and direct simulation has been performed up to times~$t \leq 4$. In a subsequent work~\cite{Brachet1992}, a maximum resolution of~$864^3$ (exploiting symmetry) has been reached, and again direct simulation has been performed up to times~$t \leq 4$. A comparsion of a spectral method and WENO finite difference method can be found in~\cite{Shu2005}, where a maximum resolution of~$368^3$ (exploiting symmetry) and simulation up to times~$t \leq 6$ has been considered. A modal discontinuous Galerkin method has been studied in~\cite{Chapelier2012}, with the simulations performed up to times~$t\approx 5-7$ until the simulations became unstable, with the maximum resolution of around~$96^3$ for different polynomial degrees from~$k=1$ to~$k=5$. DG discretizations used to study the inviscid TGV problem have also been analyzed in~\cite{Moura2017, Moura2017setting, Fernandez2018arxiv, Manzanero2020, Schroeder2019}, but with a focus on LES modeling. The study~\cite{Cichowlas2005} used a spectral method with maximum resolution of~$2048^3$ with simulations performed up to times~$t \leq 4$. The highest resolution of~$4096^3$ has been achieved in~\cite{Bustamente2012} using a spectral method, and the simulations have been performed up to times~$t \leq 4$.

In these works, indications of finite-time singularities are mentioned. In~\cite{Morf1980},~$t_* = 5.2$ is obtained from power series expansions. A more accurate variant using power series expansions presented in~\cite{Brachet1983} leads to~$t_* = 4.4 \pm 0.2$. Furthermore, the study~\cite{Brachet1983} reports indirect evidence for a finite time singularity according to the direct numerical simulation results but the authors conclude that the resolution of~$256^3$ is not sufficient to investigate the occurrence of finite-time singularities for times~$t \geq 4$. The more recent study~\cite{Bustamente2012} estimates a blow-up time of~$t_* \approx 4$ and concludes that the results are not inconsistent with the occurrence of a singularity. The work~\cite{Larios2018} obtains a blow-up time of~$t_* \approx 4.2$ similar to the blow-up time in~\cite{Brachet1983}.

The early stage of the Taylor--Green vortex evolution with the formation of thin flow structures is visualized in Figure~\ref{fig:3d_tgv_visualization}. Similar results have been shown and discussed in detail already in~\cite{Brachet1983} for the same effective resolution of~$256^3$ using a spectral method. In agreement with the above results for the two-dimensional shear layer problem, we observe that the velocity field is resolved at all times, while under-resolution effects are clearly visible in the contour plots of the vorticity magnitude at later times~$t=3$ and~$t=4$ for the chosen resolution. A high resolution visualization of the thin vortex sheet shown in Figure~\ref{fig:3d_tgv_visualization} with a volume rendering of the vorticity magnitude has been shown in~\cite{Bustamente2012} for an effective resolution of~$4096^3$. Flow visualization is one possibility to trace finite-time singularities, but appears to be impracticable due to the difficulties in handling large data sets for high resolution simulations that would be necessary for such an investigation. Hence, the attention is turned to other techniques in the following.

We present numerical results of a mesh convergence study for refinement levels~$l=3,...,10$ and polynomial degree~$k=3$.
\begin{figure}[t]
 \centering
	\includegraphics[width=0.49\textwidth]{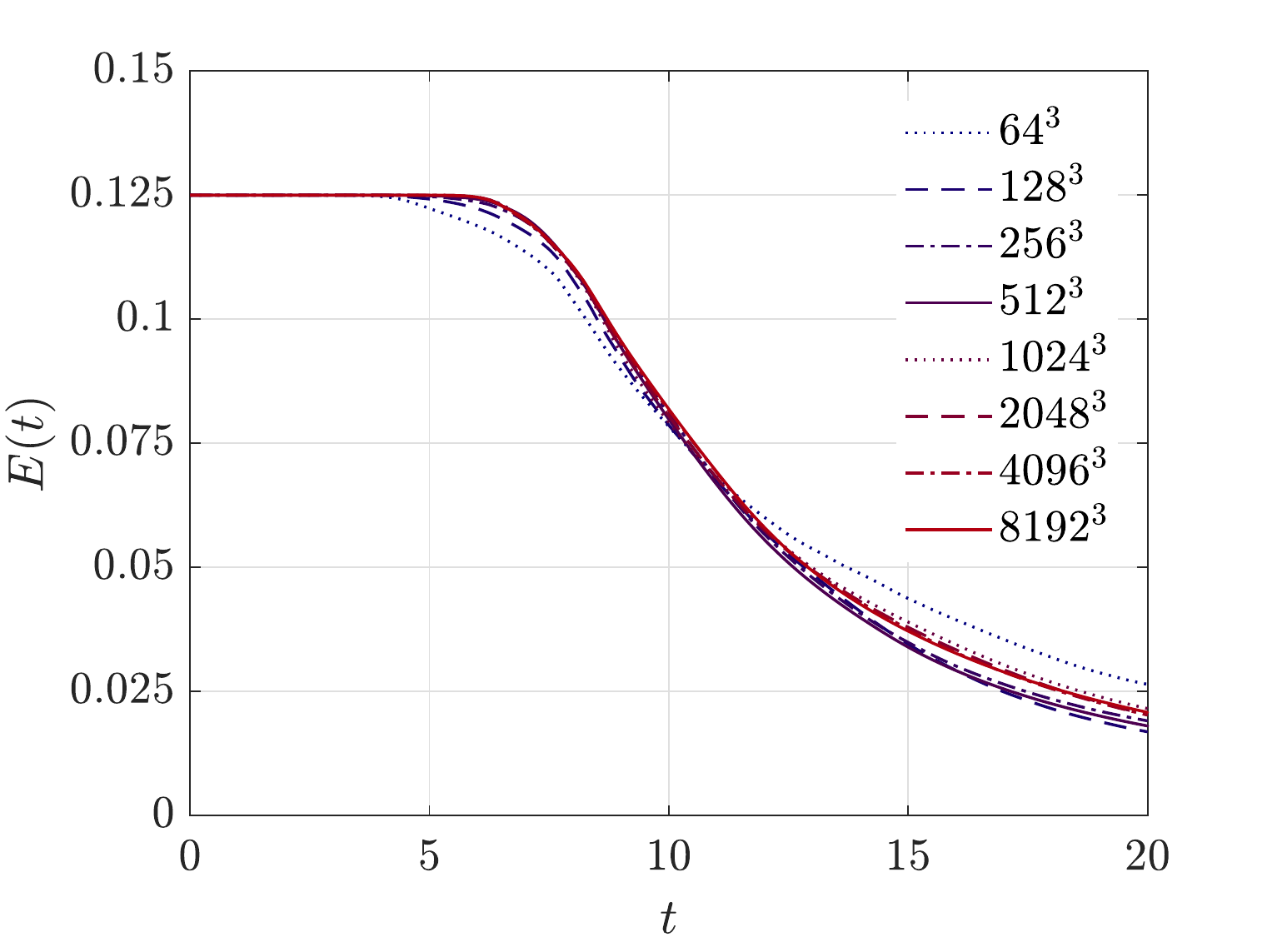}
	\includegraphics[width=0.49\textwidth]{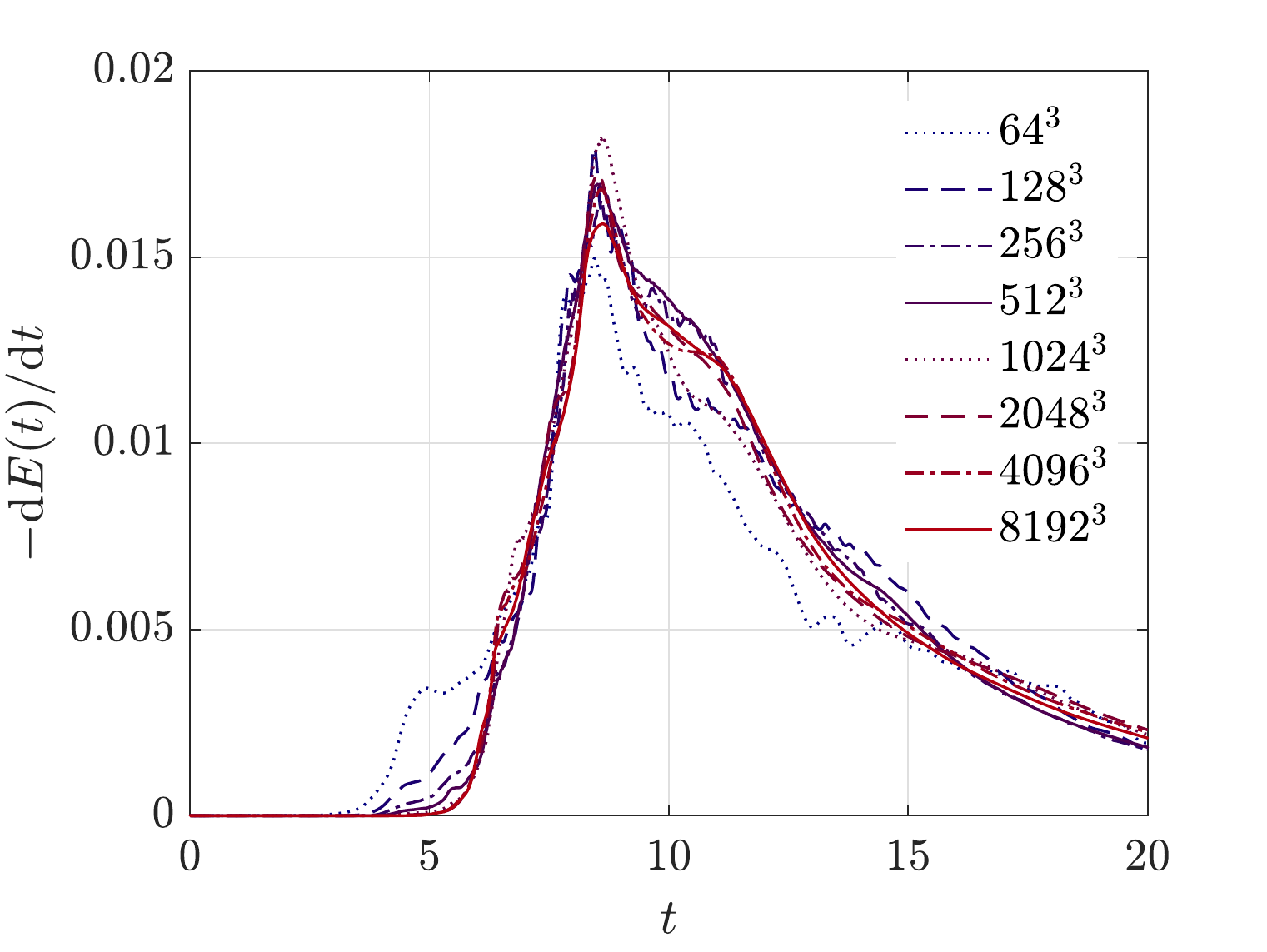}
\caption{Three-dimensional inviscid Taylor--Green problem: Temporal evolution of kinetic energy (left) and kinetic energy dissipation rate (right) for increasing spatial resolution.}
\label{fig:3d_tgv_energy_and_dissipation_rate}
\end{figure}
Figure~\ref{fig:3d_tgv_energy_and_dissipation_rate} shows the temporal evolution of the kinetic energy and the kinetic energy dissipation rate. At small times~$t$, the energy is constant and the energy dissipation rate is zero. This agrees with the expected theoretical behavior stating energy conservation as long as the solution remains smooth and has also been shown in previous works in a similar way. In this work, we do not want to terminate the simulations once we expect the simulation to become under-resolved, but instead continue the simulations until~$t=20$. Depending on the effective mesh resolution, an onset of energy dissipation can be observed that shifts towards later times for increasing spatial resolutions. However, this time of onset of dissipation is not pushed beyond~$t\approx 5$ even for the finest spatial resolutions. Instead, the kinetic energy evolution and its dissipation rate tend to converge to a dissipative solution. As in viscous simulations of this problem, the kinetic energy dissipation rate reaches a maximum at~$t =8 - 9$ and decreases afterwards. For small times up to~$t\approx 6$ it can be observed that the simulations converge uniformly towards the solution obtained on the finest mesh. At later times~$t > 8-9$, where the flow is expected to be fully turbulent, convergence to a well-defined reference solution is less clear. Instead, the kinetic energy dissipation rate exhibits a behavior that can be described as fluctuating around some mean profile when considering the sequence of spatial resolutions. A possible explanation might be the strong nonlinearity of the inviscid problem that renders the problem more sensitive than the viscous cases, so that small influences stemming from the temporal discretization and especially the spatial discretization, which can be expected to operate at its limit of resolution for this inviscid problem, might trigger larger deviations and finally cause the solution to follow one or another path. A more rigorous quantification of convergence to a reference solution is shown in Section~\ref{sec:Convergence} below.

\begin{figure}[t]
 \centering
	\includegraphics[width=0.49\textwidth]{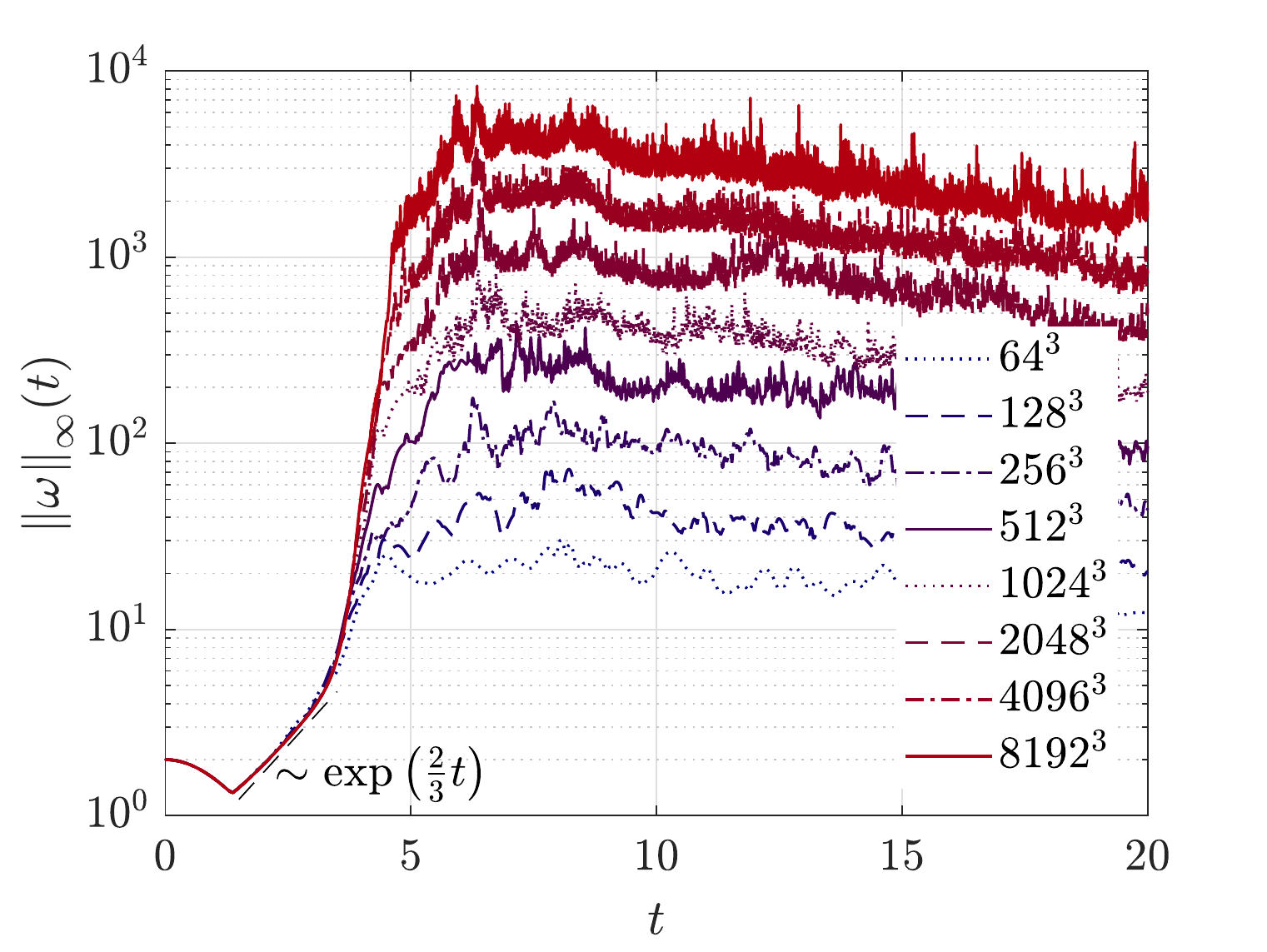}
	\includegraphics[width=0.49\textwidth]{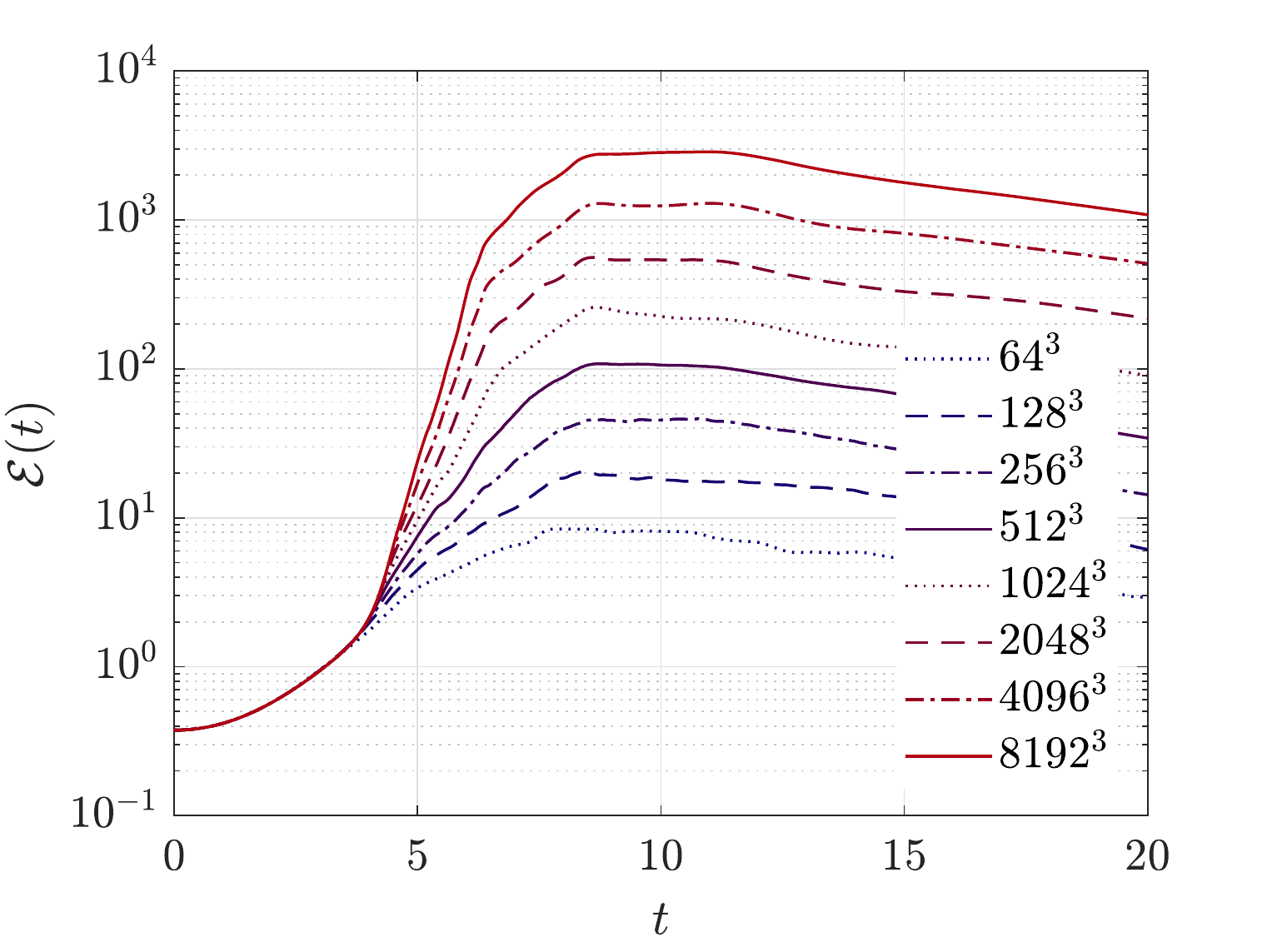}
\caption{Three-dimensional inviscid Taylor--Green problem: Temporal evolution of maximum vorticity (left) and enstrophy (right) for increasing spatial resolution.}
\label{fig:3d_tgv_maximum_vorticity_and_enstrophy}
\end{figure}

It is now examined whether this behavior is consistent with what is observed for the temporal evolution of the maximum vorticity and the enstrophy shown in Figure~\ref{fig:3d_tgv_maximum_vorticity_and_enstrophy}. For both quantities, one can immediately identify three phases: (i) a first phase up to approximately~$t \approx 3$ in which the flow is well resolved for all spatial resolutions so that the results essentially overlap for all simulations, (ii) an intermediate phase~$3 < t < 5$ in which the different simulations start to deviate from each other due to under-resolution effects depending on the spatial resolution of each simulation, and~(iii) a final phase~$t > 5$ in which the results of all simulations deviate substantially due to the different resolution capabilities of the different simulations. In the first phase, the vorticity first decreases, reaches a minimum, and then begins to grow exponentially in agreement with the results shown in~\cite[Figure 1 (b)]{Bustamente2012}. In Figure~\ref{fig:3d_tgv_maximum_vorticity_and_enstrophy}, we added a reference curve with~$\exp\left(\frac{2}{3} t\right)$ growth which describes the growth of the maximum vorticity very well in this regime. In the second phase at around~$t \approx 3.5$, the maximum vorticity begins to grow substantially faster, and the growth of vorticity essentially depends on the spatial resolution that is directly linked to the maximum velocity gradient that can be represented on a given mesh. As already mentioned in the introduction, for every simulation that does not blow up due to numerical instabilities of a discretization scheme, the maximum vorticity will remain finite no matter how fine the spatial resolution is. Therefore, the occurrence of a finite-time singularity with~$\lim_{t\rightarrow t_*} \Vert \boldsymbol{\omega} \Vert_{\infty} =  \infty$ remains speculative. Our results are clearly not inconsistent with such a vorticity blow-up scenario, but from the present results we are not able to identify a concrete blow-up time~$t = t_*$. We rather observe that the time of maximum vorticity observed in this second phase is shifted to later times also for the finest spatial resolutions. At the same time, one might argue that a finite-time blow-up at a time~$t = t_*$ with $t_* \approx 4.5-5$ would produce results similar to those shown here, with the maximum vorticity following the exact profile until the curve of a specific spatial resolution branches off due to under-resolution of the simulation. In such a scenario, one would expect the maximium vorticity to grow by a factor of~$2$ for refinement level~$l \rightarrow l+1$  due to the mesh size being reduced by a factor of~$2$, allowing numerical gradients becoming twice as large. While we observe such an increase in maximum vorticity from one refinement level to the next, it is not clear whether this suspected blow-up would happen in finite time. However, we summarize that the present results do not support a blow-up of the vorticity at times as early as~$t_*\approx 4$ or~$t_*\approx 4.2$ obtained in previous studies, since the highest resolution simulations traverse this point without a particular event. In the third phase, the maximum vorticity reaches a global maximum between~$t=6-7$ for each resolution before it starts to decrease slowly. In this phase, the maximum vorticity is offset by a factor of approximately~$2$ from one refine level to the next. This is a clear indication that none of the simulations is able to resolve the finest structures, and it is plausible that a factor of~$2$ in mesh size also gives a factor of~$2$ in maximum vorticity. Finally, the maximum vorticity shows a strongly fluctuating behavior in the third phase. This behavior is consistent with the observation that no clear convergence could be identified in the kinetic energy dissipation rate for times around and beyond the dissipation maximum. 

The temporal evolution of the enstrophy is overall similar to that of the maximum vorticity. An important difference is that the enstrophy does not reach a local minimum at early times as observed for the maximum vorticity. In the second phase, the growth of enstrophy is more moderate compared to the maximum vorticity. In the third phase, the enstrophy curves can be described essentially as smoothed variants of the maximum vorticity from which the high-frequency content has been removed. A possible explanation for the enstrophy behavior in the second and third phase is that the enstrophy is not a local quantity, but an average in space over the computational domain. Further, the maximum vorticity is determined numerically by taking the maximum over all quadrature points, i.e., the vorticity field is sampled in discrete points. This effect is negligible for well-resolved scenarios but can be expected to give a highly oscillatory behavior in case a local maximum travels through the domain. Again, a growth in enstrophy by a factor of two from one mesh level to the next is observed at later times, which is consistent with an enstrophy evolution theoretically taking infinite values, or taking finite values but much larger than those obtained numerically in Figure~\ref{fig:3d_tgv_maximum_vorticity_and_enstrophy}. Considering Onsager's conjecture as valid, it is clear that one can not expect convergence for the temporal evolution of maximum vorticity and enstrophy. Theoretically, convergence can then only be expected for the kinetic energy evolution and to some extent for kinetic energy spectra discussed in the following.

\begin{figure}[!ht]
 \centering
	\subfloat[Effective resolution of~$128^3$]{\includegraphics[width=0.9\textwidth]{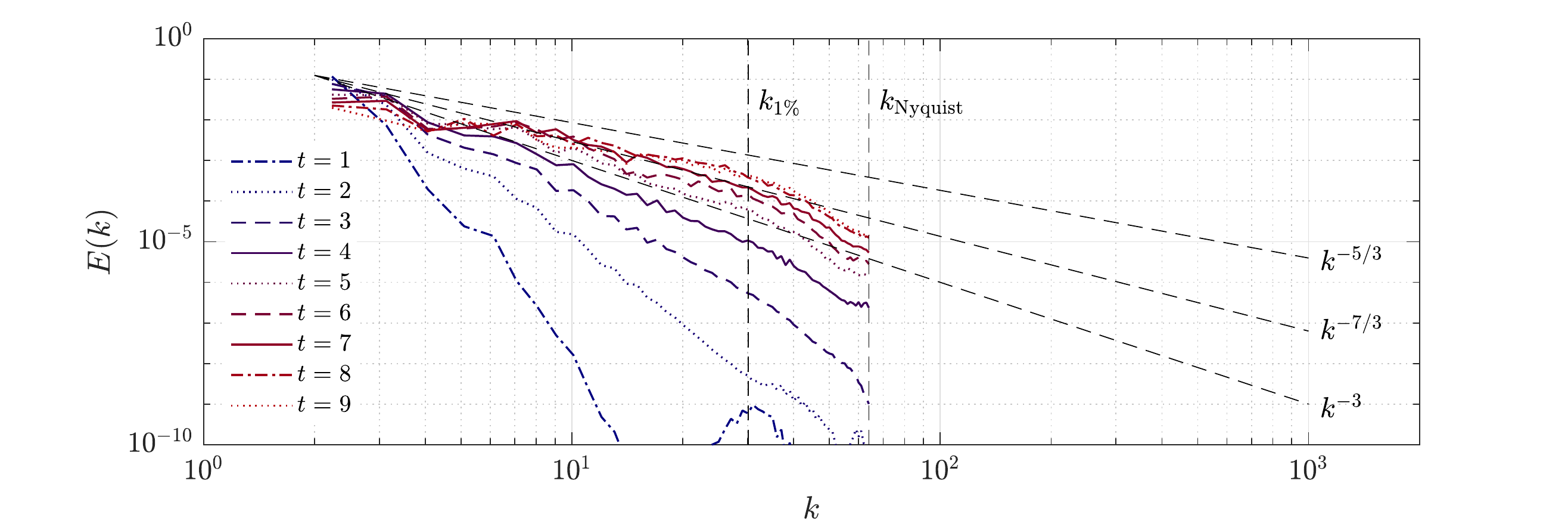}}
	\qquad
	\subfloat[Effective resolution of~$256^3$]{\includegraphics[width=0.9\textwidth]{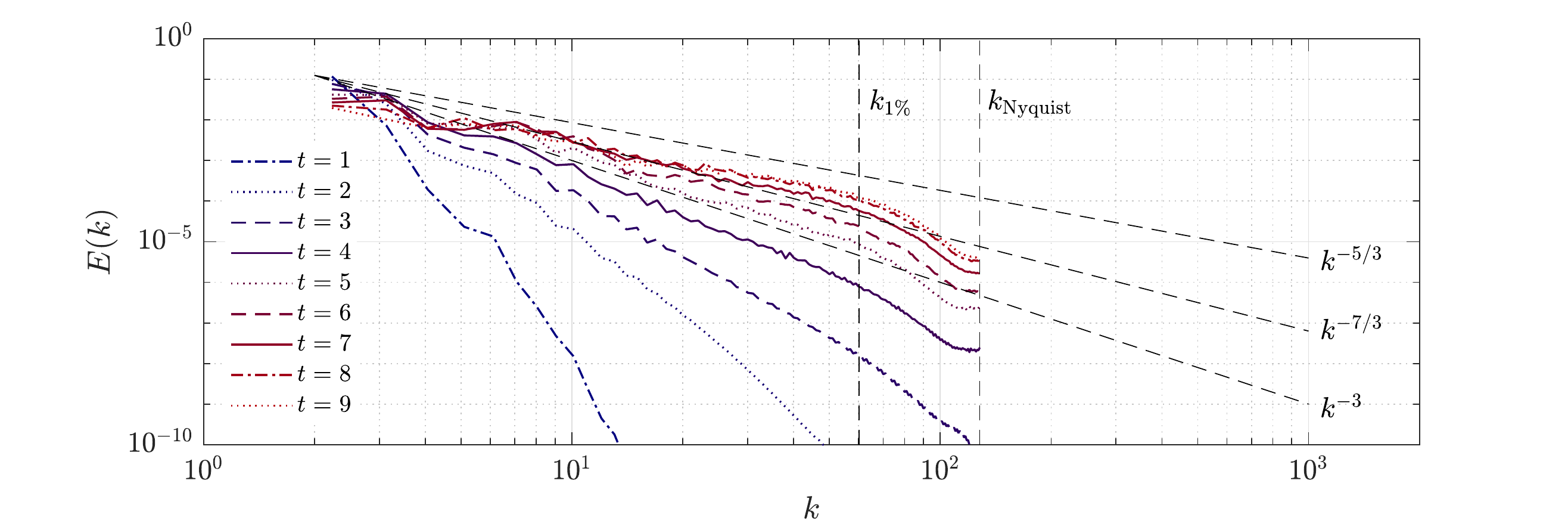}}
	\qquad
	\subfloat[Effective resolution of~$512^3$]{\includegraphics[width=0.9\textwidth]{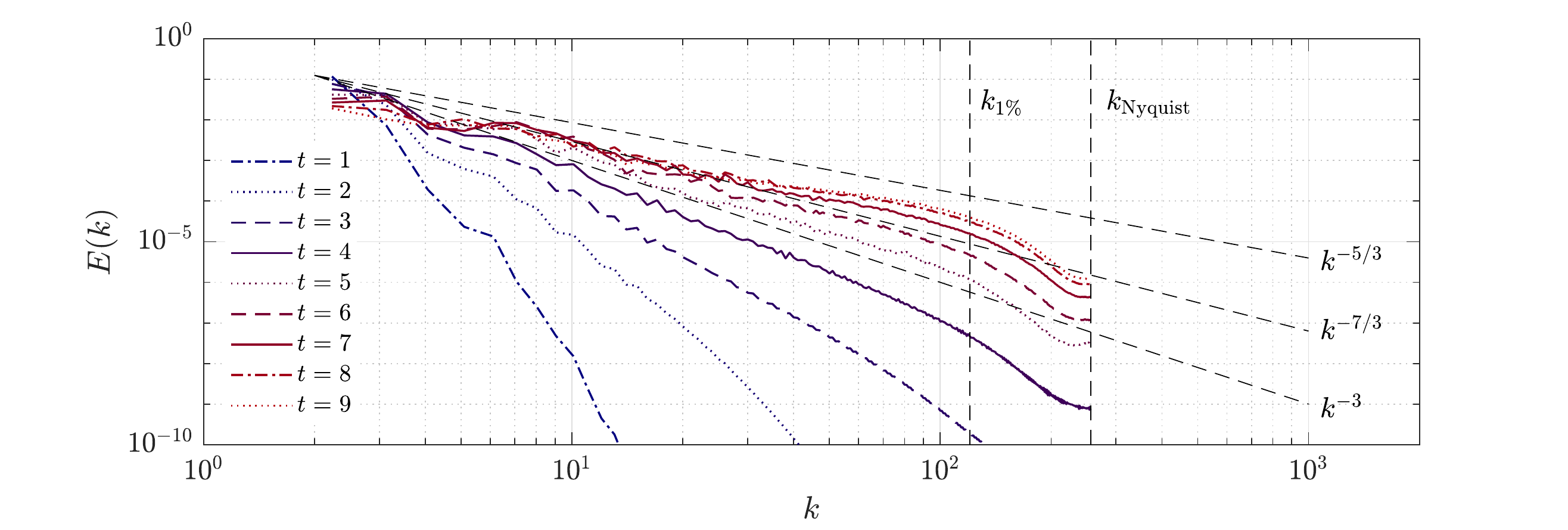}}
\caption{Three-dimensional inviscid Taylor--Green problem: Kinetic energy spectra for resolutions of~$128^3, 256^3, 512^3$ at times~$t=1,2,...,9$.}
\label{fig:3d_tgv_energy_spectra}
\end{figure}

\begin{figure}[!ht]
\ContinuedFloat
 \centering
	\subfloat[Effective resolution of~$1024^3$]{\includegraphics[width=0.85\textwidth]{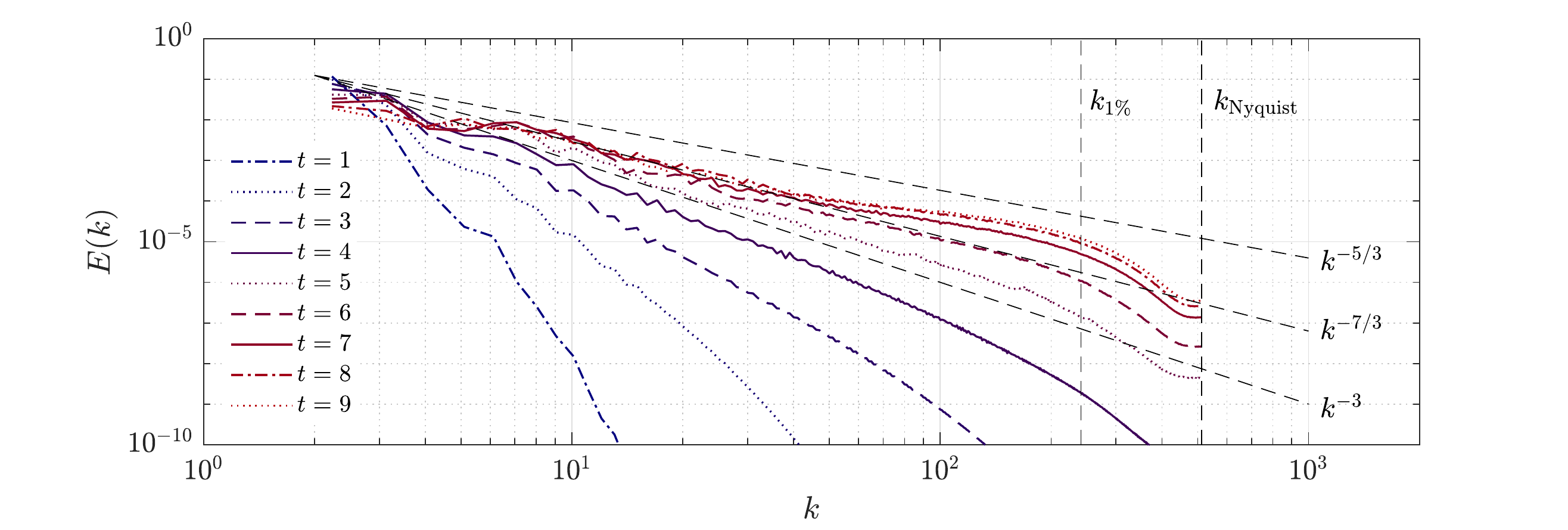}}
	\qquad
	\subfloat[Effective resolution of~$2048^3$]{\includegraphics[width=0.85\textwidth]{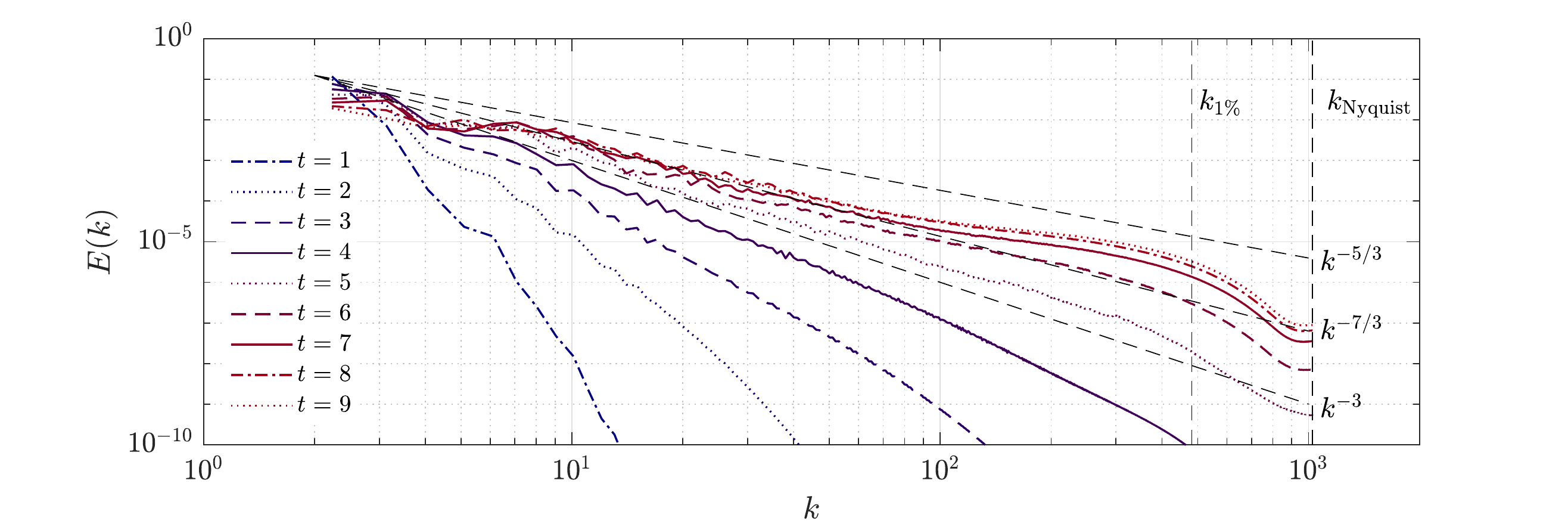}}
\caption{Three-dimensional inviscid Taylor--Green problem: Kinetic energy spectra for resolutions of~$1024^3, 2048^3$ at times~$t=1,2,...,9$.}
\label{fig:3d_tgv_energy_spectra}
\end{figure}

Figure~\ref{fig:3d_tgv_energy_spectra} shows kinetic energy spectra for increasing spatial resolution at various instances of time, namely at~$t=1,\hdots,9$ in steps of width~$1$. Results are shown for resolutions of~$128^3$ to~$2048^3$. High computational costs and memory requirements of the FFT part of our simulations did not allow us to perform the spectral analysis for the highest resolutions of~$4096^3$ and~$8192^3$. For a discussion of the general behavior of energy spectra as a function of time regarding the early time behavior~$t\leq 4$ we refer to the works of Brachet et al.~\cite{Brachet1983, Cichowlas2005, Bustamente2012}, where it is shown how the energy spectra can be fitted to function of the form~$E(k,t) = C(t) k^{-n(t)}\exp\left( - 2k \delta (t)\right)$ and where values obtained for~$n(t)$ and~$\delta (t)$ are discussed in detail. To verify the present results, we include references curves of slope~$n = 3$ (blow-up of enstrophy) and~$n = 5/3$ (Kolmogorov's inertial scaling law). The energy spectra are compared against the slope~$n = 3$ as a means to investigate the plausibility of a potentially singular behavior and to identify a time~$t = t_*$ at which such a blow-up could occur, as motivated at the beginning of this section. Once the flow has transitioned to a fully turbulent state, it can be expected that the energy spectrum exhibits some form of Kolmogorov scaling. For this purpose, we consider the energy spectra at times~$t=8$ and~$9$ where the maximum dissipation rate occurs, and compare their slopes against Kolmogorov's~$k^{-5/3}$ scaling. To quantify the resolution capabilities of the present discretization, we also plot the Nyquist wavenumber~$k_{\mathrm{Nyquist}}$ as well as the wavenumber~$k_{1\%}$ according to the~$1\%$-rule by Moura et al.~\cite{Moura2017} that aims at obtaining an accurate resolution limit for upwind-like DG discretizations for a specific polynomial degree of the function space. 

Figure~\ref{fig:3d_tgv_energy_spectra} shows that the range of scales resolved by the numerical method increases with increasing spatial resolution as expected theoretically, and that the resolution limit for polynomial degree~$3$ is described very well by the~$1\%$-rule corresponding to this degree. The energy spectra reach a slope of~$-3$ between~$t=4$ and~$t=5$, making our results consistent with the occurrence of singular behavior around that time ($4 < t_* < 5$), in agreement with the rapid growth of the maximum vorticity observed in the same time interval, see Figure~\ref{fig:3d_tgv_maximum_vorticity_and_enstrophy}. The spectra in the inertial range show a decay slightly stronger than~$k^{-5/3}$ and better agreement is achieved when compared to a~$k^{-7/3}$ scaling that is also shown in Figure~\ref{fig:3d_tgv_energy_spectra}, see~\cite{Piatkowski2019} for similar observations for viscous Taylor--Green vortex simulations. Towards the Nyquist wavenumber, a moderate pile-up of energy can be observed by comparison against the~$-5/3$ and~$-7/3$ references slopes before the energy falls off very rapidly. The energy pile up is characteristic of this type of high-order discontinuous Galerkin approach and becomes stronger for higher polynomial degrees, see~\cite{Moura2017, Moura2017setting} and references therein. This particular behavior is often the primary target when optimizing discretization schemes, see for example the recent studies~\cite{Flad2017,Winters2018,Manzanero2020} and references therein, and it is suggested to counteract this energy bump behavior by explicit sub-grid scale modeling. However, the recent study~\cite{Fernandez2018arxiv} shows that this topic is delicate and improvements in some quantities by the use of explicit sub-grid models cause deviations in other quantities such as the temporal evolution of the kinetic energy dissipation rate. In our opinion, the challenge lies in improving the spectral behavior and at the same time not giving up the improved resolution capabilities (per degrees of freedom) of high-order discretizations. The overall goal can be formulated as achieving a discretization method that is accurate w.r.t.~both the spectral behavior and the behavior in physical space, e.g., the temporal evolution of kinetic energy and its dissipation rate. We want to note that from such a holistic view it is not clear a priori whether an explicit sub-grid scale model optimizing the energy spectrum according to the inertial~$k^{-5/3}$ law is advantageous overall. We take up this point again in the outlook in Section~\ref{Discussion} discussing possible directions of future research. The energy spectra shown in~\cite[figure~9.15]{Schroeder2019} for an exactly energy-conserving discretization scheme illustrate that such a scheme leads to physical inconsistencies for the~$E(k)$-curves in the inertial range.

\subsection{Does the numerical dissipation have artificial or predictive character?}
\begin{figure}[t]
 \centering
	\includegraphics[width=0.49\textwidth]{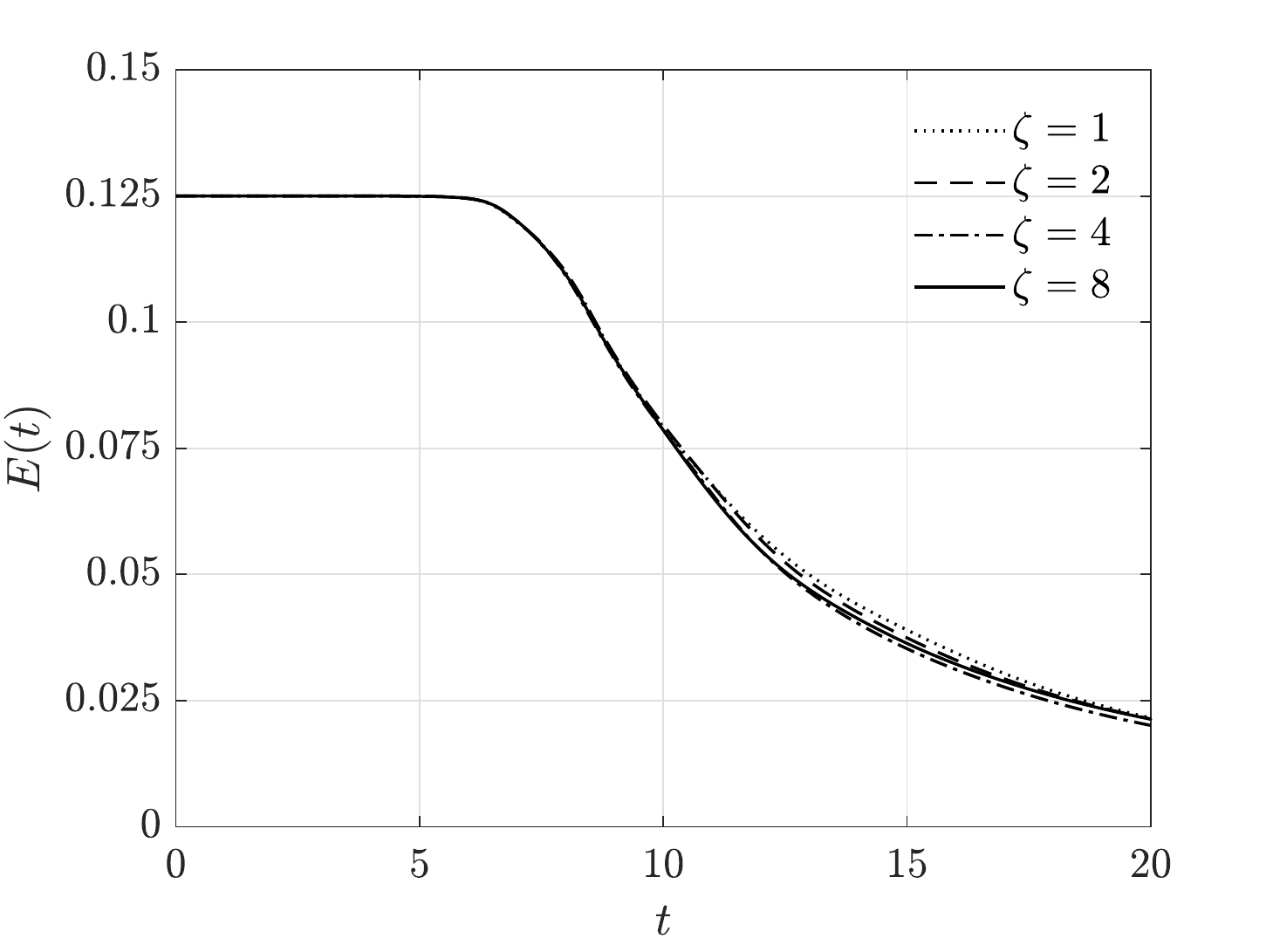}
	\includegraphics[width=0.49\textwidth]{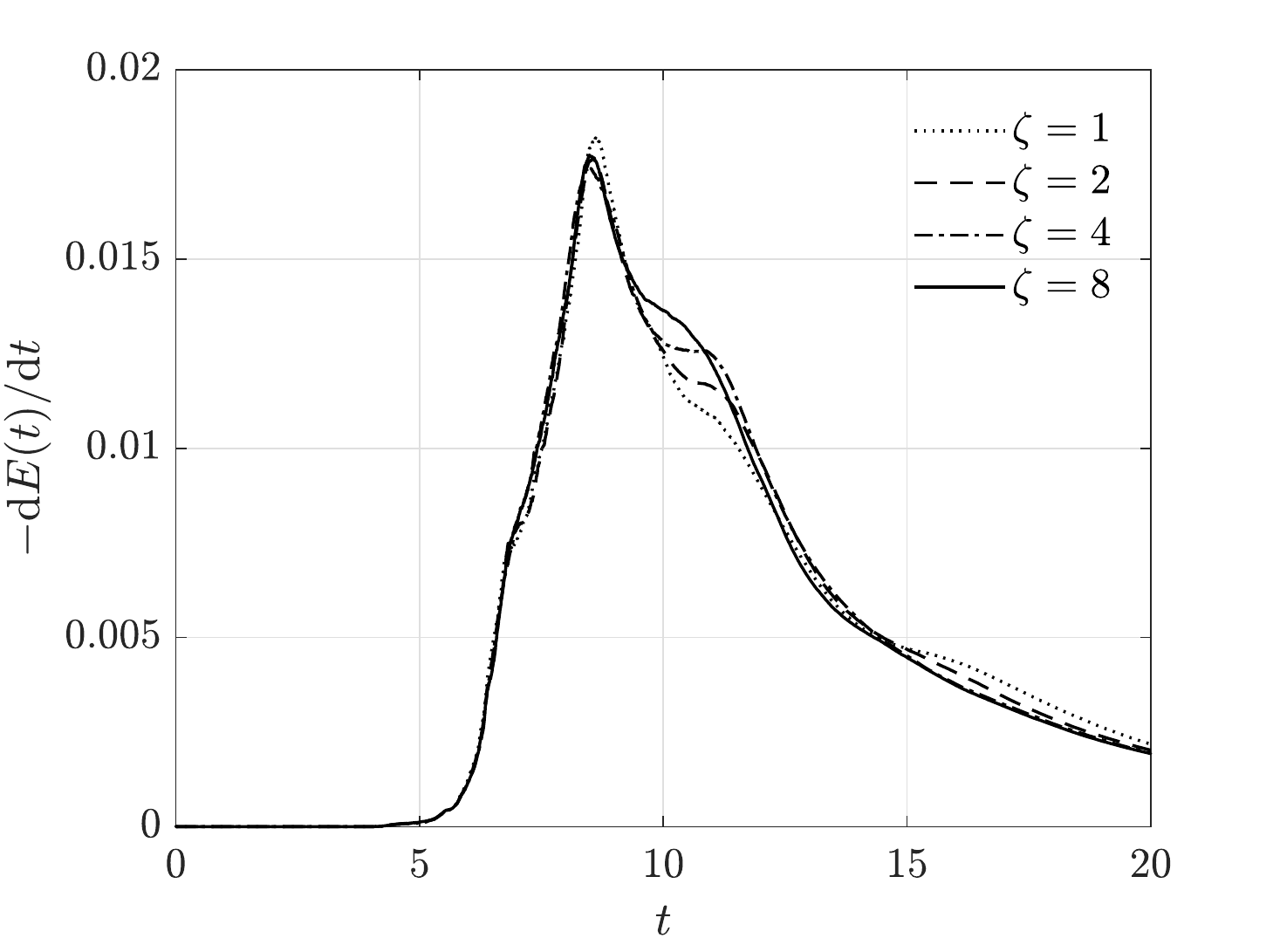}
\caption{Three-dimensional inviscid Taylor--Green problem: variation of penalty factor~$\zeta$ and its influence on the temporal evolution of the kinetic energy (left) and the kinetic energy dissipation rate (right) for~$1024^3$ resolution.}
\label{fig:3d_tgv_penalty_study}
\end{figure}
The dissipation of kinetic energy observed for the inviscid Taylor--Green simulations originates from the numerical method. At first sight, one might argue that changing the discretization scheme by choosing another numerical flux or varying certain parameters leads to results that are more or less dissipative, i.e., that the amount of dissipation is artificial and is determined by the discretization parameters. The results of the convergence study shown above do not support this point of view, and we want to give further insights putting confidence into the predictive character of these dissipative numerical solutions. In this context, it is illustrative to study the temporal evolution of the kinetic energy and its dissipation rate under a variation of parameters of the discretization scheme. Figure~\ref{fig:3d_tgv_penalty_study} shows a parameters study of the penalty factor~$\zeta$ of the divergence and continuity penalty terms of the present discretization, considering values of~$\zeta = 1,2,4,8$ by the example of the~$1024^3$ spatial resolution. Note that this parameter is the crucial one in stabilizing the method in the under-resolved and high-Reynolds regime, see~\cite{Fehn2018a, Fehn2019Hdiv}. We observe that the overall amount of dissipation as well as the dissipation maximum are essentially unaltered by a variation of this parameter. It is worth noting that the time of onset of dissipation is also not affected by the value of~$\zeta$. This has important implications. Assuming that a non-dissipative (energy-conserving) solution is the correct physical behavior and that the numerical dissipation is artificial, one might expect that an increase of the penalty parameter affects the numerical solution more strongly, e.g., changes the amount of overall dissipation or leads to a delayed onset of dissipation due to a better fulfillment of the divergence-free constraint for example. As we do not observe such a behavior, these results underline the predictive character of the present discretization scheme in our opinion. It is clear that we can not expect the results to be identical for different penalty factors, since the results are not converged for the chosen~$1024^3$ resolution, meaning that the deviation from a converged solution is mainly determined by the discretization scheme and, hence, its parameters.

\subsection{Are the results converged?}\label{sec:Convergence}
\begin{figure}[t]
 \centering
	\includegraphics[width=0.49\textwidth]{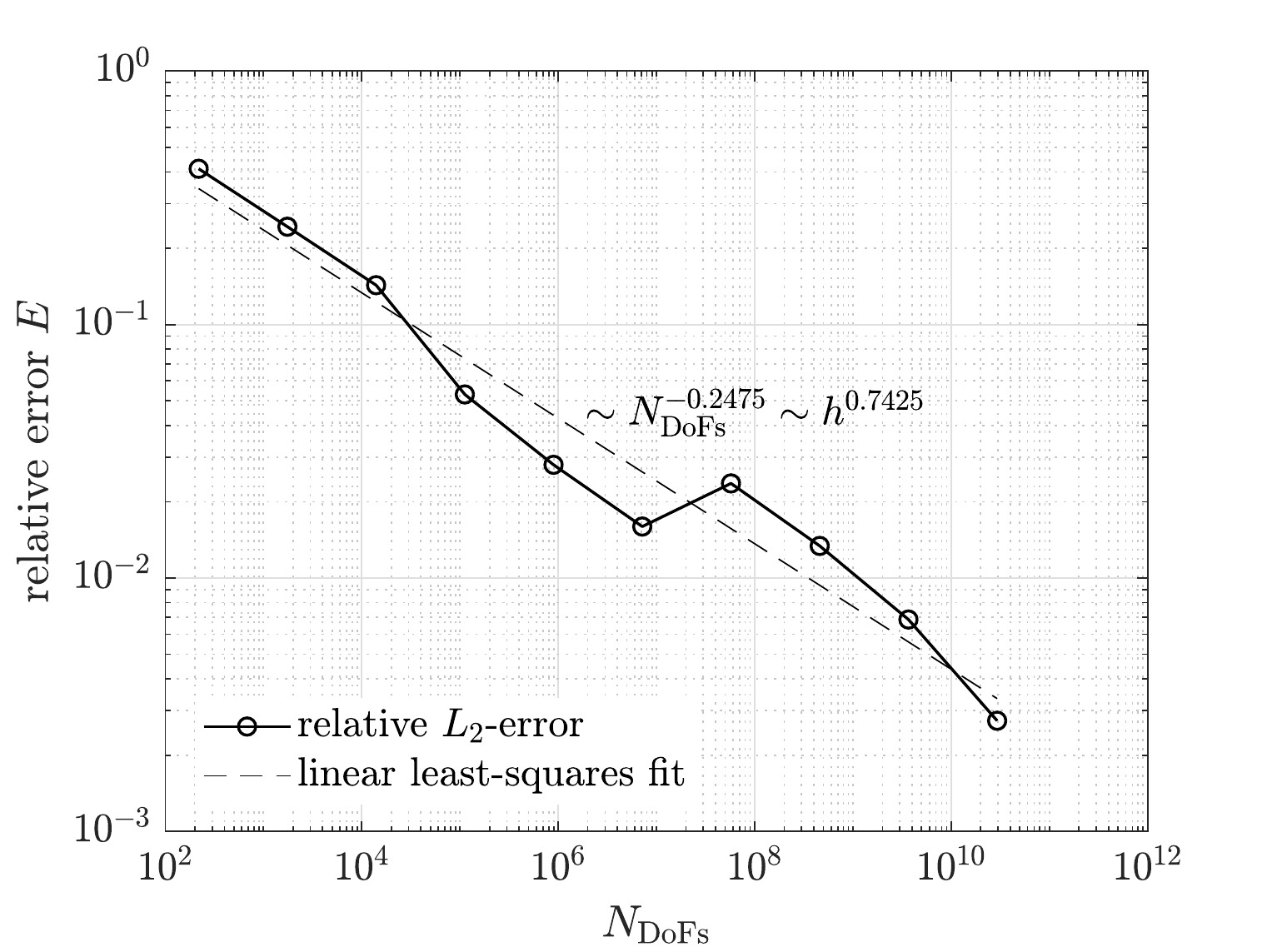}
	\includegraphics[width=0.49\textwidth]{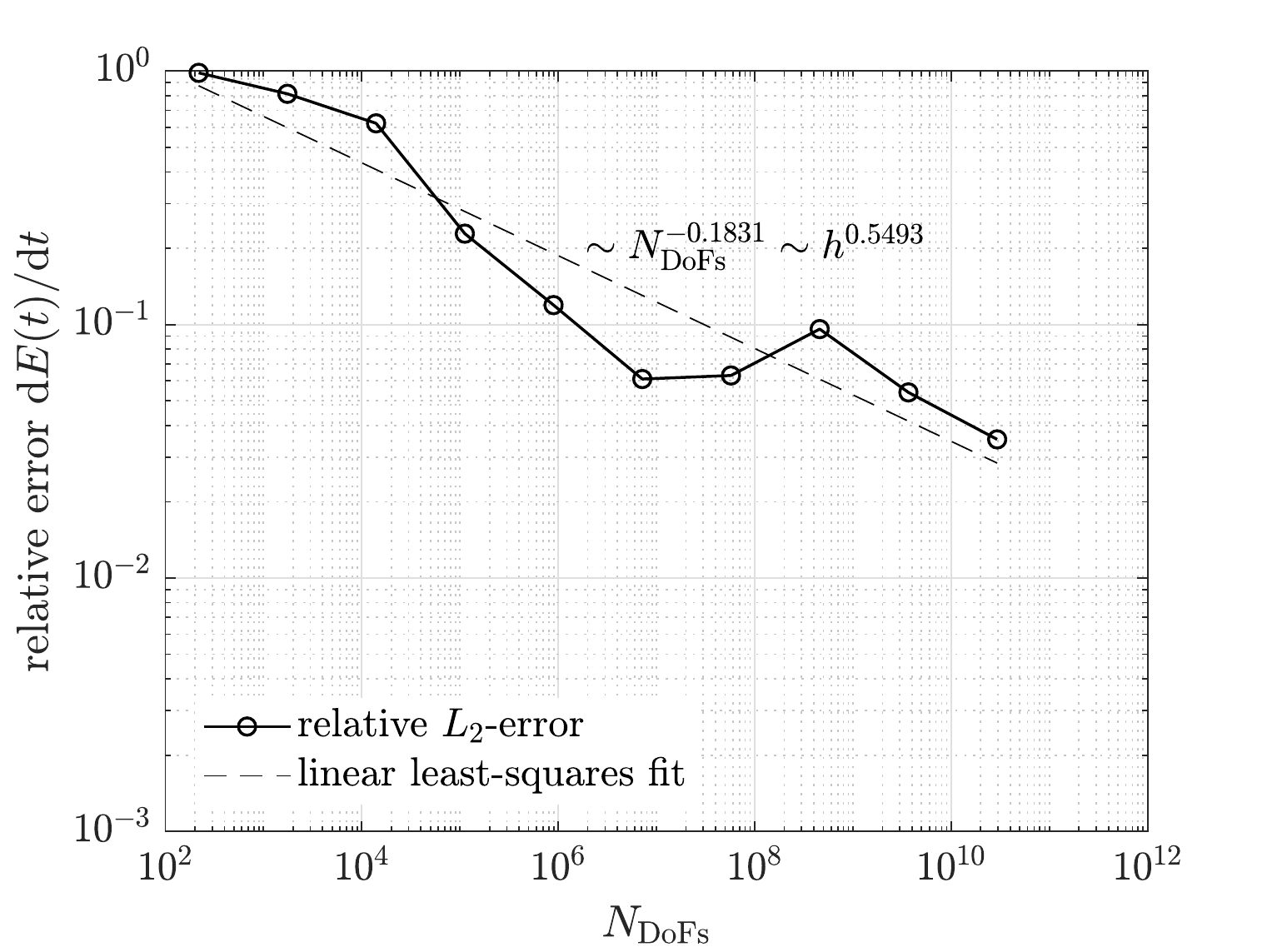}
\caption{Three-dimensional inviscid Taylor--Green problem: relative~$L_2$-errors of the temporal evolution of the kinetic energy (left) and the kinetic energy dissipation rate (right) for resolution ranging from~$8^3$ to~$4096^3$.}
\label{fig:3d_tgv_convergence}
\end{figure}
We finally address the question whether and to which extent the results presented for inviscid Taylor--Green simulations can be considered as converged. For this purpose, we compute relative~$L_2$-errors for the kinetic energy evolution and the kinetic energy dissipation rate
\begin{align*}
e_{E}^2 = \frac{\int_{t=0}^{T} \left(E(t) - E_{\mathrm{ref}}(t) \right)^{2} \mathrm{d}t}{\int_{t=0}^{T} \left( E_{\mathrm{ref}}(t) \right)^{2} \mathrm{d}t} \;\; , \;\;
e_{\mathrm{d}E/\mathrm{d}t}^2 = \frac{\int_{t=0}^{T} \left(  \frac{\mathrm{d}E(t)}{\mathrm{d}t} - \frac{\mathrm{d}E_{\mathrm{ref}}(t)}{\mathrm{d}t} \right)^{2} \mathrm{d}t}{\int_{t=0}^{T} \left( \frac{\mathrm{d}E_{\mathrm{ref}}(t)}{\mathrm{d}t} \right)^{2} \mathrm{d}t} \;\; .
\end{align*}
Since no analytical solution is available, the errors are measured using the finest resolution of~$8192^3$ as a reference~(ref). This implies that we can not compute the error for the~$8192^3$ resolution. The error of this simulation can only be roughly estimated by extrapolating the convergence trend observed for the coarser resolutions and assuming that this convergence behavior continues for the finest resolution. Defining a simulation with a relative error of~$1 \%$ or less as grid-converged, the results in Figure~\ref{fig:3d_tgv_convergence} reveal that we achieve grid-convergence in the kinetic energy evolution with errors below~$1 \%$. For the second finest resolution (the last data point in Figure~\ref{fig:3d_tgv_convergence}), the measured error is~$0.27 \%$. For the kinetic energy dissipation rate, the error is significantly larger, demonstrating that this quantity is more sensitive, in agreement with what has been observed in Figure~\ref{fig:3d_tgv_Re_study}. For the second finest resolution, the measured error is~$3.52 \%$. While the errors can be expected to be smaller for the finest resolution~$8192^3$ that is used as a reference solution here, we conclude that even finer resolutions would be required to achieve grid-convergence also for the kinetic energy dissipation rate in terms of the~$1 \%$ error level. Figure~\ref{fig:3d_tgv_convergence} also shows linear least-squares fits (e.g., $\log (E) \approx a \log (N_{\mathrm{DoFs}}) + b$ for the kinetic energy) to the data obtained from the numerical experiments, where a mean converge rate of approximately~$h^{3/4}$ is obtained for the kinetic energy and~$h^{1/2}$ for the dissipation rate.

\section{Conclusion and outlook}\label{Discussion}
Hunting for finite-time singularities in incompressible Euler flows is a challenging discipline. Searching for singular behavior in visualizations of three-dimensional simulation results is a challenge, and evokes the picture of finding a needle in the haystack, as it can be expected that singularities are very localized, will never be resolved by a numerical scheme, and the amount of data for large-scale three-dimensional simulations soon becomes cumbersome. For this reason, most previous studies focused on quantities such as the maximum vorticity, the analyticity strip width, fitting energy spectra to power law behavior, and related blow-up criteria. The present work raises the question whether a shift from local to global quantities such as the temporal evolution of the kinetic energy might ease such investigations, on the one hand to avoid geometrical complexities in visualization, and on the other hand to reduce computational costs due to a reduced sensitivity of certain quantities compared to others. We call this approach \textit{indirect} since it exploits the connection between singular behavior and anomalous energy dissipation according to Onsager's conjecture. A decisive point is that this technique requires suitable discretization schemes that remain robust in the presence of singularities and provide mechanisms of dissipation in case no viscous dissipation is present, which is a challenge in itself. Then, the idea is that observing an energy dissipating behavior for a sequence of mesh refinement levels provides insight into the physical dissipation behavior of the problem under investigation. 

We applied this technique to one-, two-, and three-dimensional problems. Results consistent with theory have been obtained in one and two space dimensions. Subsequently, this technique has been used to study the complex three-dimensional inviscid Taylor--Green problem, for which an energy dissipating behavior consistent with the high Reynolds number limit of viscous simulations available in the literature and consistent with Onsager's conjecture has been observed. Further confidence is put into the reliability of the numerical results for this challenging problem by the circumstance that the numerical method applied here is generic in the sense that it is a robust discretization scheme that remains numerically stable in the inviscid limit for all spatial resolutions that have been investigated, without the necessity to introduce new terms such as artificial dissipation to deal with the inviscid limit. This is an important prerequisite to draw conclusions about a potential physical blow-up. In contrast, a numerical simulation that blows up in finite time does not allow any conclusions, since the observed blow-up is due to numerical instabilities. In other words, a physical blow-up does not imply a numerical blow-up for a finite resolution numerical simulation.

The present study is able to demonstrate convergence to a dissipative solution for the three-dimensional inviscid Taylor--Green problem with a measured relative~$L_2$-error of~$0.27 \%$ for the kinetic energy and~$3.52 \%$ for the kinetic energy dissipation rate. The results for the temporal evolution of the kinetic energy can therefore be considered as grid-converged and might serve as accurate reference solution for future studies. Regarding the temporal evolution of the maximum vorticity and the enstrophy, we are able to resolve an increase of almost four orders of magnitude for both quantities. To the best of our knowledge, these are the highest resolution results published to date for the three-dimensional inviscid Taylor--Green vortex problem. The present study therefore complements theoretical and experimental works on Onsager's conjecture on anomalous energy dissipation and the related implications on finite-time singularities for three-dimensional incompressible Euler flows.

Applicability of high-order DG discretizations to large-scale problems in turbulence research, for which spectral methods are currently the state-of-the-art due to their accuracy and computational efficiency, has been demonstrated. Finally, the present study gives indications in terms of what a promising large-eddy simulation strategy might be, and contributes to the long-lasting and difficult discussion on explicit versus implicit sub-grid scale models. High-order discretizations that can be described as implicit LES have shown very promising results for moderate Reynolds number flows, but it is often argued that such techniques can be expected to finally need explicit sub-grid scale modeling once they are applied in the limit~$\mathrm{Re} \rightarrow \infty$. The present work contributes to this discussion by investigating a high-order discontinuous Galerkin discretization without explicit model in the inviscid limit. A purely numerical (implicit) approach based on a powerful discretization method does not introduce assumptions like homogeneous isotropic turbulence on which classical eddy-viscosity sub-grid scale models rely or for which such models have been calibrated. Assuming that such an implicit approach gives physical results in the inviscid limit, e.g., consistent with Onsager's conjecture on anomalous energy dissipation, it makes one more confident that such a method is able to naturally account for more complex physical mechanisms in turbulence beyond K41 theory~\cite{Dubrulle2019}. Taking as an alternative an energy-conserving numerical method with explicit sub-grid scale model, the anomalous energy dissipation has to be realized by the sub-grid model. We therefore believe it is an interesting future research direction to take the results shown in the present study for the inviscid Taylor--Green problem as a reference for further validation or to perform comparative studies between explicit and implicit LES techniques for the highest Reynolds number case, the inviscid limit.

\appendix

\section*{Acknowledgments}
The authors thank Gert Lube and Oliver Neumann for inspiring discussions on this topic. The research presented in this paper was partly funded by the German Research Foundation (DFG) under the project ``High-order discontinuous Galerkin for the EXA-scale'' (ExaDG) within the priority program ``Software for Exascale Computing'' (SPPEXA), grant agreement no. KR4661/2-1 and WA1521/18-1. The authors gratefully acknowledge the Gauss Centre for Supercomputing e.V.~(\texttt{www.gauss-centre.eu}) for funding this project by providing computing time on the GCS Supercomputer SuperMUC-NG at Leibniz Supercomputing Centre (LRZ, \texttt{www.lrz.de}) through project id pr83te. Further, collaboration with and contributions of the~\texttt{deal.II} community is gratefully acknowledged.


\bibliography{paper}

\end{document}